\documentclass[11pt,preprint]{aastex}
\usepackage{graphicx}
\usepackage[rotateright]{rotating}
\shorttitle{Catalogue of High PS Regions}
 \shortauthors{Li et al.}

\begin{document}

\title{Catalogue of High Protostellar Surface Density Regions in Nearby Embedded Clusters }

\author{Juan Li\altaffilmark{1, 2}, Philip C. Myers\altaffilmark{3}, Helen Kirk\altaffilmark{4, 5}, Robert A. Gutermuth\altaffilmark{6}, Michael M. Dunham\altaffilmark{7, 3}, Riwaj Pokhrel\altaffilmark{6, 3}}

\altaffiltext{1}{Shanghai Astronomical Observatory, Chinese Academy of Sciences, 80 Nandan RD, Shanghai 200030, China; lijuan@shao.ac.cn}
\altaffiltext{2}{Key laboratory for Radio Astronomy, Chinese Academy of Science, 2 West Beijing Road, Nanjing, JiangSu 210008, China}
\altaffiltext{3}{Harvard-Smithsonian Center for Astrophysics, 60 Garden Street, Cambridge, MA 02138, USA}
\altaffiltext{4}{Department of Physics and Astronomy, University of Victoria, 3800 Finnerty Road, Victoria, BC, Canada V8P 5C2}
\altaffiltext{5}{Department of Physics and Astronomy, University of Victoria, Victoria, BC, V8P 1A1, Canada}
\altaffiltext{6}{Department of Astronomy, University of Massachusetts, Amherst, MA 01003, USA }
\altaffiltext{7}{Department of Physics, State University of New York at Fredonia, 280 Central Ave, Fredonia, NY 14063, USA }

\begin{abstract}

We analyze high-quality stellar catalogs for 24 young and nearby (within 1 kpc) embedded clusters and present a catalogue of 32 groups which have a high concentration of protostars. The median effective radius of these groups is 0.17 pc. The median protostellar and pre-main sequence star surface densities are 46 M$_{\odot}$ pc$^{-2}$ and 11 M$_{\odot}$ pc$^{-2}$, respectively. We estimate the age of these groups using a model of constant birthrate and random accretion stopping and find a median value of 0.25 Myr. Some groups in Aquila and Serpens, Corona Australia and Ophichus L1688 show high protostellar surface density and high molecular gas surface density, which seem to be undergoing vigorous star formation. These groups provide an excellent opportunity to study initial conditions of clustered star formation. Comparison of protostellar and pre-main-sequence stellar surface densities reveal continuous low-mass star formation of these groups over several Myr in some clouds. For groups with typical protostellar separations of less than 0.4 pc, we find that these separations agree well with the thermal Jeans fragmentation scale.  On the other hand, for groups with typical protostellar separations larger than 0.4 pc, these separations are always larger than the associated Jeans length.
\end{abstract}

\keywords{infrared: stars - stars: formation - stars: pre-main sequence}

\section{Introduction}

It is now commonly accepted that most stars form in clusters of hundreds of stars (Reipurth et al. 2014; Lada \& Lada 2003). Understanding the process of forming stars in clusters is of considerable importance. Although we have an increasingly detailed picture of relatively isolated star formation in nearby dark clouds, such as those in the Taurus complex, many gaps remain in our understanding of clustered star-formation, such as the elusive initial conditions (Myers 2010), as well as the domainant fragmentation process (e.g. Zhang et al. 2015; Busquet et al. 2016; Pokhrel et al. 2018). 

The embedded clusters, in which the mass of the clusters are dominated by the mass of their natal molecular clouds, are the primary laboratory for research into the question of the physical origin of stellar clusters (Megeath et al. 2016; Friesen et al. 2016; Gutermuth et al. 2009; Lada 2010). In particular, regions with high protostellar (PS) surface density and PS fraction are very young, and therefore, they could be useful objects for understanding the star formation and evolution in star clusters. Through observations of gas associated with these regions, the initial conditions for the clustered star formation can be constrained. 

Our understanding of the formation and early evolution of young stellar clusters has been greatly hindered by observational challenges, including their distance, their spatial density, and their association with high column density molecular clouds (Gutermuth et al. 2005). High angular resolution and high sensitivity are required to resolve individual stars, detect embedded sources and identify members against a field of background stars. With the generation of mid-IR telescopes especially the {\it Spitzer} Space Telescope (Werner et al. 2004), observations can finally probe nearby young clusters with the sensitivity to detect objects well below the hydrogen-burning limit and the angular resolution to resolve high-density groupings of star (Allen et al. 2007). Furthermore, {\it Spitzer} has been providing detailed images of young clusters in the mid-IR, which for the first time allows to identify young stars with disks (pre-main-sequence objects, hereafter PMS objects) and infalling envelopes (PS objects) efficiently in clusters out to 1 kpc and beyond (e.g., Gutermuth et al. 2009; Evans et al. 2009; Kryukova et al. 2012; Dunham et al. 2015). In addition, the {\it Herschel} Space Observatory (Pilbratt et al. 2010) also provide dust column density maps for nearby star-forming regions with high sensitivity and angular resolution (Harvey et al. 2013; Pokhrel et al. 2016), making it possible to study the relationship between young stellar clusters and their natal gas in detail. Extensive studies have been carried out to study the relation between the surface density of young stellar objects (YSOs) and gas density with the {\it Spitzer} and {\it Herschel} data, and power-law correlation has been reported (e.g. Evans et al. 2009; Gutermuth et al. 2009; 2011; Heiderman et al. 2010; Harvey et al. 2013).

Jeans fragmentation is known to be an important phenomenon in star-forming regions (Jeans 1929). The detailed fragmentation mechanism is a topic that continues to be under debate, with possibilities including purely thermal Jeans fragmentation,  as well as Jeans fragmentation where thermal and non-thermal motions play a role (e.g., Palau et al. 2015; Busquet et al. 2016). Some studies of massive IRDCs found that the fragments have masses much larger than the thermal Jeans mass and seem to be consistent with the non-thermal Jeans mass (e.g., Zhang et al. 2009, 2015; Pillai et al. 2011; Wang et al. 2014). On the contrary, Palau et al. (2015) found that thermal Jeans fragmentation seems to be the dominant factor that determines the fragmentation level of relatively nearby star-forming massive dense cores at a 0.1 pc scale. They proposed that the inconsistency between mass and thermal Jeans mass in other studies could be caused by the low sensitivity and poor spatial resolution due to the large distance, e.g., the mass sensitivity is above the Jeans mass ($>$ 2 M$_{\odot}$), and the spatial resolution is $>$ 5000 au for most of IRDCs. In Palau et al. (2015) the massive dense cores were observed with mass sensitivities $<$ 1 M$_{\odot}$, and spatial resolutions of about 1000 au. Busquet et al. (2016) assessed the fragmentation level in a IRDC with the SMA combined data that is sensitive to structures of 3000-10,000 au, and also sensitive to flattened condensations. They also found that the observed fragmentation in the hub of an IRDC is more consistent with thermal Jeans fragmentation. 

Recently Pokhrel et al. (2018) studied the hierarchical structure in the Perseus molecular cloud from the scale of the entire cloud to protostellar objects. This study is carried over five scales of hierarchy-cloud, clumps, cores, envelopes and protostellar objects. Their results provide clues that the thermal Jeans fragmentation begins to dominate at the scale of cores fragmenting into envelopes. More young stellar groups are needed for further investigate the fragmentation mechanism in the star-formation process. In this paper, we present a catalogue of high PS fraction groups in nearby embedded clusters using a catalogue of YSOs extracted from the {\it Spitzer} c2d and Gould Belt Legacy surveys (Evans et al. 2009; Gutermuth et al. 2018, in preparation). In Section 2, we describe the YSO catalogs we use in our analysis, and our procedure for identifying groups with high PS surface densities and PS ratios. We analyze the properties of the groups identified in Section 3, then discuss the implications in Section 4, and conclude in Section 5.

\section{The Sample and Methodologies}
\label{sample}

\subsection{The Sample}

The Gould Belt (GB) is a ring of nearby O-type stars inclined approximately 20$^{\circ}$ with respect to the Galactic Plane (Herschel 1847; Gould 1879), in which most of the current star formation within 500 pc of the Sun occurs. All of the nearby, well-studied molecular clouds are located within this ring. The GB ring has been surveyed by the {\it Spitzer} Space Telescope "cores to disks" (c2d; Evans et al. 2003, 2009) and "Gould Belt" (GB) Legacy surveys  (Dunham et al. 2015). The {\it Spitzer} c2d survey (PI: N.J. Evans) imaged five large, nearby molecular clouds in the GB, including Serpens, Perseus, Ophiuchus, Lupus, and Chamaeleon II, as well as approximately 100 isolated dense molecular cores (Evans et al. 2003, 2009). The {\it Spitzer} GB survey (PI: L.E. Allen) is a follow-up program that imaged the additional 11 molecular clouds in the GB, completing most of the remaining clouds in the GB except for the Taurus and Orion molecular clouds. Both surveys imaged molecular clouds at 3.6-8.0 $\mu$m with the {\it Spitzer} Infrared Array Camera (IRAC; Fazio et al. 2004), and at 24-160 $\mu$m images with the Multiband Imaging Photometer (MIPS; Rieke et al. 2004). Here we use YSOs catalogues classified by R. A. Gutermuth et al. (2018, in preparation), in which YSOs were identified following methods described in Gutermuth et al. (2008, 2009) with many improvements (Winston et al. 2018). The classification of YSOs are identified based on the spectral index of their spectral energy distribution (Gutermuth et al. 2018, in preparation). These data provide a comprehensive sample of clusters in the solar neighbourhood and good opportunity to analyze regions with high PS surface density and fraction. In this paper, both Class 0/I and Flat Class are referred as PS stars, while Class II and transition disk (TD) are referred as PMS stars. 

\subsection{The Methodologies}

In order to select high PS fraction regions, an objective way is required to identify such regions. One simple method which does not rely on the definition of stellar groups is to use the local surface density, $\Sigma$ (Kirk \& Myers 2012). If the projected separation from the star to its $n$th nearest neighbor is $r_n$, then the local stellar surface density is
\begin{equation}
\Sigma=\frac{n-1}{\pi r^2_n}.
\end{equation}
The fractional uncertainty in $\Sigma$ varies as $(n-2)^{0.5}$; higher values of $n$ give a lower spatial resolution, but smaller fractional uncertainty (Casertano \& Hut 1985, Gutermuth et al. 2005). We adopted $n=4$ in this paper to give a good compromise between resolution and uncertainty. We calculate the PS surface density, $\Sigma_{PS}$, at every pixel in the map, using the distance to the fourth nearest protostar, $d_{PS}$, using equation 1. The mean mass of YSOs are assumed to be 0.5 M$_{\odot}$ (Evans et al. 2009). 

%The way to calculate PS surface density is: for each pixel, calculate its distance from each PS, find the fifth nearest distance $d_{PS}$, thus $\Sigma_{PS}= (5-1)/(\pi d_{PS}^2)$. The way to calculate PS fraction: Firstly, calculate YSO surface density. For each pixel, calculate its distance from each YSO, find the fifth nearest distance $d_{YSO}$, thus $\Sigma_{YSO}= (5-1)/(\pi d_{YSO}^2)$; Secondly, for each pixel, find $n_{PS}$, the number of PS located within $d_{YSO}$, then $pf=n_{PS}/n_0$.

Since the survey areas are very large, we first identify regions with abundant PS objects visually for each cloud, then perform surface density analysis to identify groups with high PS fraction. Information for these groups are listed in Table 1.

\section{Results}
\label{results}

\subsection{Identification of high PS surface density groups}

 We analyzed the local surface density of PS and PMS objects toward the sample of Gutermuth et al. (2018, in preparation). Figures 1-17 present PS objects surface density maps and PS surface density overlaid on the extinction map or column density map. In all figures, the contours represent the surface density of the PS stars. The PS, Class II and TD stars are overlaid in red circles, blue circles and yellow circles, respectively. We are interested in zones which have both a high PS fraction and a high concentration of protostars. Using the PS surface density map, we examine contours at 20\%, 10\%, 5\% and 2.5\% of the peak PS surface density, and select clusters of interest using the smallest contour value (i.e., largest area) which visually encompasses only one local concentration of protostars. We then measure the area, the number and average surface density of PS and PMS objects, and the average surface density and mass of gas contained within the identified regions. We identify a total of 32 PS-groups using this method. There are five groups that have only PS objects, including Aquila-b5, Musca, Perseus-b2, Serpens-a1, and Serpens-b2. These clusters should be very young. 

In Table 2, we present a catalogue of high PS groups, which stand for clusters with both the high PS surface density and PS fractions. The location, the peak surface density of PS objects, the contour adopted, the effective radius, the number of PS objects, the percentage of PS objects formed within high PS groups, and the number of PMS objects within high PS groups are present. The values are measured as follows. The position of the density center was defined as the density-weighted average of the positions of the stars, in a similar way as von Hoerner (1963):
\begin{equation}
\textbf{x}_{d,j} =\frac{\sum_i \textbf{x}_{i}\Sigma_{j}^{(i)}}{\sum_i \Sigma_{j}^{(i)}},
\end{equation}
where $\Sigma_{j}^{(i)}$ is the density estimator of order $j$ around the $i$th particle, and $\textbf{x}_{i}$ is the two-dimensional position vector of the $i$th star. 
The contour is the surface density fraction used to select the high PS surface density groups (i.e., 20\%, 10\%, 5\%, or 2.5\% ). The effective radius of the high PS surface density groups $r_{eff}$, is the square root of the area (divided by $\pi$) of the selected region. The effective radius of the groups range from 0.02 to 1.10 pc.

\subsection{Relationship of the PS surface density with gas}

To investigate relationship of the PS groups and gas, we made use of {\it Herschel} Gould Belt survey to derive the gas column density for most groups (Andr$\acute{e}$ et al. 2010). The column densities and temperatures are derived on a pixel-by-pixel basis by fitting a greybody model to the resolved emission at 160, 250, 350, and 500 $\mu$m with a fixed power law (Pokhrel et al. 2016). The two exceptions are Auriga/CMC and Aquila-1.The Auriga/CMC data also come from the {\it Herschel} observations (Harvey et al. 2013). The Two Micron All Sky Survey (2MASS) derived extinction map of Lombardi (2009) and Lombardi \& Alves (2001) is used to derive the column density of molecular gas for Aquila-1. The average column density of molecular gas within selected regions, $\Sigma_{gas}$, is calculated using $\Sigma_{gas}=\overline{A_K}\times 8.5 \times 4.40\times10^{-3}$ g~cm$^{-2}$ (Kirk et al. 2006; Boogert et al. 2013). Pokhrel et al. (2016) compared 2MASS Av maps with the {\it Herschel} column density and found good agreement over A$_V$=1$\sim$ 8 mags. We only used 2MASS for Aquila-1 (where Herschel data was not available), in which A$_K$=0.81, and A$_V\sim$ 6.88, so the column densities in this regime should be reasonably well probed by both the 2MASS and {\it Herschel} observations (Pokhrel et al. 2016). For the remaining regions, {\it Herschel} data is better because of its unprecedented angular resolution and sensitivity in the far-IR (Andre et al. 2010). In addition, {\it Herschel} is much better at tracing higher column densities where A$_V$ maps saturate.
We compute the position of nearby gas column density peak, average extinction, and average molecular gas column density density of identified groups and present them in Table 3. All of these parameters are obtained in Starlink (Currie et al 2014). The uncertainty of average column density is the statistical standard deviation obtained using Starlink GAIA. The gas and dust tend to be distributed in filament, which much more compact than selected region, therefore the standard deviation in the column density values are large.

We can see from the column density maps that the PS groups are always aligned with dense filaments. However, the PS surface density peaks do not coincide with the gas column density peaks in most cases. We measure the angular separation between the PS surface density peak and its nearest neighbor gas column density peak, and present the result in Table 3. The uncertainty of the separation listed is the pixel size of the PS surface density map. The largest separations between PS surface density peaks and their nearby gas column density peaks comes from Chamaeleon I, in which the separation is up to $\sim12$ arcmin. The separation between two peaks are larger than twice the uncertainty for most PS-clusters. From a visual inspection, we note that the regions with a high PS fraction are overdense in stars relative to the molecular gas density. Such a phenomenon has been noted by Gutermuth et al. (2011) and Kirk et al. (2011). They proposed that the immediate environments of the YSOs have finished star formation process, but there were still reservoirs of material nearby which are capable of forming a significant number of new stars.  

\subsection{Properties of high PS surface density groups}

We compute the average PS and PMS surface density, PS ratio, age, free fall time, and present them in Table 4. The value of the average PS surface density, $\Sigma_{PS}$, is simply the number of PS objects divided by the area of the selected region ($N_{PS}\times 0.5/\pi r_{eff}^2$ M$_{\odot}$ pc$^{-2}$), where the mean PS mass is assumed to be 0.5M$_{\odot}$. $\Sigma_{PS}$ ranges from 2.24$\times10^{19}$ to 1.07$\times10^{23}$ cm$^{-2}$ with the highest PS surface density being associated with Aquila and Serpens. Similarly, the average PMS surface density, $\Sigma_{PMS}$, is the number of PMS stars divided by the area of the selected region. The relative uncertainty for n=4 surface density map is $\frac{1}{\sqrt{n-2}}$, $\sim$ 0.7 (Casertano \& Hut 1985). The fraction of PS objects, $f_{PS}$, is the number of PS objects divided by total number of YSOs within the high PS surface density groups. 
\begin{equation}
f_{PS}=N_{PS}/(N_{PS}+N_{PMS}).
\end{equation}
The age of the groups, $t$, is derived using the model of Myers (2012):
\begin{equation}
f_{PS}=\frac{1-exp(-t/a)}{t/a}, 
\end{equation}
where a=0.2 Myr. Myers (2012) developed a cluster evolution model for cluster age estimation. They assumed a constant PS birthrate, core-clump accretion, and equally likely accretion stopping in the model. The cluster ages could be obtained from the observed numbers of PS and PMS objects. This method of age estimation is simpler than optical spectroscopy that are derived from stars' luminosities, spectral types, and evolutionary tracks on the color-magnitude diagram (da Rio et al. 2009; Reggiani et al. 2011). In addition, the Myers 2012 age estimation method can be applied to young embedded clusters where optical spectroscopy is not possible.

Six of the groups studied here have also been analyzed by Gutermuth et al. (2009), including Auriga/CMC-3, Ophiuchus L1688, Corona Australis, Chameleon I, NGC 1333, and Serpens-a, corresponding to LkH$\alpha$101, L1688, CrA, Cha I, NGC 1333, and Serpens in Gutermuth et al. (2009). Gutermuth et al. (2009) obtained surface density maps of YSOs including Class I and II objects (see Table 8 in Gutermuth et al. 2009). The surface density map of PS objects used here seem to be consistent in their morphology with those of YSOs in Gutermuth et al. (2009). However, there are slight differences in the protostellar identification results in Gutermuth et al. (2009) and Gutermuth et al. (in preparation). For example, Gutermuth et al. (2018, in preparation) identified four PS objects in Chamaeleon I-a, while Gutermuth et al. (2009) identified only two PS objects in similar region. %The size of identified regions in Gutermuth et al. (2009) also differ from regions in this work {\bf since they identified regions with high surface density of YSO, including class I and II.} Thus the derived surface densities are not exactly the same. 

We plot the PS mass surface density versus the molecular gas mass column density in Figure 18. We classified the groups into five categories using the PS ratio. There are eleven groups with a PS ratio larger than 0.8 (34 percent), three groups with a PS ratio range from 0.6 to 0.8 (9 percent), ten groups with PS ratio range from 0.4 to 0.6 (31 percent), six groups with PS ratio range from 0.2 to 0.4 (6/32 percent), and two groups with PS ratio smaller than 0.2 (2/32 percent). We can see from Figure 20 that most of the cluster-forming regions tend to be forming stars at a fairly moderate rate, and cluster around the $\Sigma_{PS}$/$\Sigma_{gas}$=0.2 line at the bottom of the plot. Meanwhile, there are a few groups which are undergoing very vigorous star formation at the upper part of the plot and around the $\Sigma_{PS}$/$\Sigma_{gas}$=1 line, including Serpens-b3 (1989), Serpens-b2 (1592), Aquila-b2 (1382 ), Aquila-b3 (1169), Aquila-b6 (688), Corona Australis (428), Aquila-b5 (398), Oph L1688-1 (382). Aquila-b, which is also referred as Serpens South, was already known to be an interesting and unusual example because of high PS surface density and high PS fraction (Gutermuth et al. 2008).  With properties similar to Serpens and Aquila, Corona Australis and Oph L1688 also appear to be interesting targets for studies of earliest stages of clustered star formation.

\subsection{Statistical Properties of PS groups}

Statistics of PS group properties provide us information about the typical physical conditions of young stellar clusters within the nearest kiloparsec. Figure 19 presents a series of histograms showing how properties are distributed among these groups. The distribution of $r_{eff}$ is plotted in the top-left histogram of Figure 19, which shows that despite a significant tail of large values, most of the PS groups lie in a relatively narrow peak between 0.02-0.3 pc in radius, with a median value of 0.17 pc. This is smaller than median effective radius of these cluster cores (0.39 pc, Gutermuth et al. 2009), as we only calculated effective radius of the high PS fraction region. 

Figure 19(b) shows distribution of PS counts.
The PS counts are highly peaked, with a median value of 6. The major outlier on the far right end of Figure 19(b) comes from NGC1333, in which the number of PS objects is as large as 35.

Figure 19(c) shows distribution of surface density of PS objects.
In agreement with Gutermuth et al. (2009), the surface density of PS and PMS objects are skewed to low values, with median values of 46 and 11 M$_{\odot}$ pc$^{-2}$, respectively. The densest PS groups are Serpens-b3,  Serpens-b2, Aquila-b2 and Aquila-b3 , with $\Sigma_{PS}$ value of 1989, 1592, 1382 and 1169 M$_{\odot}$ pc$^{-2}$, respectively.
As is stated above, Aquila-b, which is also referred as Serpens South, was known to be an interesting cluster with high PS surface density and PS fraction (Gutermuth et al. 2008). 

Figure 19(d) shows distribution of surface density of PMS objects. The major outlier on the far right end of Figure 19(d) comes from Aquila-b5.

The PS ratio in PS groups ($f_{PS}$) is presented in Figure 19(e). The median value of the PS ratio is 0.58. The highest PS ratios come from Aquila-b3, Aquila-b4, Aquila-b6, Perseus-b2, Serpens-a1 and Serpens-b2, in which only PS objects are seen. 

The group age is presented in Figure 19(f). Most of these groups are younger than 0.5 Myr, with a median value of 0.25 Myr. The youngest groups are regions with the highest PS fractions. The major outlier on the far right end of Figure 19(f) is Auriga/CMC-3, which is the eldest group among the sample.

Figure 19(g) presents the distribution of the molecular gas column density, with a median value of 1.6$\times10^{22}$ cm$^{-2}$. The densest group is Aquila-b3, with a column density of 1.2$\times10^{23}$ cm$^{-2}$. As the histogram shows, most of the groups lie in the range 1.0-4.0$\times10^{22}$ cm$^{-2}$. The groups with the most diffuse gas are Cepheus and IC5146-b, with values of (2.75$\pm$2.96)$\times10^{21}$ and (2.85$\pm$1.76)$\times10^{21}$ cm$^{-2}$, corresponding to 51$\pm$55 and 53$\pm$35 M$_{\odot}$ pc$^{-2}$.  This result seems to be lower than previous observational and theoretical studies (e.g. Gutermuth et al. 2011; Heiderman et al. 2010; Lada, Lombardi \& Alves 2010; McKee 1989). Recently, based on {\it Herschel} data of the Lupus complex, Benedettini et al. (2018) found that most prestellar cores are found above $\sim$3$\times10^{21}$ cm$^{-2}$. They argue that the column density threshold should be interpreted more as a level over which a higher probability exists to find prestellar cores rather than a stringent limit under which star formation is inhibited.  

Figure 19(h) shows distribution of $\Sigma_{*}/\Sigma_{gas}^2$, with a median value of 0.00037 pc$^2$ M$^{-1}_{\odot}$, indicates a relative young and overdensity in gas (Gutermuth et al. 2011) of these regions. The major outlier on the far right end of Figure 19(h) comes from Aquila-b6, which should have undergone significant gas dispersal (Gutermuth et al. 2011). 

Evans et al. (2009) found that most stars form in clusters. Figure 19(i) presents the distribution of $\frac{N_{PS}}{N_{totalPS}}$, which ranges from 0.29 to 0.70, with a median value of 0.5. This means that about 50\% PS objects formed in high PS groups, as these groups always aligned with dense filaments and is rich in molecular gas. Though 50\% of protostellar objects formed outside of high PS groups, many protostellar objects formed in clusters with high PMS stars counts, which means that they still formed in clusters.

\section{Discussion}
\label{discussions}

\subsection{Nearest Neighbor Distance Distributions}

As is stated in Section 3, we adopt $N=4$ while computing the surface density of PS objects. Groups with numbers smaller than 4 will be ignored. If we had used a smaller value of $N$ such as $N=3$, some small groups with only three or four PS objects will be identified. Figure 20 shows the surface density map of PS objects for Chameleon I and Corona using $N=3$. The $N=3$ PS surface density map is similar to $N=4$ PS surface density map for Corona Australis. For Chamaeleon I, small groups with only three PS objects was identified in the $N=3$ surface density map. Since groups with large number of PS objects are better for statistical studies of cluster star-formation, we adopt $N=4$ in this paper. 

\subsection{Continuous Star Formation Activity}

We examine the star formation sequence of nearby embedded clusters with these datasets. Figure 21 shows a comparison of the PS surface density versus the PMS surface density. We found that $\Sigma_{PS}$ correlates with $\Sigma_{PMS}$ (r$_{corr}$=0.78), implying continuous and steady low-mass star formation over a period longer than the age of the Class II sources in some clouds, which is several Myr (e.g., Wilking et al. 1989; Evans et al. 2009). Such continuous star formation has also been observed for intermediate-mass stars in other galactic clusters (DeGioia-Eastwood et al. 2001). The dashed line in Figure 21 represents the best-fit linear line to data with n$_{PMS}>0$. The slope is 0.236$\pm$0.001. 

\subsection{Correlations with Gas Surface Density}

The visual similarity between the distribution of \emph{Spitzer}-identified YSOs and maps of gas structure has been noted previously (e.g., Allen et al. 2007; Evans et al. 2009; Gutermuth et al. 2009). Moreover, Gutermuth et al. (2011) found a positive power-law correlation between the YSO surface densities and the molecular gas mass column densities in eight nearby molecular clouds, with a power law index of about 2, which agrees with the star formation law $\Sigma_{SFR} = A\Sigma_{gas}^2$. We fit lines to the log $\Sigma_{PS}$ and $\Sigma_{PMS}$ versus log $\Sigma_{gas}$ data, finding them well fit with power-law indexes of 1.40$\pm$0.01 and 1.13$\pm$0.02 (Figure 22), respectively. The power-law index that we found here seems to be lower than those found by Gutermuth et al. (2011), in which the power-law indexes are 1.87$\pm$0.03 in Ophicuhus, 1.95 in Serpens, and 3.8$\pm$0.1 in Perseus. The possible reason is that Gutermuth et al. (2011) used the 2MASS extinction
maps to measure the gas column density, which might underpredict the gas column densities towards the dense regions of clusters (Pokhrel et al. 2016). The deviation of results presented here from the star formation law $\Sigma_{SFR} = A\Sigma_{gas}^2$ can be explained by gas dispersion and non-coevality within the molecular clouds (Gutermuth et al. 2011). 
 
\subsection{Jeans Analysis}

The identified PS groups and their associated gas in this work allow to perform a statistical Jeans analysis.  
The mean neighbouring between PS stars could be used to study the fragmentation of clouds during star-formation process.
We calculated the typical PS separation $\lambda_{PS}$ from $\lambda_{PS}=(\frac{r_{eff}^3}{n_{PS}})^{1/3}$, 
and the Jeans length $\lambda_J=\sigma \times \sqrt{\frac{\pi}{G\rho}}$, where $\sigma$ is the velocity dispersion for 10 K gas, $\rho$ is the mean density from $\Sigma_{gas}$ and $r_{eff}$, and $G$ is the gravitational constant. Figure 25 shows $\lambda_{PS}$ versus $\lambda_{J}$. The separation between PS stars ranges from 0.02 pc to 0.9 pc. 
We found that while taking into account the 70\% uncertainty of $\lambda_{PS}$ in the fitting, a slope of 1.017$\pm$0.007 was found, with a Pearson correlation coefficient of 0.92. Therefore, the observed fragments in PS clusters seem to be in reasonable agreement with thermal Jeans fragmentation. While the red line in Figure 23 represents $\lambda_{PS}=\lambda_J$, we can see from the figure that $\lambda_{PS}$ correlate well with $\lambda_J$ for $\lambda_{PS}<0.4$ pc, which is consistent with Pokhrel et al. (2018). For $\lambda_{PS}>0.4$ pc, most $\lambda_{PS}$ values are larger than $\lambda_J$. Gutermuth et al. (2011) presents a simple evolutionary model which is quite effective in explaining the observed star-gas correlation. They found that the correlation itself can be a direct consequence of thermal Jeans fragmentation, which agrees with our finding that the group spacings are similar to Jeans length for r$<$0.4 pc. 

\subsection{Comparison with Gas Free-fall Time}

The role that ptotostellar feedback such as ptotostellar outflows and stellar radiation play in clustered star formation is still under debate (e.g. Nakamura \& Li 2014). Two main scenarios have been proposed for this issue. In the first scenario, protostellar feedback is believed to destroy the cluster-forming clump and terminate further star formation, thus star formation in clusters should be rapid and brief (Elmegreen 2007; Hartman \& Burkert 2007). This scenario is referred as rapid star formation. In the second scenario, the protostellar feedback is believed to play the role of maintaining the internal turbulent motions of the clumps, and star formation should be slow and can last for several free-fall times or longer (Tan et al. 2006; Nakamura \& Li 2014). This scenario is referred as slow star formation. Clarifying how clustered star formation proceeds could help discriminate whether star formation is rapid or slow, and which kind of role the protostellar feedback plays in clustered star formation. 

To investigate this question we compare the age of the groups, which is derived using the model of Myers (2011, 2012), with the free-fall times of the associated molecular gas clumps. The model of Myers et al. (2011, 2012) assumes constant protostar birthrate, core-clump accretion, and equally likely accretion stopping. The cluster ages and birthrates are obtained from the observed numbers of protostars and pre-main sequence stars, and from the modal value of the protostar luminosity. 

For a sphere of mean column density $\Sigma_{gas}$ and radius $R$, the free-fall time $\tau_{ff}$ is given by (Burkert \& Hartmann 2004)
\begin{equation}
\tau_{ff} \approx \sqrt \frac{\pi R}{8G \Sigma_{gas}}.
\end{equation}
Table 4 presents the calculated  $\tau_{ff}$ value for the identified PS groups. Figure 24 shows a comparison of $\tau_{ff}$ versus the group age. We find a possible correlation between $\tau_{ff}$ and the group age. The slope of the best fit line is 1.31, with a Pearson correlation coefficient of 0.78. This suggests that groups with shorter dynamical times have a greater fraction of protostars, i.e. they are "younger". This result also suggests that star cluster formation is likely to be a relatively fast process, and favors the first scenario. In this scenario, the protostellar feedback destroys the dense cluster-forming clump, and terminates further star formation (Elmegreen 2007; Hartmann \& Burkert 2007).

\subsection{Future Applications}

In this paper we present a catalogue of 32 groups with high protostellar surface density in nearby embedded clusters. Some groups show extremely high protostellar surface density and molecular gas surface density. These sources provide ideal targets for future high-resolution spectral line and continuum observations with facilities such as the SMA or ALMA to study the physical condition of these very young groups in more detail. Through such observations, we could quantify the Jeans number as in Pokhrel et al. (2018), estimate the mass accretion rate from spectral line velocity and asymmetry, and estimate the depth of the gravitational potential well as a guide to each region's ability to attract more gas for regions with high protostellar surface density.

\section{Summary}
\label{summary}

Using data from the {\it Spitzer} c2d and GB legacy surveys and {\it Herschel} column density maps, we identified 32 groups with high protostellar surface density in nearby embedded clusters. Their properties, including their effective radius, protostellar and pre-main sequence star surface densities, ages, and average molecular gas column densities are derived. The main results of this work are summarized as follows:

1. Several groups show extremely high protostellar surface density and high molecular gas surface density, including Serpens-b3, Serpens-b2, Aquila-b2, Aquila-b3, Aquila-b6, Corona Australis, Aquila-b5, and Oph L1688-1. These groups seem to be undergoing vigorous star formation activity, and will be good targets for future high-resolution spectral line and continuum observations to study the fragmentation process. 

2. The median molecular gas column density of these groups is 1.6$\times10^{22}$ cm$^{-2}$, corresponding to 296 M$_{\odot}$ pc$^{-2}$. The lowest gas column density of these sub-clusters is about (2.75$\pm$2.96)$\times 10^{21}$ and (2.85$\pm$1.76)$\times10^{21}$ cm$^{-2}$, corresponding to 51$\pm$55 and 53$\pm$35 M$_{\odot}$ pc$^{-2}$.

3. We found possible correlation between $\Sigma_{PS}$ and $\Sigma_{PMS}$ ($r_{corr}=0.78$), implying continuous and steady low-mass star formation over several Myr.

4. We found positive power-law correlation between the YSO surface densities and the molecular gas column densities. The power-law indexes were 1.62$\pm$0.01 and 1.42$\pm$0.01 for ptotostellar and pre-main sequence stars, respectively. 

5. The average separation between protostellar sources seems to agree well with thermal Jeans fragmentation for sub-clusters with $\lambda_{PS}\leq0.4$ pc, and $\lambda_{PS}$ is always larger than $\lambda_J$ for $\lambda_{PS}\geq0.4$ pc. These results support the picture that the thermal Jeans fragmentation dominates at smaller scales.

6. The calculated gas free fall time of these sub-clusters seem to correlate with the cluster age derived with theoretical model ($r_{corr}=0.65$), suggests that regions with shorter dynamical times have a greater protostellar fraction. This result also suggests that star cluster formation is likely to be a relatively fast process. A possible correlation was found between the group age and $\tau_{ff}$: $age= (1.31\pm0.02) \tau_{ff}$ ($r_{corr}=0.78$).

This research has made use of Starlink and GAIA software.
The Starlink software (Currie et al 2014) is currently supported by the East Asian Observatory. GAIA is a derivative of the Skycat catalogue and image display tool, developed as part of the VLT project at ESO. This research has made use of data from the Herschel Gould Belt survey (HGBS) project (http://gouldbelt-herschel.cea.fr). The HGBS is a Herschel Key Programme jointly carried out by SPIRE Specialist Astronomy Group 3 (SAG 3), scientists of several institutes in the PACS Consortium (CEA Saclay, INAF-IFSI Rome and INAF-Arcetri, KU Leuven, MPIA Heidelberg), and scientists of the Herschel Science Center (HSC).
JL would like to thank the staff of Smithsonian Astrophysical Observatory (SAO) for supporting my visits, and Xingwu Zheng and Jim Moran for helping to arrange the visits. We also thank James Lane for assistant with using Marco Lombardi's extinction mapping. We are grateful to Marco Lombardi, who worked to make his programs available on the web. We also thank the referee for the very helpful report which improved our paper. This work was supported in part by the National Natural Science Foundation of China (11590780, 11590784, and 11773054), and Key Laboratory for Radio Astronomy, CAS. RAG's participation in this project was supported by NASA ADAP grants NNX11AD14G, NNX15AF05G, and NNX17AF24G, NASA JPL/Caltech contract 1489384, and NSF grant AST 1636621 in support of TolTEC, the next generation mm-wave camera for LMT.

\begin{table}
\scriptsize
    \begin{center}
%      \begin{minipage}{105mm}
      \caption{List of sub-regions identified.}\label{tab:sour2}
      \begin{tabular}{lccccccccccc}
      \\
    \hline
    \hline
Source Name  &   RA (J2000)     &   DEC(J2000)   	&  D    & Dist. Ref.  	 \\
                       &                           &                             & (pc)  &	          	           \\
\hline
Aquila-a                 &    18:38:00        &    00:12:00          &  260    &	 1    	           \\
Aquila-b                 &    18:30:00        &   -02:00:00         &  260    &	  1     	           \\
Aquila-c                 &     18:28:00       &   -03:48:00          &  260    &	   1  	           \\
Aquila-d                 &    18:29:00        &   -01:39:00    &  260    &	  1     	           \\
Aquila-e                 &    18:31:36        &   -02:14:00      &  260    &	  1     	           \\
 Auriga/CMC          &      04:29:36       &    35:42:00           &  450    &	   2               \\
%  Cepheus-a             &     22:34:00        &    75:15:00        &    200    &	     3               \\
    Cepheus             &    21:02:00        &   68:12:00     &    288    &	     3              \\
  Chamaeleon I        &   11:08:48        &   -77:00:00    &    150         &	       4           \\
%  Chamaeleon II       &         -                  &      -                  &      178       &	  5       	           \\       
 % Chamaeleon III       &       -                  &        -               &       150      &	   4       	           \\
  Corona Australis       &   19:02:00     &     -36:57:00     &     130     &	         5   	           \\      
%    IC5146-a     &     21:44:36                &	47:39:00          &	460       &      6      \\   
      IC5146     &   21:53:12               &	 47:15:00        &	460       &      6        \\                           
 % Lupus I              &        -                   &	  -        &	 150          &      7      \\                               
% Lupus III            &  16:10:24     &	39:00:00    &	200         &       7  \\
% Lupus IV        &  16:01:00    &	-41:40:00   &	 150        &     7        \\        
% Lupus V          &     -        &	-          &	 150        &      8       \\
% Lupus VI          &     -        &	-          &	 150        &       8      \\         
Ophiunchus      &  16:29:12   &  -24:30:00    &	 125        &       7       \\      
% sOphiunchus North   &     -        &	-          &	130         &     11        \\        
  Perseus-a         &  03:43:36   & 32:00:00     &	 250        &      8       \\
  Perseus-b        &   03:29:12   &  30:48:00     &	   250      &       8       \\         
  Serpens-a       &  18:30:24     &   01:14:24   &	   429      &      9         \\
  Serpens-b        &   18:29:24    & 00:36:00    &	    429     &       9       \\                                                                                                  
             \hline
      \end{tabular}
  \end{center}
  Notes. \\
  References for the distances quoted in this work: (1) Maury et al. (2011), (2) Lada et al. (2009), (3) Kirk et al. (2009), (4) Belloche et al. (2011), (5) Neuh$\ddot{a}$user \& Forbrich (2008), (6) Arzoumanian et al. (2011), (7) Wilking et al. (2008), (8) Enoch et al. (2006), (9) Dzib et al. (2011).
\end{table}

\clearpage

\begin{table}
\scriptsize
    \begin{center}
%      \begin{minipage}{105mm}
      \caption{Properties of Clumps Identified.}\label{tab:sour0}
      \begin{tabular}{lcccccccccccccc}
      \\
    \hline
    \hline
Region$^a$	& RA$^a$     & Dec$^a$  & $\Sigma_{peak}$$^a$   & contour$^a$ &	r$_{eff}$$^b$  & N$_{PS}$$^b$   & $\frac{N_{PS}}{N_{total PS}}$$^c$  & N$_{II}$$^b$   &  TD$^b$    \\
                         & (J2000)    &  (J2000)     &       ( deg$^{-2}$)          &	                     &	(pc)	               &	                    &         	                                               &	                       &	     \\
\hline
Aquila-a             &   18:39:20.001    &   00:33:59.93      &   120        &  20\%         &  0.53      &	  4              &    0.29      &	7                    &	3               \\                          
Aquila-b1            &  18:29:37.989     &   -01:51:02.99     &  123120 	 &  1.25\%      &  0.10    &        6          &            &	4	             &	0                \\	                  
Aquila-b2            &  18:29:58.999      &  -02:01:18.00    &	                   &  20\%         & 0.024  &       5         &                                     &	1	             &	 0           \\	                  
Aquila-b3           &  18:30:03.002     &   -02:03:03.00     &	                  &  20\%         &	  0.035  &          9        &                                   &	0	            &	0           \\	     
Aquila-b4           &  18:30:09.973      &   -02:06:03.50      &	                  &  5\%       &    0.044   &    4                 &                            &	0	            &	0            \\	     
Aquila-b5           &  18:30:25.973      &   -02:11:03.67       &	                 &  5\%     & 0.049   &    6                  &                             &  7          &	0            \\	   
 Aquila-b6           &  18:30:47.982      &   -01:56:17.28       &	         &  2.5\%   & 0.034    &    5                  &                             &	0	            &	0            \\	              
 Aquila-c	         &   18:27:51.747   &   -03:46:03.53   &  282             &  20\%      & 0.44         &	5                  &                            &	 12	            &	0             \\
 Aquila-d	          &  18:28:56.104   &   -01:37:55.26  &    5071      &  20\%         & 0.12         &	   6                 &                        &	1	             &	0              \\
 Aquila-e            &   18:31:37.943  &    -02:13:30.88      &  1628      &  20\%       & 0.18        &	    4               &                          &	3	              &  	0              \\	           	  
 \hline                
Auriga/CMC-1     &  04:28:36.876    &   36:28:05.55    & 893                  & 5\%            &	1.10     &	 7          &            0.48                 &	7              &	0              \\
Auriga/CMC-2    &  04:30:34.632  &    35:47:52.44    &	                      &  10\%         &	0.92      &	     7        &                            &	15	               &	1            \\
Auriga/CMC-3     &  04:30:14.646  &  35:16:06.29   &	                             &  10\%         &	0.90    &	       9      &                             &     66               &	         7            \\	                  
Auriga/CMC-4     &  04:30:55.690  &  34:56:05.25   &	                            &  40\%         &	0.36   &	  7           &                            &	    5                 &	 	0             \\	   
\hline               
%Cepheus-a        &  22:34:58.789  &   75:16:55.38    &	   985                    &  20\%         &	0.21        &	        5               &            35\%                &	7	                   &	0               \\	                  
Cepheus        &  21:01:36.00  &  68:14:15.00   &	280                 &   20\%         &	  0.62        &	    7                      &        0.54                 &	 24	            &	     1          \\
\hline
 Chamaeleon I   &  11:05:46.231    &   -77:20:28.63     &	120        &  20\%         &	0.36               &	     4                   &          0.36                &	4	    &	0             \\	                  	   
 \hline               
Corona Australis   &  19:01:57.839  &   -36:57:04.09     &	15304         &  10\%         &  0.051    &	      7       &           0.5      &	   2        &	  0                      \\	 
\hline                 
 %IC5146-a1          &  21:47:10.871    &   47:32:53.95    &     3051           &   10\%         &	               &	7              &                                &	6	                   &	0              \\
 % IC5146-a2          &  21:44:48.339   &  47:42:52.06       &	                &   10\%         &	                 &	                &                                               &	7	                   &	0              \\
 IC5146           & 21:53:36.066       &   47:19:01.43   &    1513             &  20\%         & 1.01    &	    7          &       0.57                   &	68	           &	0             \\      
 %\hline     
 %Musca               & 12:23:34.204   &  -71:55:59.89  &        53             &  20\%        &    0.68          &     5          &   0.625            &      0                 &      0      \\    
 \hline                      
Ophiuchus-a	 &   16:31:53.420      &   -24:02:31.30          &   176     & 20\%    &	 0.25        & 	 4                &       0.32     &	5	          &	0              \\
Ophiuchus-b     &  16:31:52.301      &   -24:56:01.50  &     803               & 20\%         &	0.11     &	       4            &                                              &	8	           &	0            \\	 
L1688-1	 &     16:26:22.668      &   -24:23:29.72        &     5447             & 20\%        &     0.05      &	       6           &                                               &	3	                   &	0              \\
L1688-2      &  16:27:28.611    &    -24:39:59.83           &	                       &  5\%       &      0.16    &	     11             &                                              &	14	                   &	0             \\     
  \hline           
  IC348	&    03:43:54.877     &    32:02:59.91     &	 4018                      & 20\%         &  0.15    &	  7            &     0.53       &	6	  &	0               \\        
 Perseus-a  &   03:42:06.579   & 31:47:58.05         &                               & 5\%            &  0.19     &	  4             &                   &  6          &	0               \\          
NGC1333       &  03:29:02.634      &     31:19:59.92    &	5151              &  2.5\%         &	0.71     &	  35       &                       &	 87	               &	2              \\
Perseus-b1     &  03:25:37.879    &   30:45:49.01    &	                                  &   5\%         &	0.28     &	   5         &                       &	   0	                 &	0              \\
Perseus-b2     &   03:33:24.314    &   31:07:44.62   &	                                 &  10\%         &	0.14     &	    5         &                       &	1	        &	 0            \\                 
\hline
Serpens-a1        &   18:29:48.198    &   01:16:31.50    &	114657                    &   2.5\%     & 0.14  &	8      &   0.70     &	  0	                   &  	0          \\
Serpens-a2         &   18:29:57.701     &  01:13:01.50    &	                              &  2.5\%      & 0.20  &	 16     &             &	11	                   &	0          \\	                  
Serpens-b1         &   18:28:45.996    & 00:52:29.98    &	18029                  &   2.5\%     &  0.23  &	  4     &            &	1	                   &	 0          \\	     
Serpens-b2         &   18:29:07.999   &  00:31:00.00    &	                           &   20\%     &    0.02  &	 4     &            &	               0                  &	 0          \\	     
Serpens-b3         &  18:28:55.999   &  00:30:00.00   &	                           &   40\%     &     0.02  &	 5     &            &                1               &	 0          \\	     
 \hline
      \end{tabular}
  \end{center}
  Notes. $^a$ Regions identified, the position of the center, peak surface density of PS objects, and contours used to select regions. \\
  $^b$ Effective radius, number of PS, Class II and TD objects within identified regions. \\
  $^c$ Percentage of PS stars within the identified regions compared to the total number of PS stars in the entire region.
\end{table}

\clearpage

\begin{table}
\scriptsize
    \begin{center}
%      \begin{minipage}{105mm}
      \caption{Properties derived from dust associated with the identified clumps.}\label{tab:sour1}
      \begin{tabular}{lcccccccccccccc}
      \\
    \hline
    \hline
Region$^a$	& RA$^a$     & Dec$^a$  &     Distance    &  $\overline{A_K}$	  &  $\Sigma_{gas}$        &      $\Sigma_{gas}$                    \\
                         & (J2000)    &  (J2000)     &   (arcmin)   	&                       &   ( cm$^{-2}$)       &       (M$_{\odot}$  pc$^{-2}$)                           \\
\hline
Aquila-a$^*$             &  18:38:56.001     &  00:33:59.95   &  6.0$\pm$2.0     & 0.81$\pm$0.44         &   (7.28$\pm$4.01)$\times 10^{21}$    &     134 $\pm$76                               \\                          
Aquila-b1             &  18:29:41.991    &  -01:50:18.00       &   8.33$\pm$0.25             &    &  (3.62$\pm$0.11)$\times 10^{22}$            &     670$\pm$20                              \\	                  
Aquila-b2             &  18:29:58.916       &   -02:01:04.25     &   0.32$\pm$0.25	      &     &   (5.64$\pm$0.94)$\times 10^{22}$          &    1044$\pm$174                                 \\	                  
Aquila-b3            &  18:30:04.003       &   -02:03:03.00    &	   9.92$\pm$0.25          &        &   (1.20$\pm$0.29)$\times 10^{23}$       &   2222$\pm$537	                               \\	     
 Aquila-b4             &  18:30:13.009       &   -02:06:48.00   &	    0.90$\pm$0.25        &        &   (4.13$\pm$0.74)$\times 10^{22}$          &  765$\pm$137                                 \\	                  
Aquila-b5          &  18:30:25.980       &   -02:10:48.56    &	0.25$\pm$0.25       &     &   (1.83$\pm$0.17)$\times 10^{22}$      &    339$\pm$31	                               \\	   
Aquila-b6          &  18:30:50.028       &   -01:56:02.95    &	  0.56$\pm$0.25       &      &  (1.70$\pm$0.45)$\times 10^{22}$      &    315$\pm$83	                               \\	             
 Aquila-c	         &  18:28:08.463       &   -03:48:20.0         &   4.8$\pm$2.0   &       &            (1.14$\pm$0.64)$\times 10^{22}$          &    211$\pm$119                                 \\
 Aquila-d	          &  18:29:04.107     &   -01:38:58.42         &   2.26$\pm$0.25  &       &             (1.54$\pm$0.36)$\times 10^{22}$   &   285$\pm$67                               \\
 Aquila-e            &   18:31:31.997     &   -02:15:00.0         &   2.10$\pm$0.06    &        &      (9.37$\pm$4.1)$\times 10^{21}$  &    174$\pm$76                            \\	           	  
 \hline                
Auriga/CMC-1     &   04:28:46.799     &    36:29:52.49         &   3.05$\pm$0.13	 &        &     (6.25$\pm$2.66)$\times 10^{21}$     &    116$\pm$49                         \\
Auriga/CMC-2    &    04:30:35.806      &    35:54:05.74        &  6.23$\pm$0.13	    &        &      (9.74$\pm$5.87)$\times 10^{21}$              &       180$\pm$109	                                 \\
Auriga/CMC-3     &   04:30:15.704      &    35:12:06.31        &	4.01$\pm$0.13       &        &       (1.01$\pm$0.72)$\times 10^{22}$             &     187$\pm$133 	                                \\	                  
Auriga/CMC-4     &    04:30:54.598     &    34:56:05.25       &    0.27$\pm$0.13	     &      &       (1.10$\pm$0.76)$\times 10^{22}$                &       204$\pm$141	                     	             \\	   
\hline               
Cepheus        & 21:01:36.00          &     68:12:07.50        &	  2.125$\pm$0.50   &          &       (2.75$\pm$2.96)$\times 10^{21}$        &       51$\pm$55	                                        \\	                  
\hline
 Chamaeleon I   &  11:06:31.337      &   -77:23:30.30      &  11.68$\pm$0.50   &      &      (6.14$\pm$5.96)$\times10^{21}$ &    114$\pm$110   \\	                  
% Chamaeleon I-2  &  11:06:57.206     &   -77:23:01.55       &   1.4$\pm$0.5    &     255$\pm$80        &  3.2 - $\pm$0.5 \\	   
 \hline               
Corona Australis   &  19:01:55.335  &   -36:58:01.36     &	1.14$\pm$0.25    &         &      (4.30$\pm$2.56)$\times10^{22}$   &    861$\pm$506                          \\	 
\hline                 
% IC5146-a1          &  21:47:11.411    &   47:32:59.39    &     0.6$\pm$0.4                  &      102$\pm$77                &      0.2$\pm$1.0	                            \\
 % IC5146-a2          &  21:45:00.873   &  47:38:58.12       &	0.5$\pm$0.3                &       82$\pm$55                    &      5.0$\pm$1.0                          \\
 IC5146           & 21:53:36.206       &   47:21:11.85   &    2.17$\pm$0.25       &     & (2.85$\pm$1.76)$\times10^{21}$         &      53$\pm$35 	                          \\      
% \hline      
% Musca      & 12:24:25.659  &  -71:49:59.89      &                        &  (1.99$\pm$1.39)$\times10^{21}$              &     37$\pm$26  	                          \\ 
 \hline                
Ophiuchus-a	 &   16:31:42.469      &   -24:00:44.99     &   3.26$\pm$0.50         &	          &  (5.39$\pm$4.54)$\times 10^{21}$      &       100$\pm$84	                          \\
Ophiuchus-b     &    16:31:56.825    &   -24:58:04.98        &   2.28$\pm$0.25      &	          &   (1.32$\pm$0.70)$\times 10^{22}$    &   239$\pm$137	                        \\	 
L1688-1	 &     16:26:27.058      &   -24:23:59.78        &   1.21$\pm$0.50        &	                 &  (4.16$\pm$3.66)$\times 10^{22}$  &      770$\pm$678	                        \\
%L1688-2      &   16:27:13.182     &    -24:27:09.56     &	  2.9$\pm$1.0                    &        511$\pm$173      &      	 3.0$\pm$0.5                    \\	
L1688-2      &  16:27:24.211    &    -24:40:29.88       &	1.21$\pm$0.50       &         &      (1.88$\pm$0.64)$\times 10^{22}$     &     348$\pm$119                    \\     
  \hline           
  IC348	&    03:43:59.26     &    32:02:55.58     &	1.10$\pm$0.50      &         &    (1.99$\pm$0.85)$\times 10^{22}$       &      369$\pm$157                          \\        
 Perseus-a  &   03:42:06.916   & 31:47:53.75         &  0.11$\pm$0.50     &        &    (4.52$\pm$2.27)$\times 10^{21}$      &      85$\pm$42                   \\          
%Perseus-b1    &  03:33:32.889  &    31:05:34.99      &	         &       1.2$\pm$0.1$\times 10^{                   &              $\pm$1.0                           \\
NGC1333       &  03:29:02.241    &  31:15:59.94        &	4.0$\pm$1.0   &       &        (1.05$\pm$1.14)$\times 10^{22}$   &    194$\pm$211                          \\
Perseus-b1     &  03:25:37.879    & 30:45:49.01       &	0         &       &       (2.18$\pm$1.37)$\times 10^{22}$     &      404$\pm$254 	                         \\
Perseus-b2     &   03:33:14.971    &  31:07:45.74     &	2.3$\pm$1.0       &       &    (2.98$\pm$1.01)$\times 10^{22}$       &      552$\pm$187 	                          \\                 
\hline
Serpens-a1        &   18:29:48.198    &   01:16:46.50    &	0.36$\pm$0.13     &        &        (5.73$\pm$2.04)$\times 10^{22}$    &      1054$\pm$389               \\
Serpens-a2         &   18:29:56.70     &  01:13:09.00    &	0.29$\pm$0.13     &    &  (4.62$\pm$2.3)$\times10^{22}$    &      856$\pm$426	                         \\	                  
Serpens-b1         &   18:28:43.796    &   00:52:56.98  &	0.71$\pm$0.50       &        &      (1.00$\pm$0.47)$\times 10^{22}$   &      185$\pm$87	                          \\
Serpens-b2       &  18:29:05.80   & 00:30:27.00    &    0.78$\pm$0.50      &         &                (5.83$\pm$0.26)$\times 10^{22}$       &       1080$\pm$48	                          \\ 
Serpens-b3       &  18:28:53.80   &   00:28:57.00  &   1.19$\pm$0.50       &            &            (4.20$\pm$0.55)$\times 10^{22}$        &      778$\pm$ 102	                          \\ 	                  	                  	             
             \hline 
      \end{tabular}
  \end{center}
  Notes. $^a$ Position of the extinction peak nearest to the identified clusters. \\
  $^*$ The $\Sigma_{gas}$ for Aquila-a is derived from 2MASS data, while  $\Sigma_{gas}$ for other clusters are derived using the {\it Herschel} data. \\
\end{table}

\begin{table}
\scriptsize
    \begin{center}
%      \begin{minipage}{105mm}
      \caption{Derived physical properties of Clumps Identified.}\label{tab:sour4}
      \begin{tabular}{lcccccccccccccc}
      \\
    \hline
    \hline
Region	& $\Sigma_{PS}$        & $\Sigma_{PMS}$ &   $f_{PS}$  &    age   	   &	$\tau_{ff}$         \\
                 &   (M$_{\odot}$  pc$^{-2}$)             &     (M$_{\odot}$  pc$^{-2}$)    &      	    &       (Myr)     &	   (Myr)          \\
\hline
Aquila-a             &  2.3       &  5.7      &      0.30         &  0.64          &    0.53          \\                          
Aquila-b1            &  95      &   64      &     0.6             &   0.22   &       0.10       \\	                  
Aquila-b2            &  1382      &  276    &  0.83           &    0.08     &      0.04             \\	                  
Aquila-b3           &  1169     &   0   &	  1       &     -     &      0.11                \\	       
Aquila-b4         &   329     &   0     &    1     &     -     &        0.06              \\
Aquila-b5         &  398     &   464   &	  0.46      &   0.37     &         0.10            \\     
Aquila-b6         &   688   &       0   &   1          &       -          &     0.09       \\
 Aquila-c	      &   4.1    &  9.9      &     0.29     &      0.66       &         0.39        \\
 Aquila-d	       &  66   &   11   &	    0.86        &      0.06     &        0.17          \\
 Aquila-e	      &   20    &  15     &	     0.57         &    0.25       &       0.27              \\       	  
 \hline                
Auriga/CMC-1    &  0.9    &   0.9   &	    0.50           &    0.32       &       0.82            \\
Auriga/CMC-2    &  1.3     &  3.0   &	    0.30        &      0.63      &        0.60                \\	 
Auriga/CMC-3     &   1.8   &   14    &	0.11          &    1.81    &       0.59          \\	                  
Auriga/CMC-4     &  8.6    &   6.1   &	0.58          &    0.24     &     0.36              \\	     
\hline                               
Cepheus       &  2.9      &  10    &	      0.22           &  0.9      &       0.93                \\
\hline
 Chamaeleon I     &   4.9     &   4.9   &          0.50       &  0.32       &     0.48                   \\                      
 \hline               
Corona Australis  	 &   428    &   122     &      0.78      &  0.1     &        0.068            \\ 
\hline                 
 IC5146         &   0.9   &   11    &       0.09         & 2.2      &       1.17         \\	 
% \hline    
% Musca        &    1.7       &    0                              &       1                  &	-        &     1.15           \\
 \hline                        
Ophiuchus-a	&  10    &    13   &	0.44            &     0.39      &       0.42                     \\    
Ophiuchus-b    &    53   &   105    &	  0.33            &     0.57      &          0.18                 \\  
L1688-1	    &   382    &   191    &      0.67                &    0.17      &          0.068             \\ 
L1688-2         &  68   &   87    &	   0.44                &    0.39       &        0.18             \\
  \hline           
  IC348	      &   50   &   42    &	 0.54                 &      0.28          &         0.17          \\
 Perseus-a        &   18   &   26   &	 0.4               &        0.45            &         0.40        \\        
NGC1333       &  11      &   28   &	0.29               &        0.70             &      0.51           \\
Perseus-b1    &  10      &   0       &	    1                     &          -       &     0.22               \\  
Perseus-b2     &  41      &  8.1    &	     0.83        &       0.08      &     0.13              \\            
\hline
Serpens-a1    &  65      &   0    &	  1             &        -      &     0.10               \\
Serpens-a2     &  64      &   44    &	     0.59      &     0.23      &     0.13               \\	                  
Serpens-b1       &   12     &   3.0    &	    0.80    &     0.09       &      0.30                  \\	  
Serpens-b2     &  1592    &   0       &	    1               &      -         &     0.036               \\	                  
Serpens-b3     &  1989     &   398   &  	0.83     &      0.07       &      0.043                  \\	                	                  	             
             \hline
      \end{tabular}
  \end{center}
\end{table}

\clearpage

\begin{figure}[htbp]
   \centering
\includegraphics[scale=0.4]{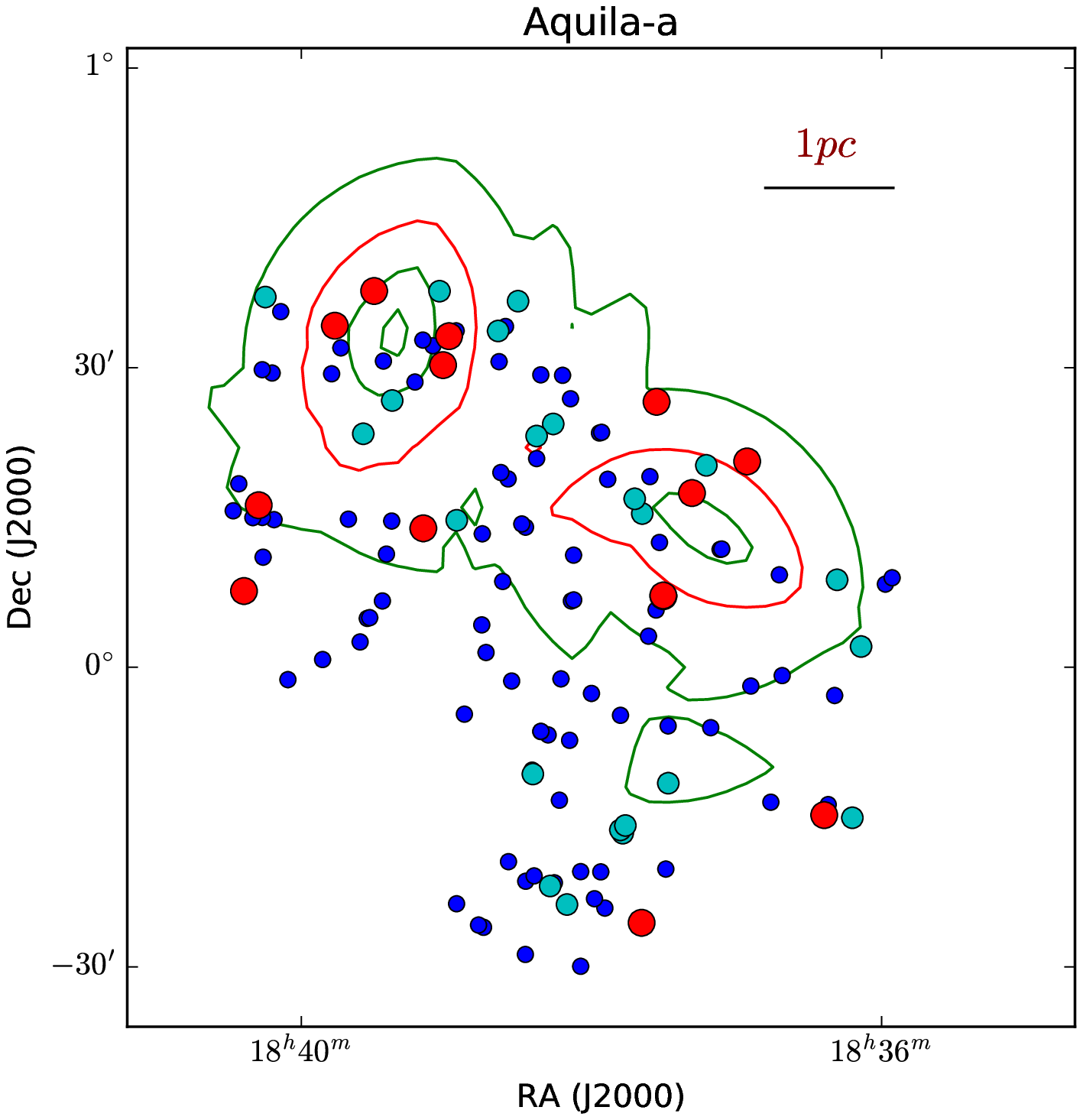}
\includegraphics[scale=0.4]{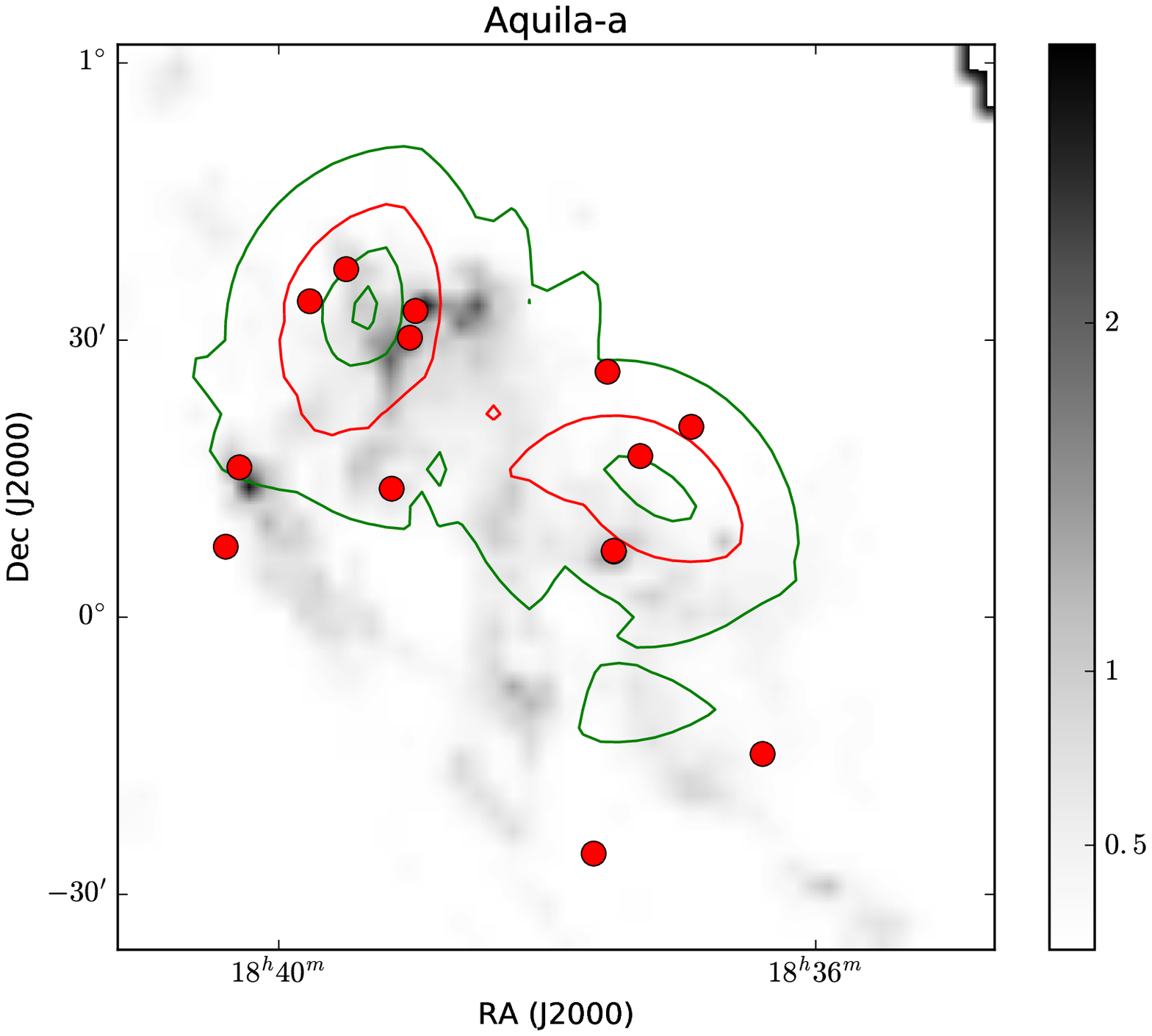}
\caption{Maps of the Aquila-a region. Left: The $N=4$ surface density of PS objects for the clusters with positions of PS (red dots), class II (blue dots), and TD (cyan dots) objects overlaid. Contours represent the PS surface density, shown at 10\%, 20\%, 40\%, and 80\% of the peak value. The 20\% contour is shown in red, while other contours are shown in green. Right: The 2MASS $A_K$ extinction map of the region with the $N=4$ surface density of PS and positions of PS objects (red dots) overlaid. }
\end{figure}

\begin{figure}[htbp]
   \centering
\includegraphics[scale=0.4]{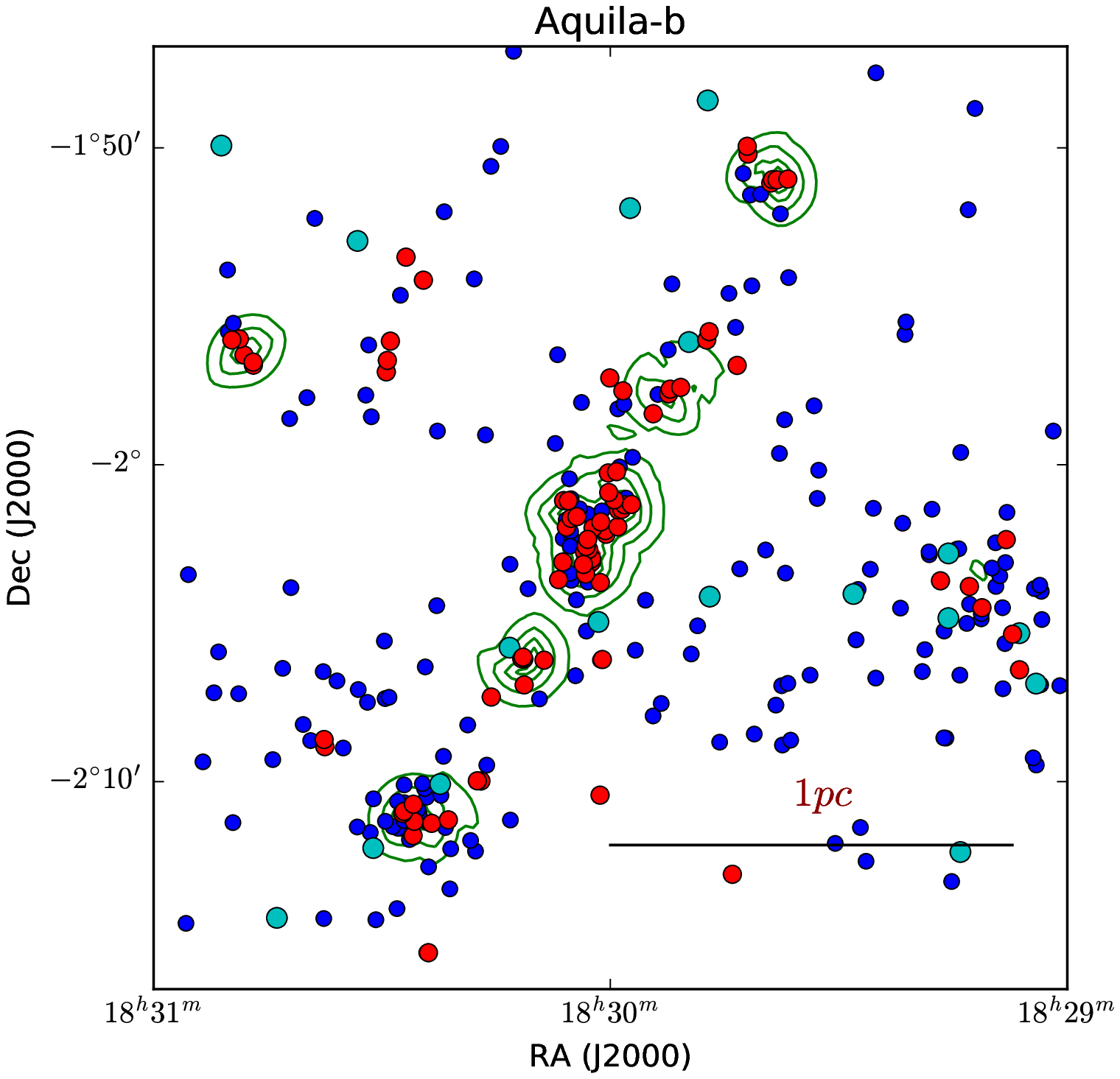}
\includegraphics[scale=0.4]{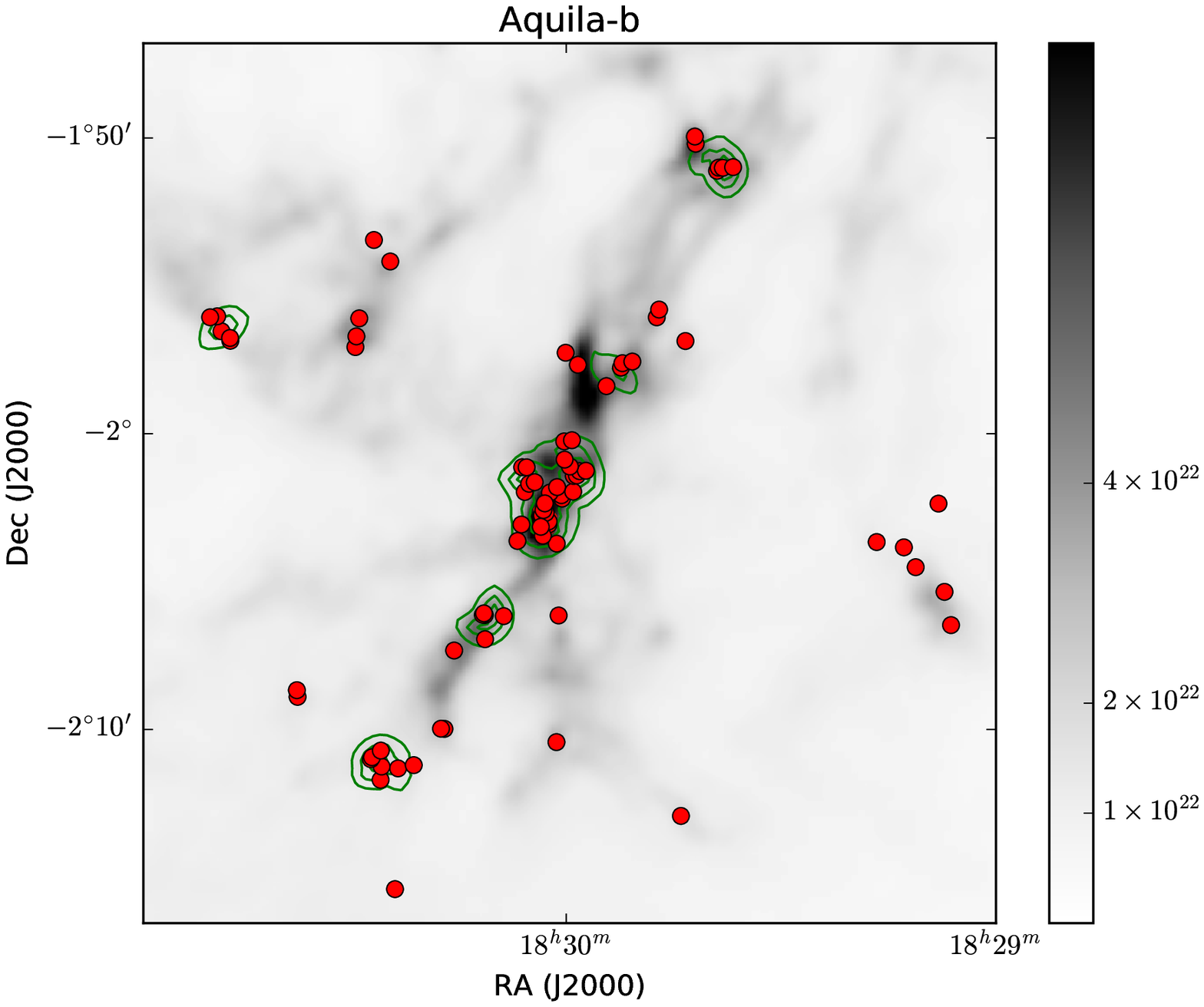}
\caption{Maps of the Aquila-b region. Left: The $N=4$ surface density of PS objects for the clusters with positions of PS (red dots), class II (blue dots), and TD (yellow dots) objects overlaid. Contours represent the PS surface density, shown at 2.5\%, 5\%, 10\%, 20\%, 40\%, and 80\% of peak value. The 20\% contour is shown in red, while other contours are shown in green. Right: The {\it Herschel} column density map of the region with the $N=4$ surface density of PS and positions of PS objects (red dots) overlaid. The color scale is given in units of cm$^{-2}$.}
\end{figure}

\begin{figure}[htbp]
   \centering
\includegraphics[scale=0.4]{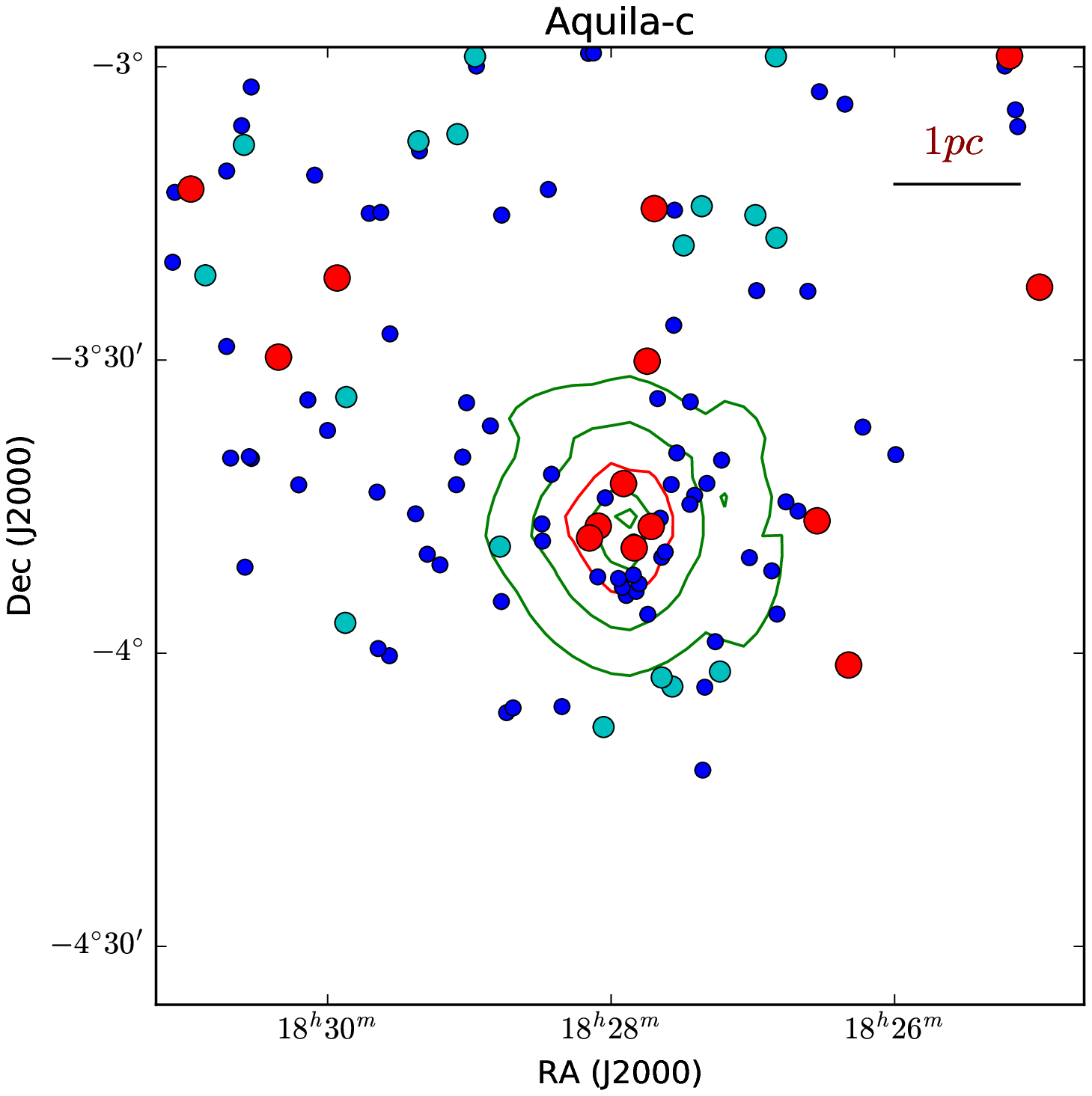}
\includegraphics[scale=0.4]{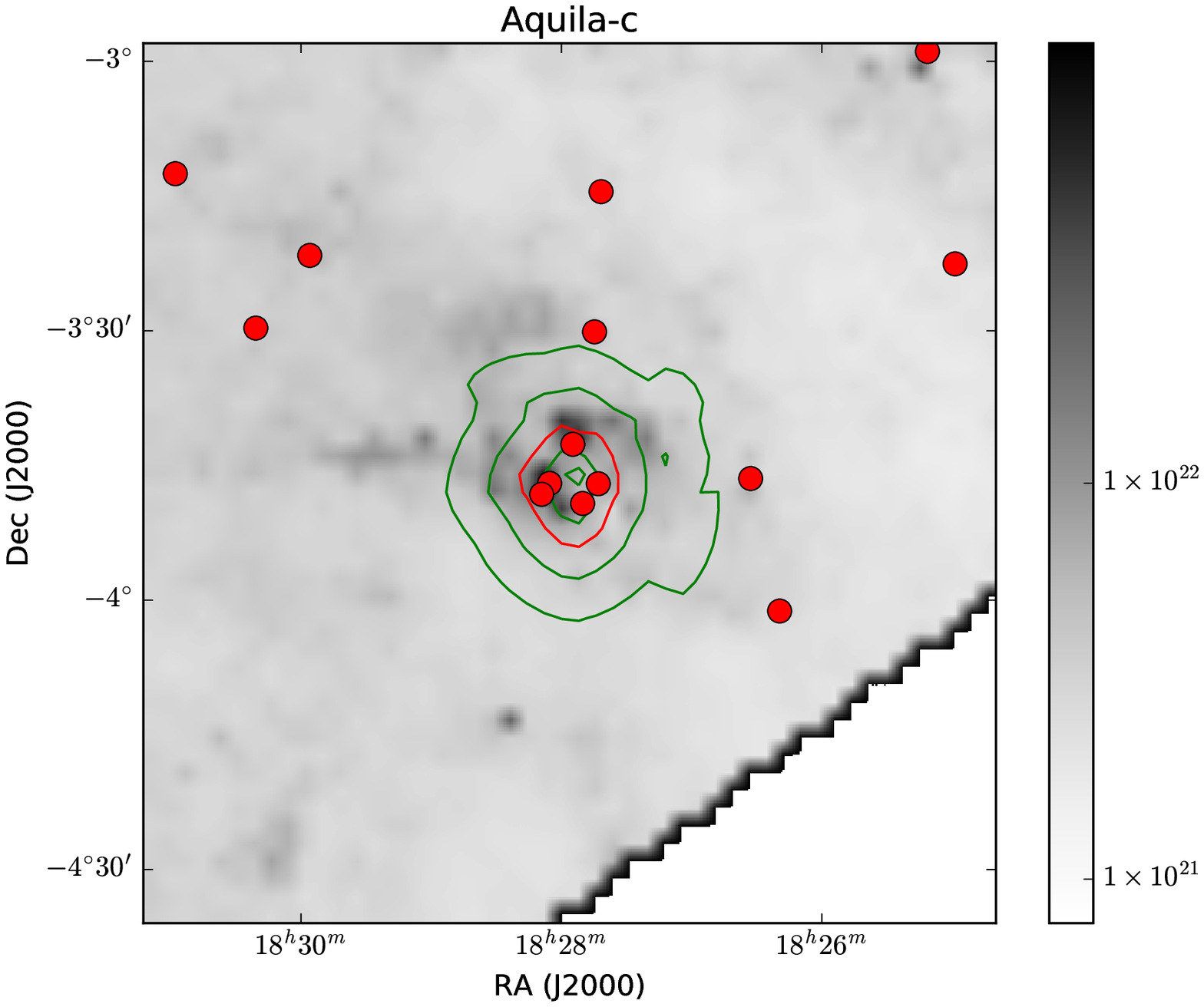}
\caption{Maps of the Aquila-c region. Contours represent the PS surface density, shown at 5\%, 10\%, 20\%, 40\%, and 80\% of peak value. The right panel shows the the {\it Herschel} column density map of the region with the $N=4$ surface density of PS and positions of PS objects (red dots) overlaid.}
\end{figure}

\begin{figure}[htbp]
   \centering
\includegraphics[scale=0.4]{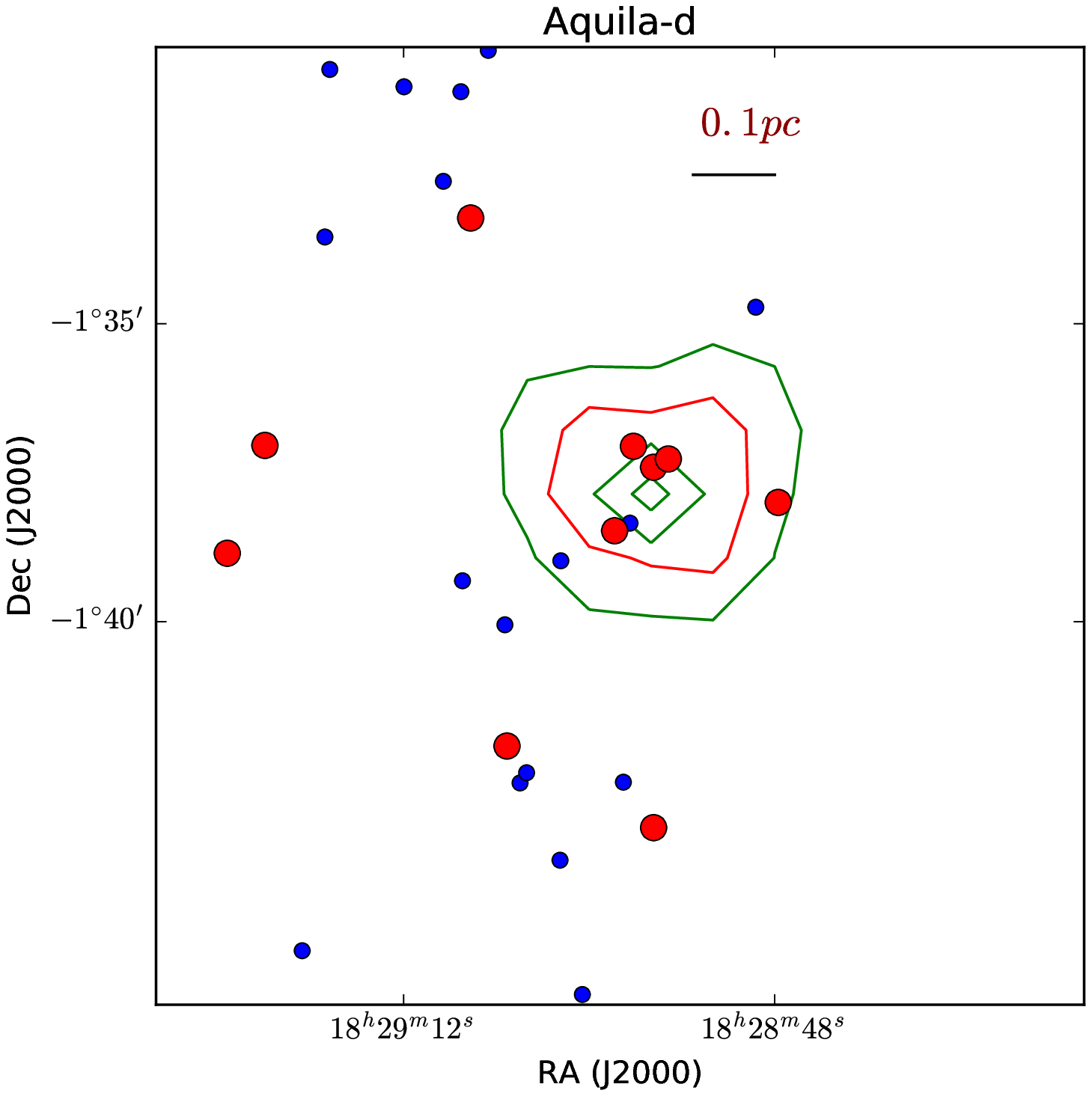}
\includegraphics[scale=0.4]{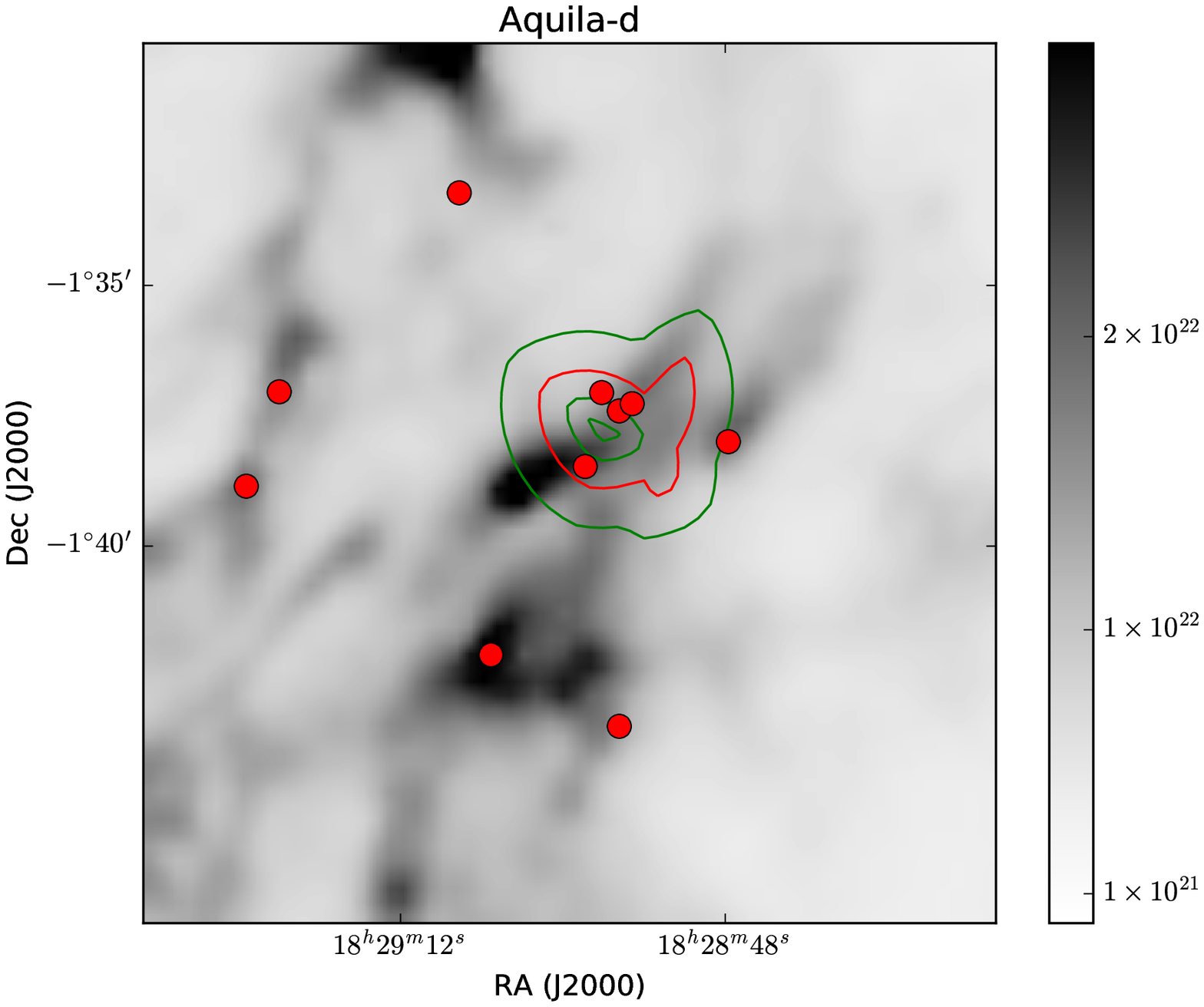}
\caption{Maps of the Aquila-d region. Contours represent the PS surface density, shown at 10\%, 20\%, 40\%, and 80\% of peak value. The right panel shows the the {\it Herschel} column density map of the region with the $N=4$ surface density of PS and positions of PS objects (red dots) overlaid.}
\end{figure}

\begin{figure}[htbp]
   \centering
\includegraphics[scale=0.4]{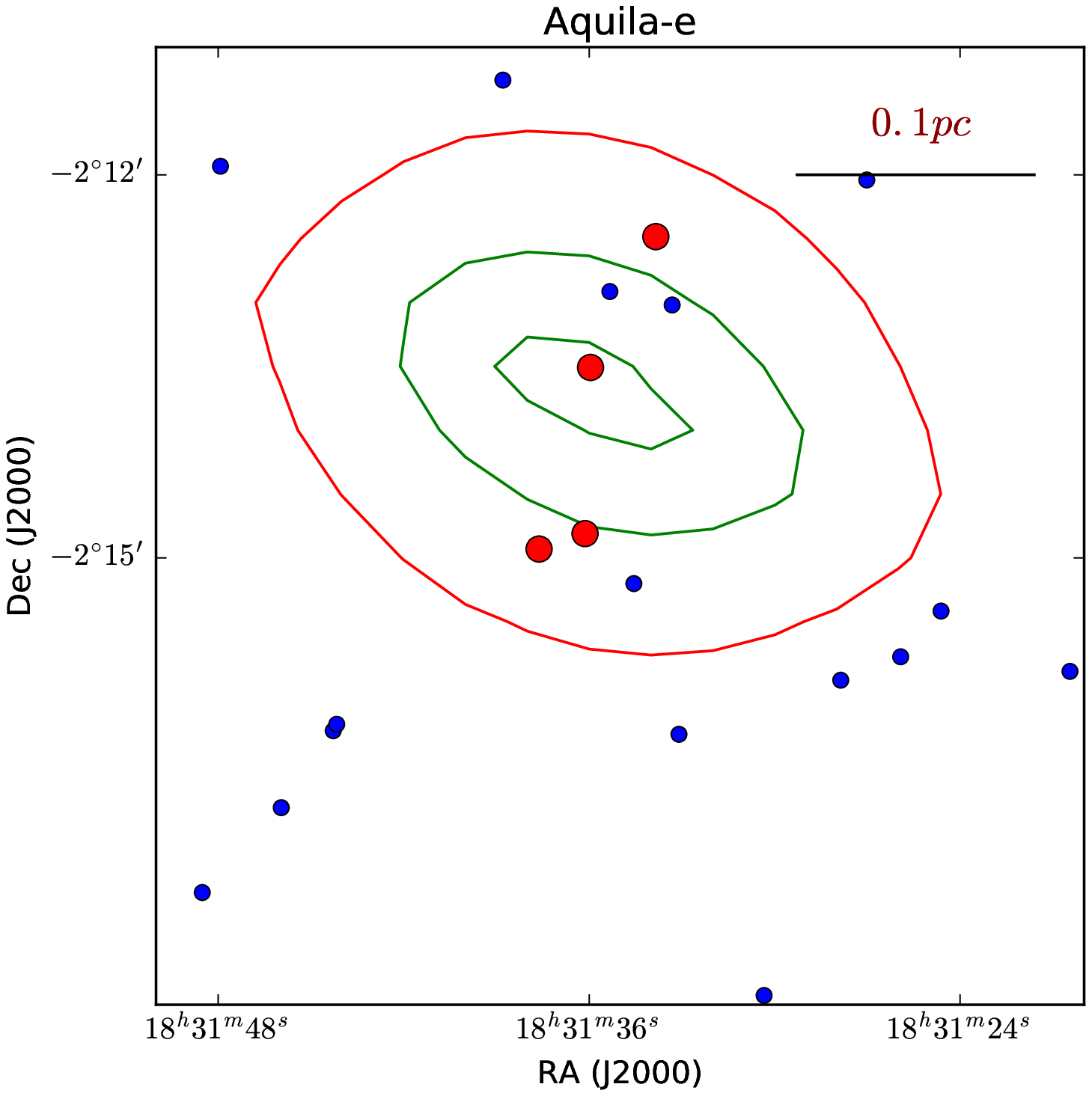}
\includegraphics[scale=0.4]{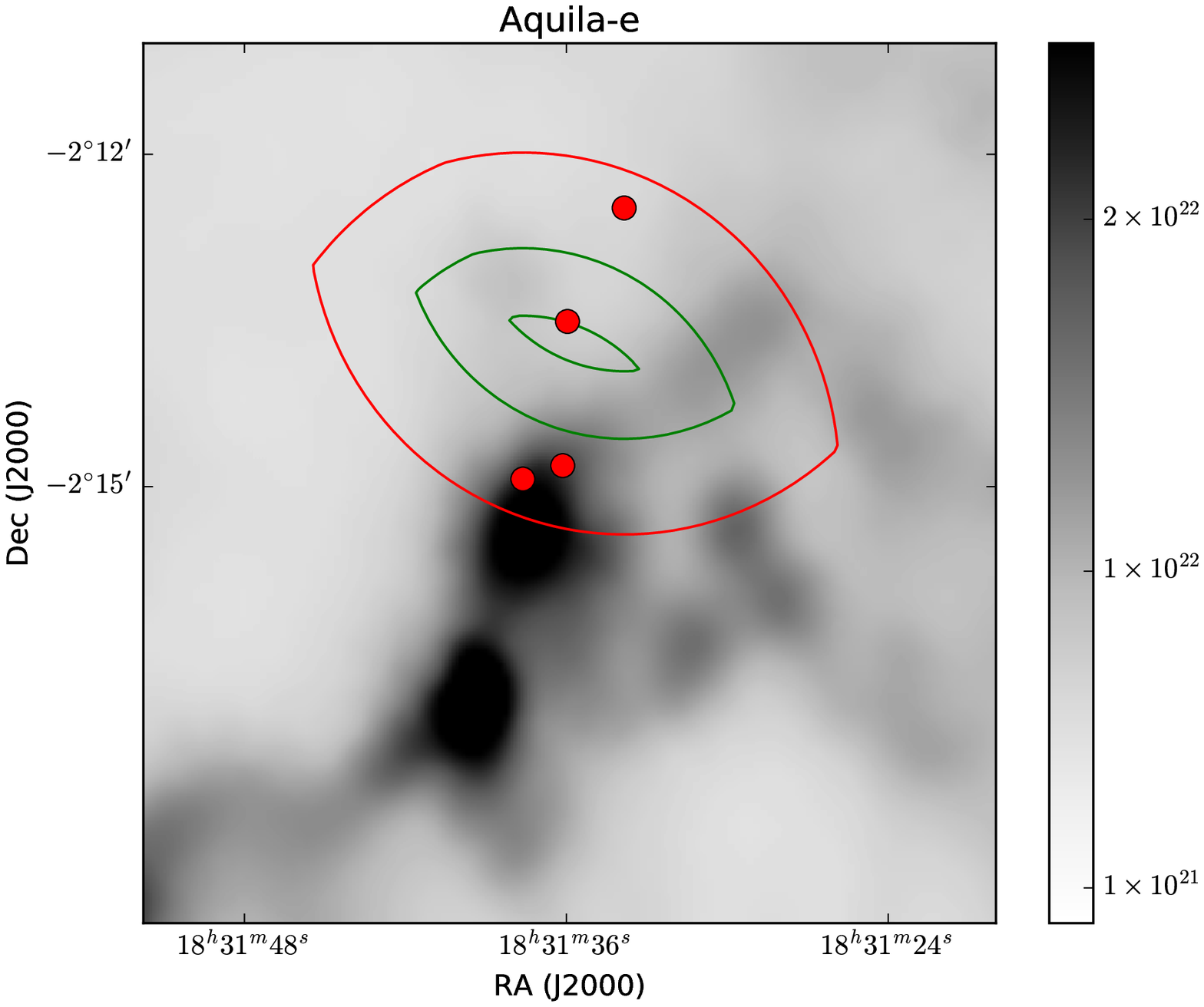}
\caption{Maps of the Aquila-e region. Contours represent the PS surface density, shown at 20\%, 40\%, and 80\% of peak value. The right panel shows the the {\it Herschel} column density map of the region with the $N=4$ surface density of PS and positions of PS objects (red dots) overlaid.}
\end{figure}

\begin{figure}[htbp]
   \centering
\includegraphics[scale=0.4]{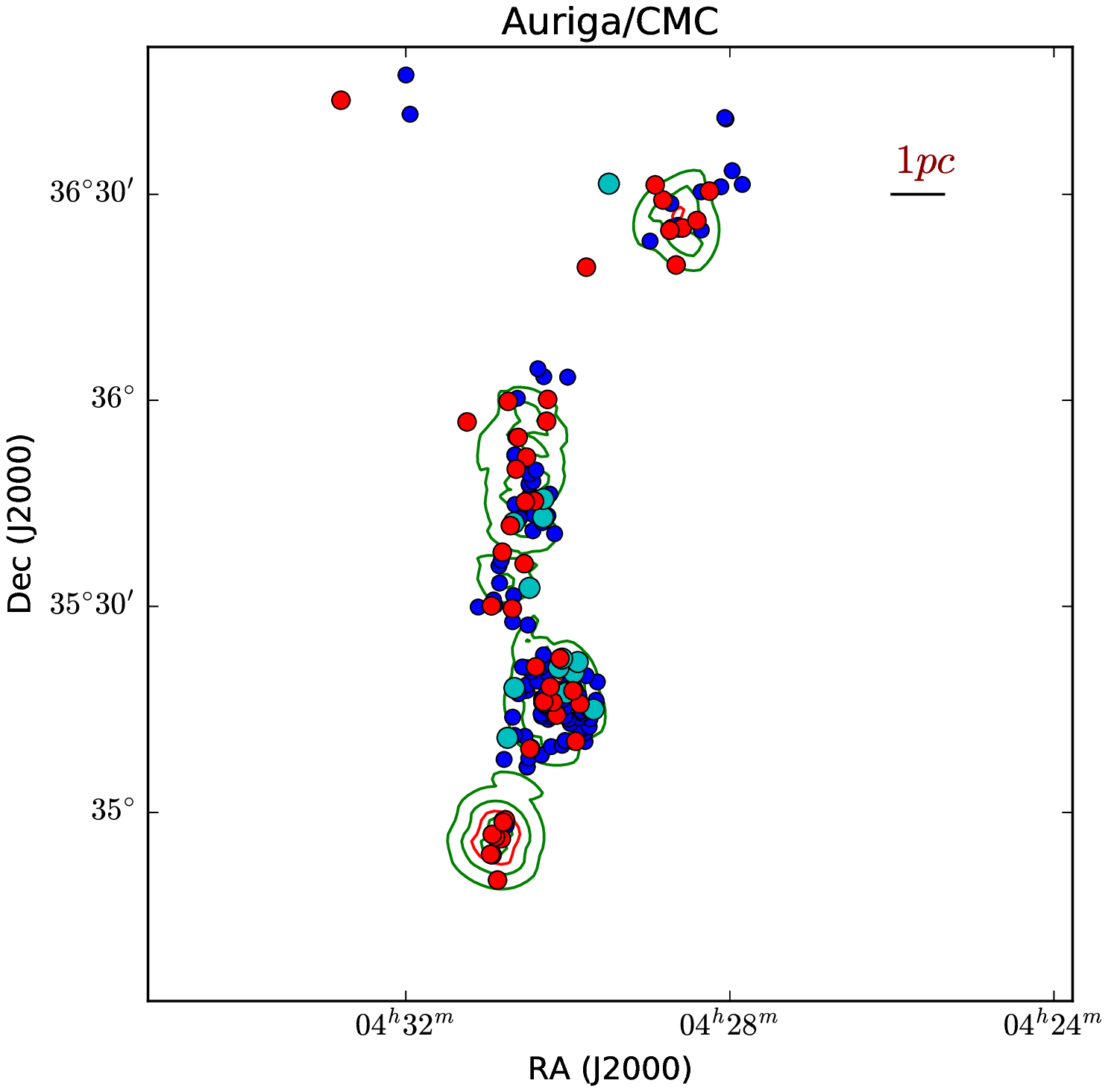}
\includegraphics[scale=0.4]{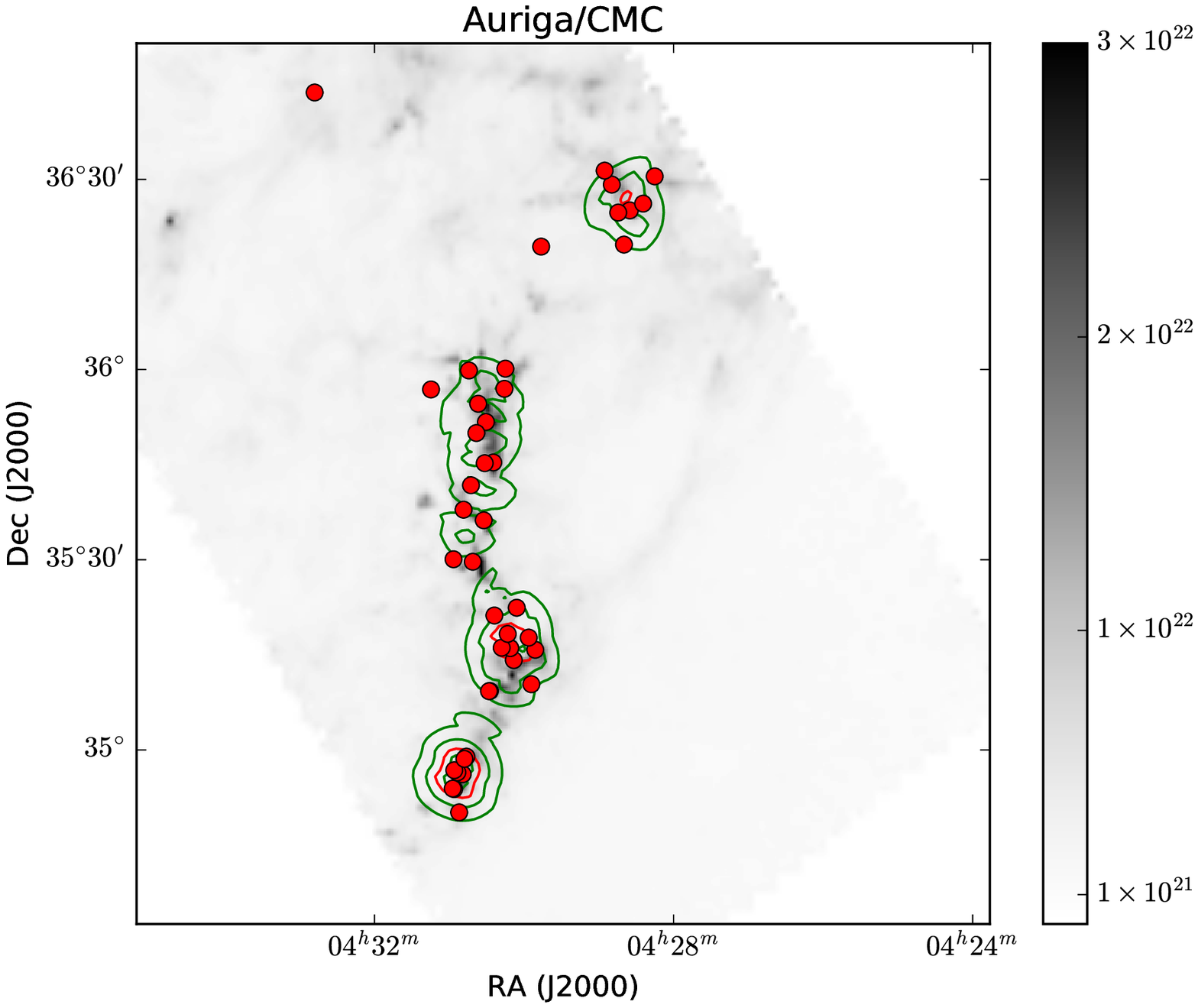}
\caption{Maps of the Auriga region. Contours represent the PS surface density, shown at 5\%, 10\%, 20\%, 40\%, and 80\% of peak value. The right panel shows the the {\it Herschel} column density map of the region with the $N=4$ surface density of PS and positions of PS objects (red dots) overlaid.}
\end{figure}

%\begin{figure}[htbp]
 %  \centering
%\includegraphics[scale=0.5]{cep-a.ps}
%\includegraphics[scale=0.5]{av-b-cep-a.ps}
%\caption{Maps of the Cepheus-a region. Contours represent the PS surface density, shown at 2.5\%, 5\%, 10\%, 20\%, 40\%, and 80\% of peak value. A$_K$ ranges from 0.1 mag to 1.29 mag.}
%\end{figure}

\begin{figure}[htbp]
   \centering
\includegraphics[scale=0.4]{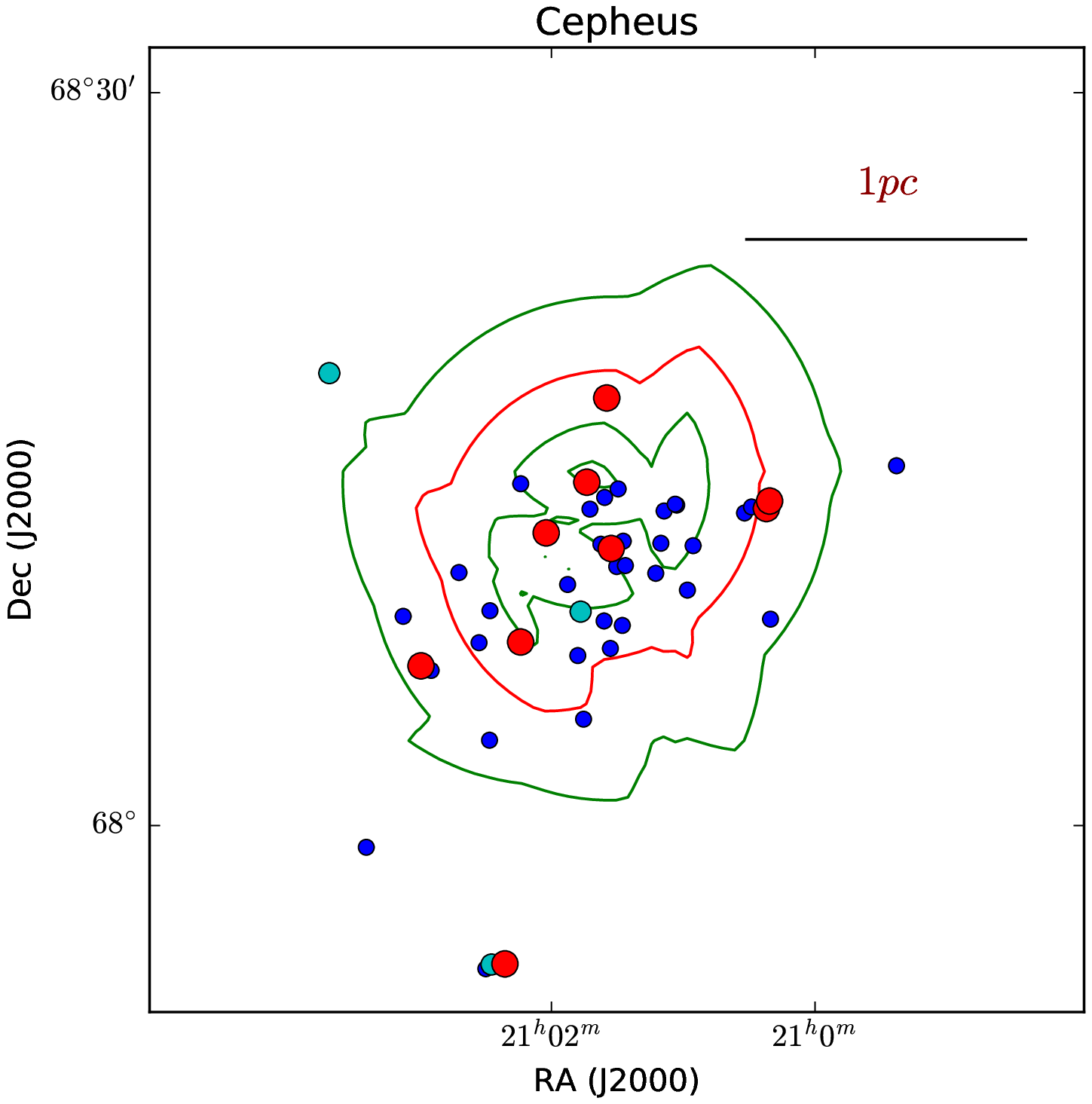}
\includegraphics[scale=0.4]{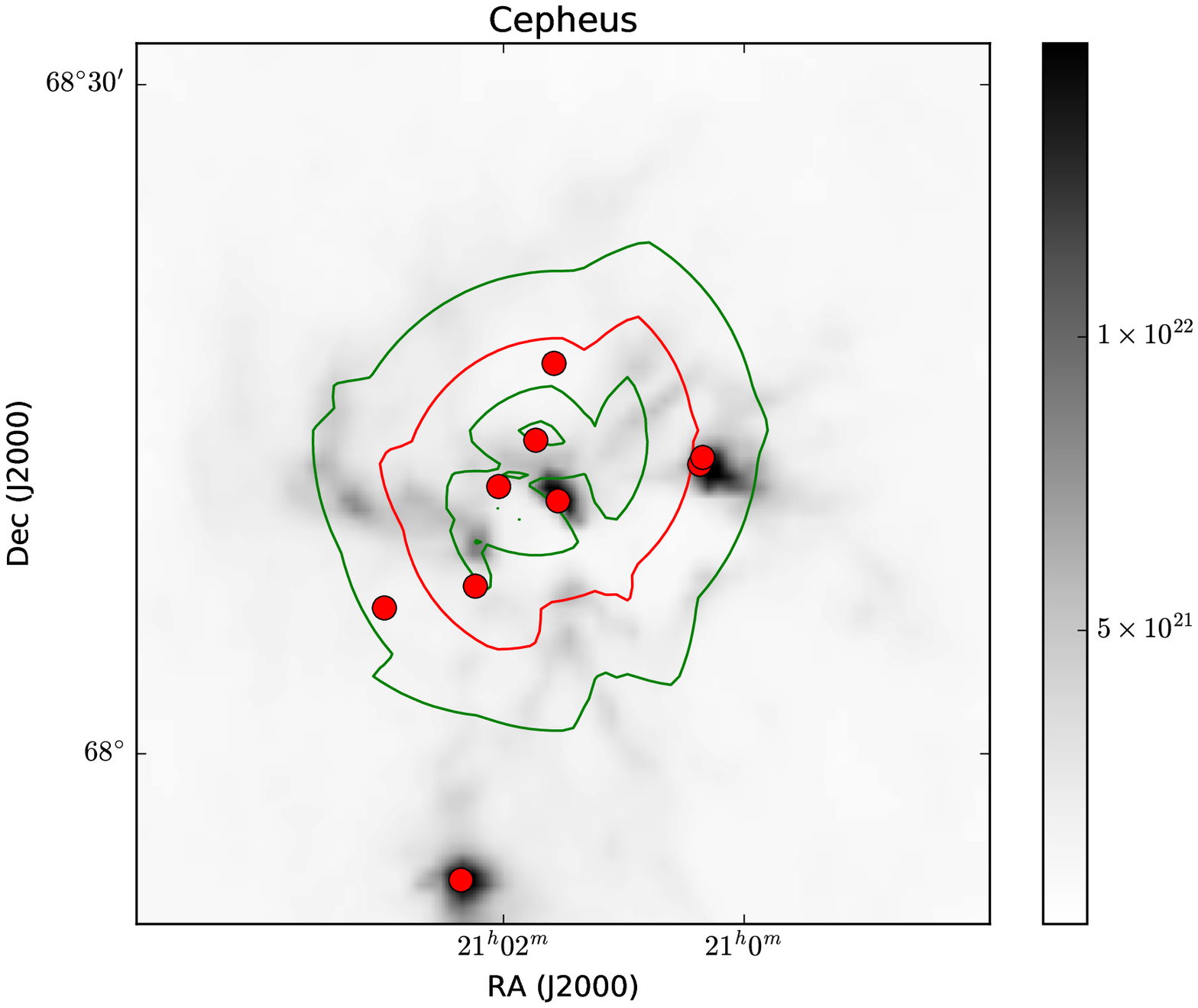}
\caption{Maps of the Cepheus-b region. Contours represent the PS surface density, shown at 20\%, 40\%, and 80\% of peak value. The right panel shows the the {\it Herschel} column density map of the region with the $N=4$ surface density of PS and positions of PS objects (red dots) overlaid.}
\end{figure}

\begin{figure}[htbp]
   \centering
\includegraphics[scale=0.4]{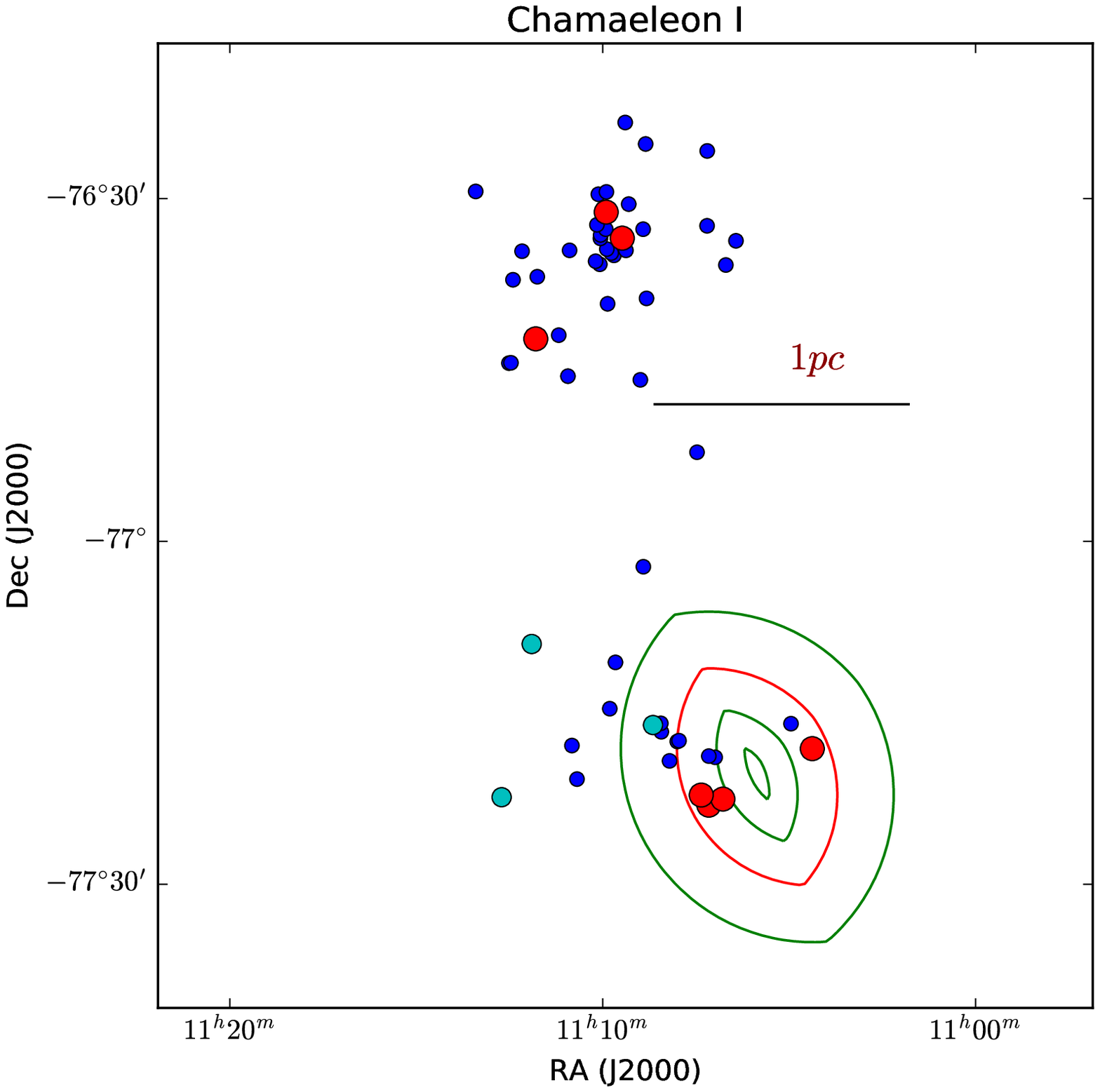}
\includegraphics[scale=0.4]{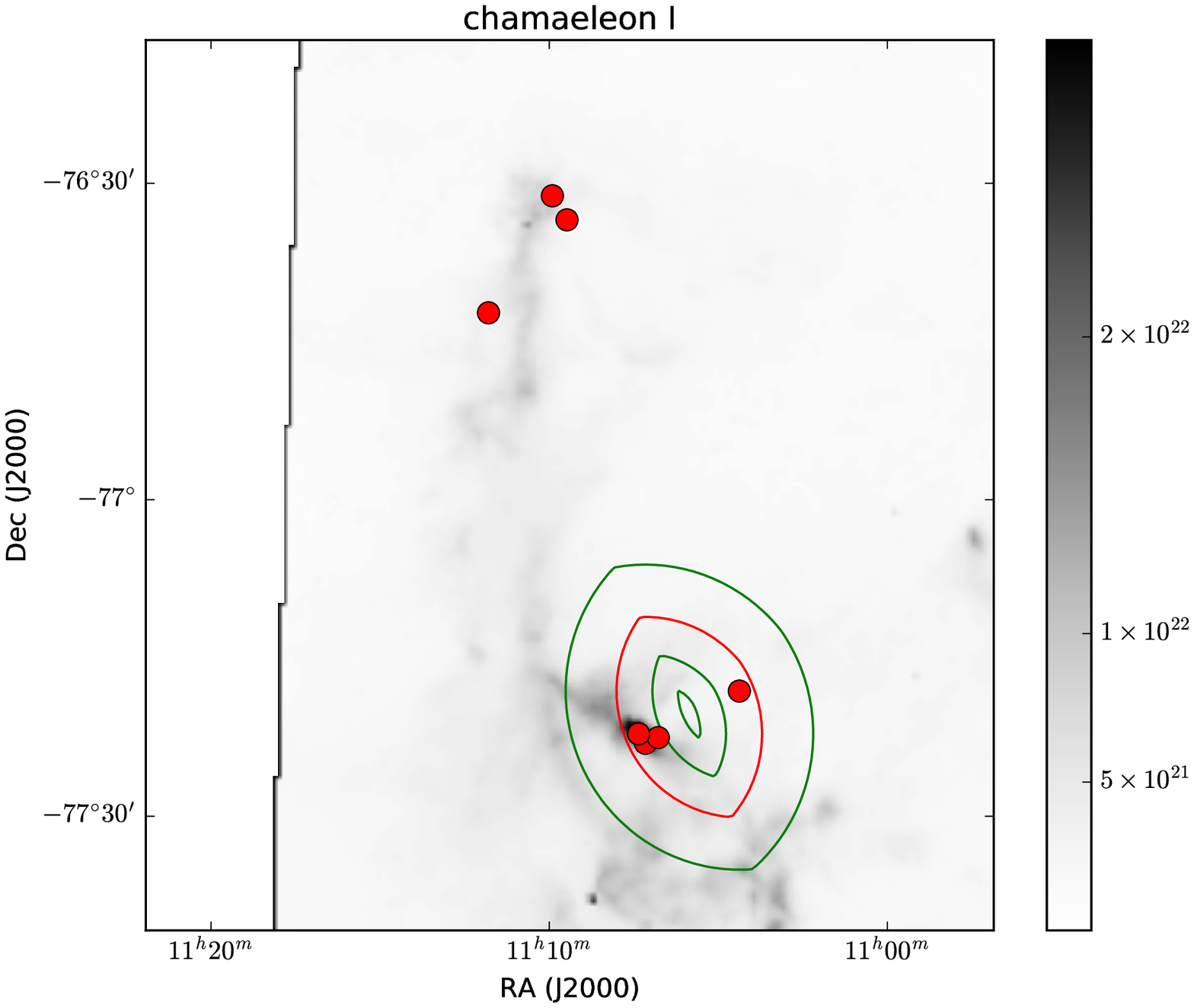}
\caption{Maps of the Chamaeleon I region. Contours represent the PS surface density, shown at 10\%, 20\%, 40\%, and 80\% of peak value. The right panel shows the the {\it Herschel} column density map of the region with the $N=4$ surface density of PS and positions of PS objects (red dots) overlaid.}
\end{figure}

\begin{figure}[htbp]
   \centering
\includegraphics[scale=0.4]{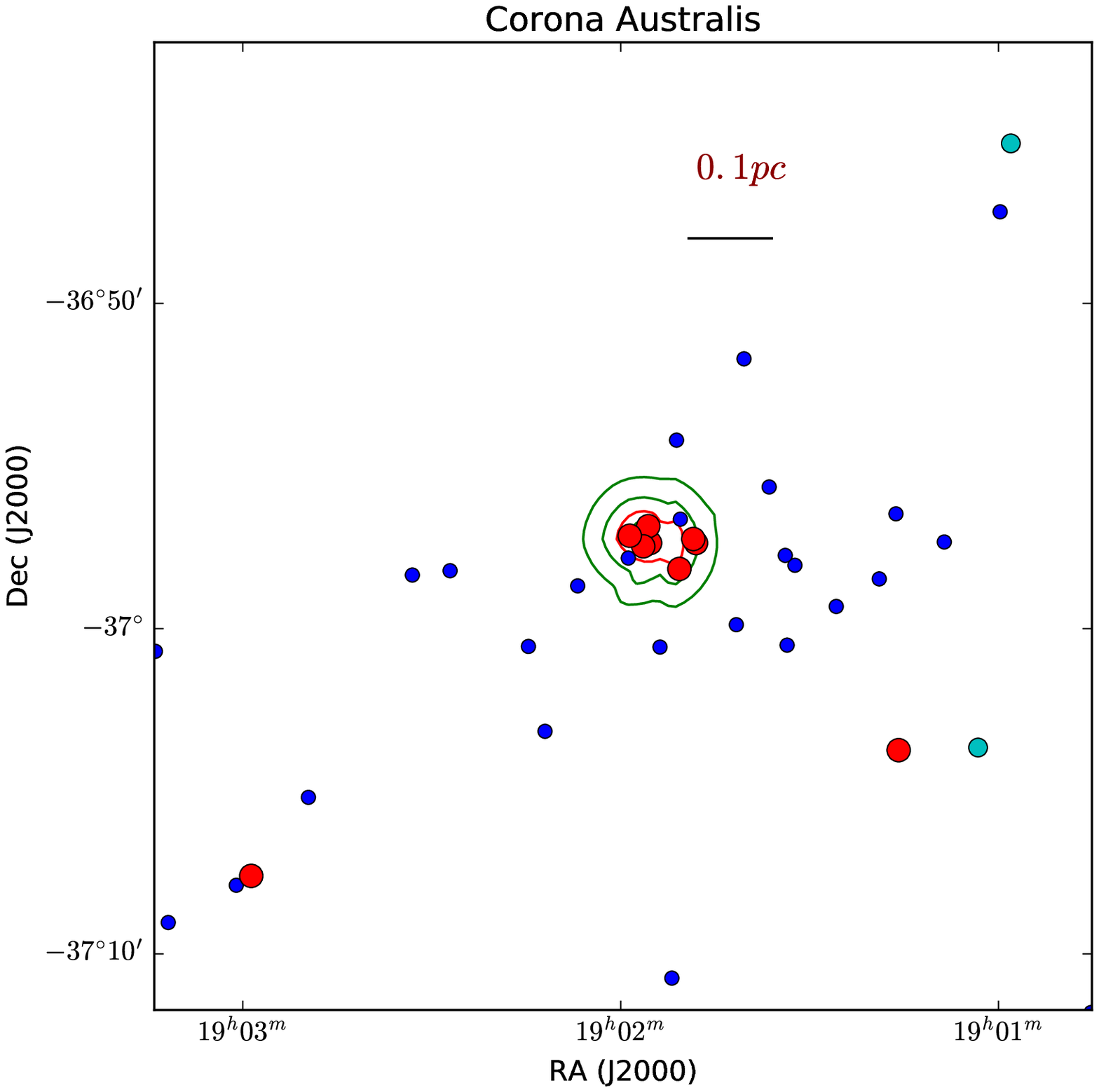}
\includegraphics[scale=0.4]{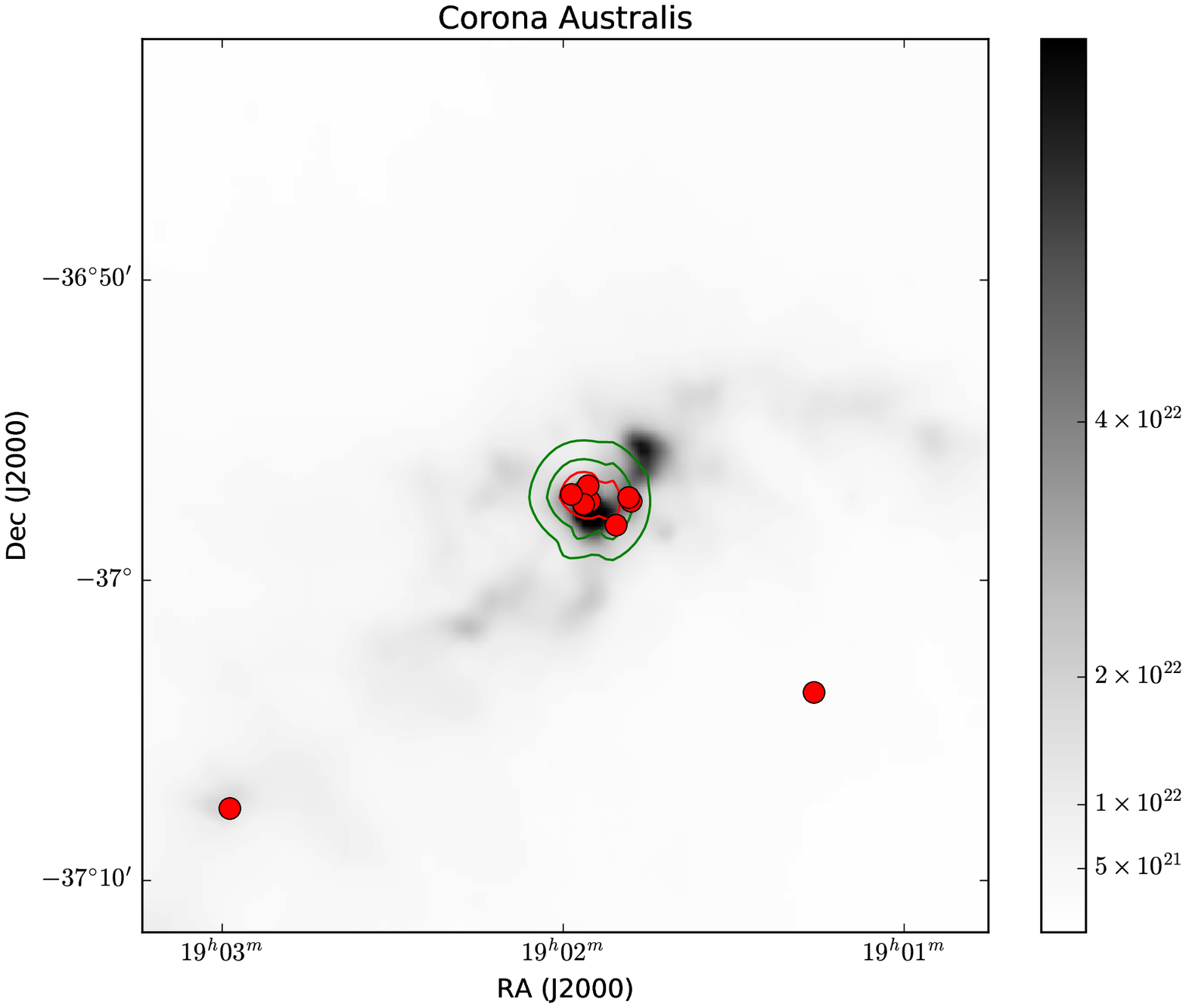}
\caption{Maps of the Corona Australis region. Contours represent the PS surface density, shown at 5\%, 10\%, 20\%, 40\%, and 80\% of peak value. The right panel shows the the {\it Herschel} column density map of the region with the $N=4$ surface density of PS and positions of PS objects (red dots) overlaid.} 
\end{figure}

\begin{figure}[htbp]
   \centering
\includegraphics[scale=0.4]{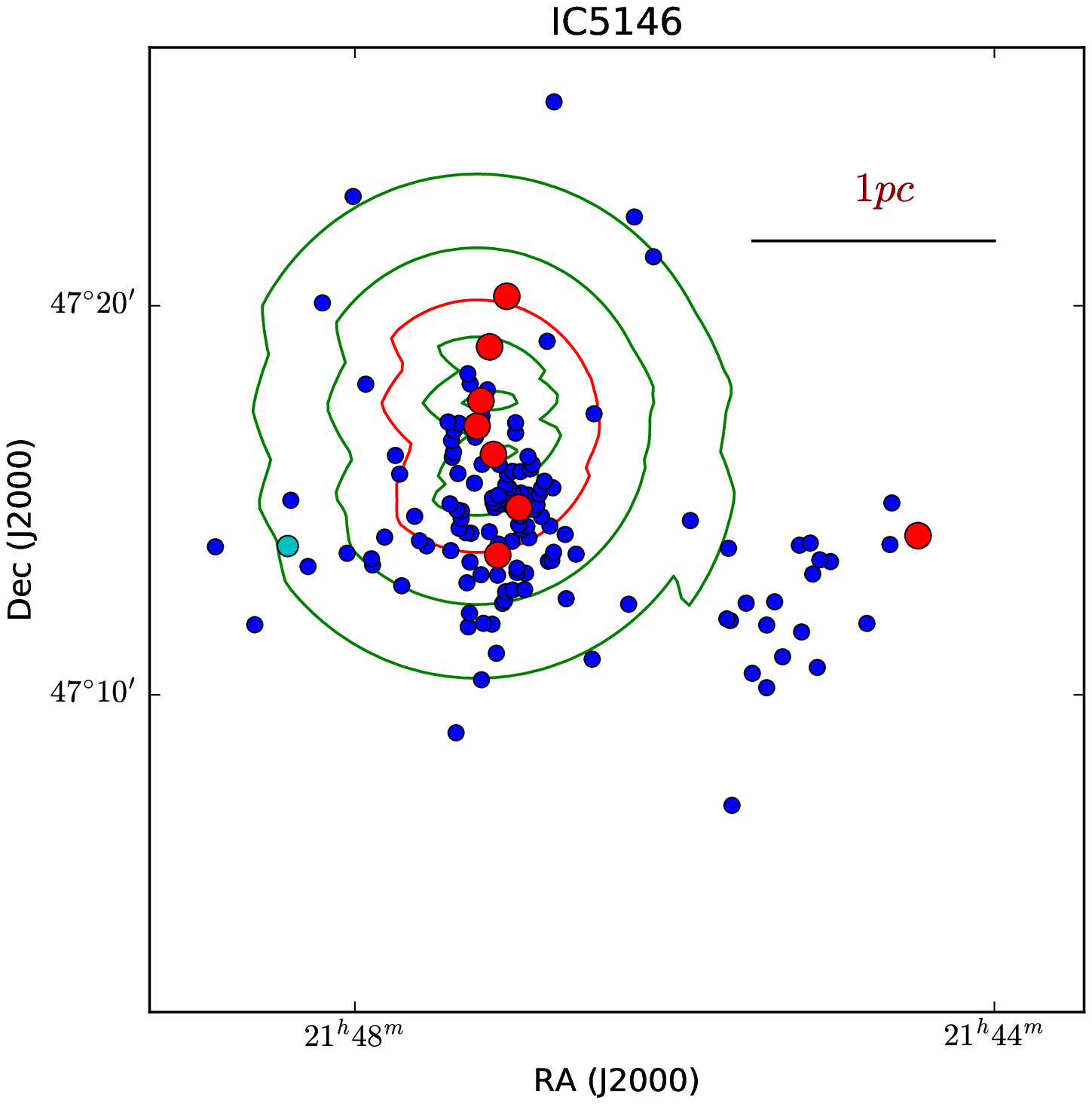}
\includegraphics[scale=0.4]{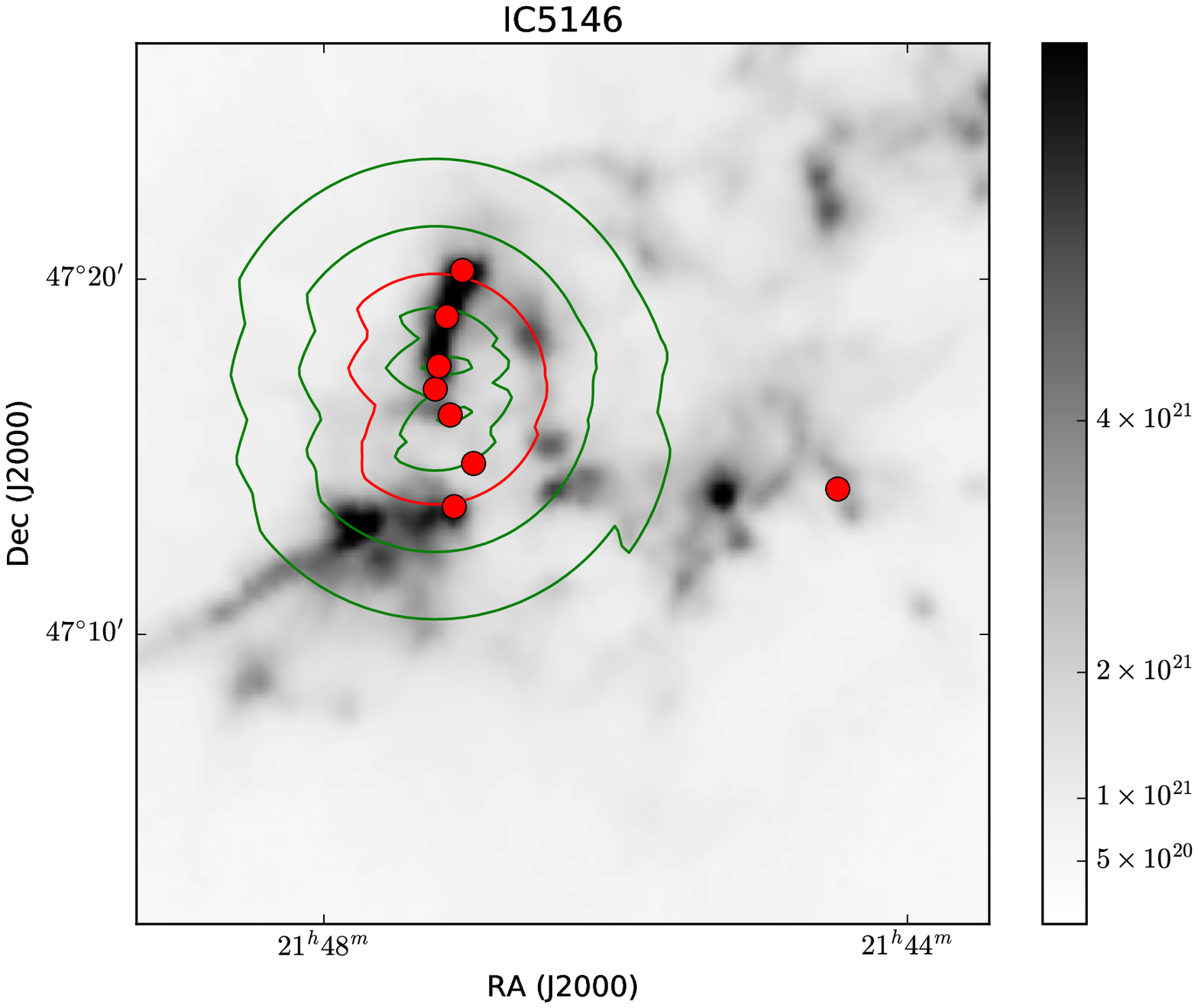}
\caption{Maps of the IC5146-b region. Contours represent the PS surface density, shown at 5\%, 10\%, 20\%, 40\%, and 80\% of peak value. The right panel shows the the {\it Herschel} column density map of the region with the $N=4$ surface density of PS and positions of PS objects (red dots) overlaid.} 
\end{figure}

\begin{figure}[htbp]
   \centering
\includegraphics[scale=0.4]{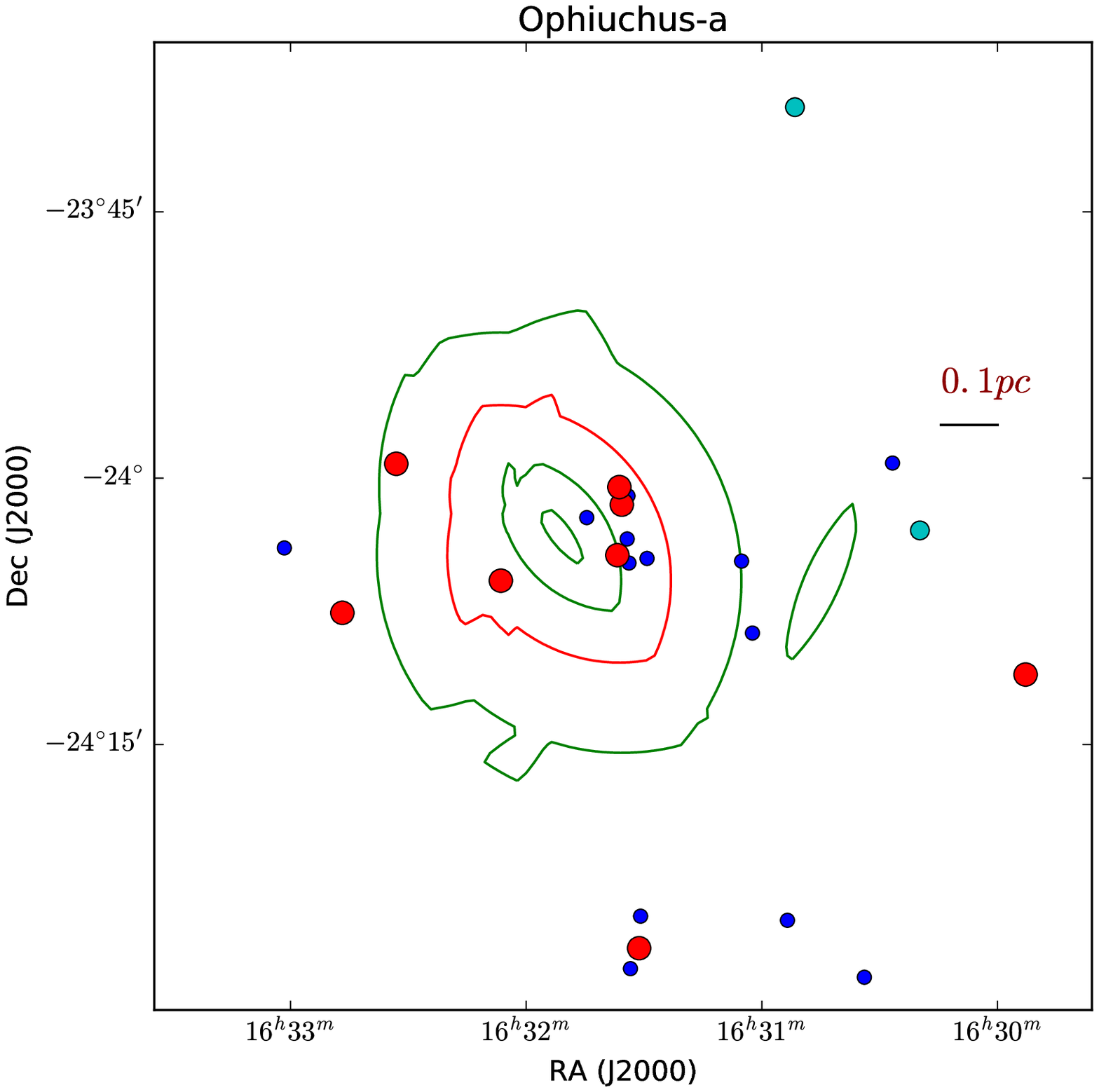}
\includegraphics[scale=0.4]{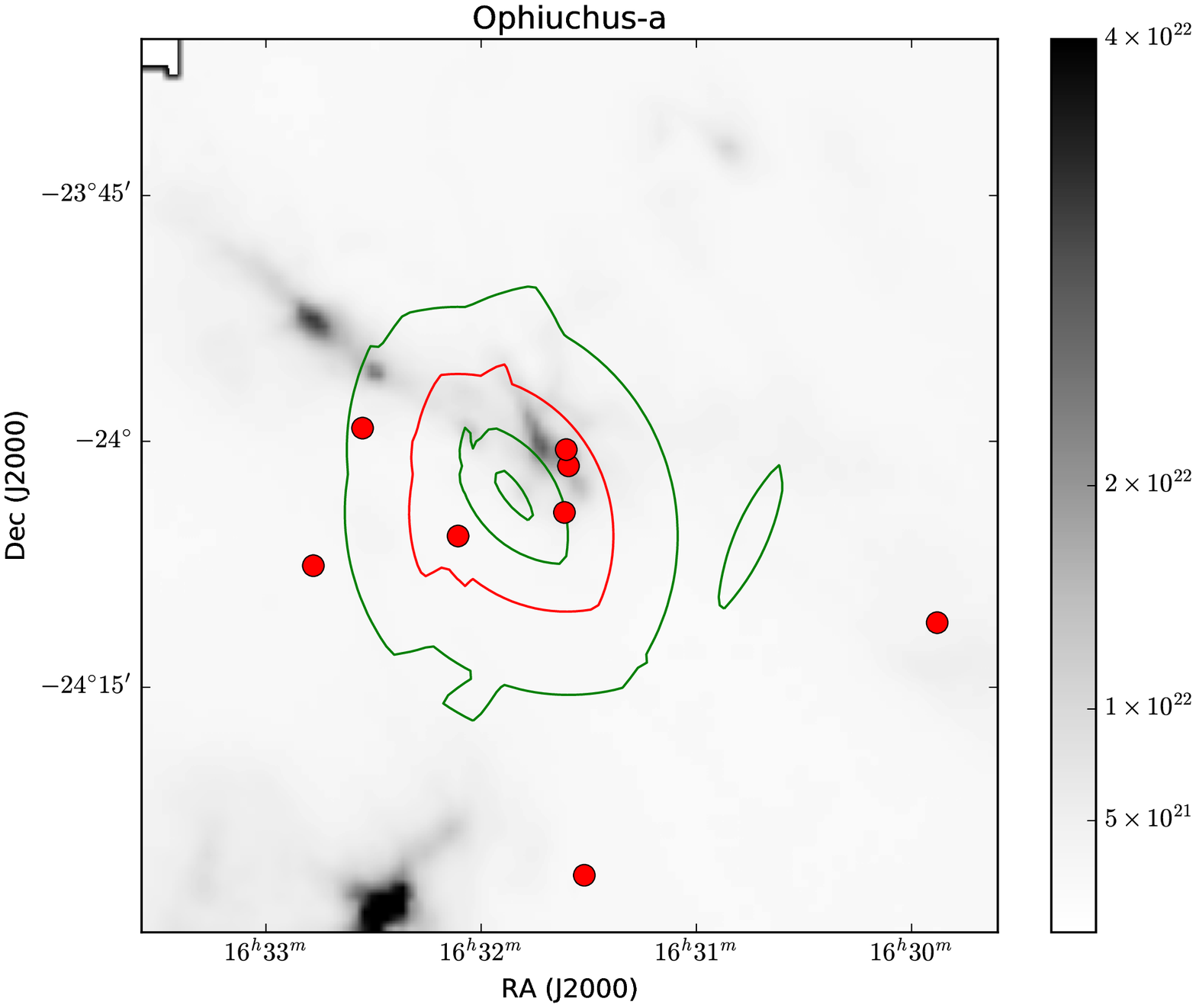}
\caption{Maps of the Ophiuchus-a region. Contours represent the PS surface density, shown at 10\%, 20\%, 40\%, and 80\% of peak value. The right panel shows the the {\it Herschel} column density map of the region with the $N=4$ surface density of PS and positions of PS objects (red dots) overlaid.}
\end{figure}

\begin{figure}[htbp]
   \centering
\includegraphics[scale=0.4]{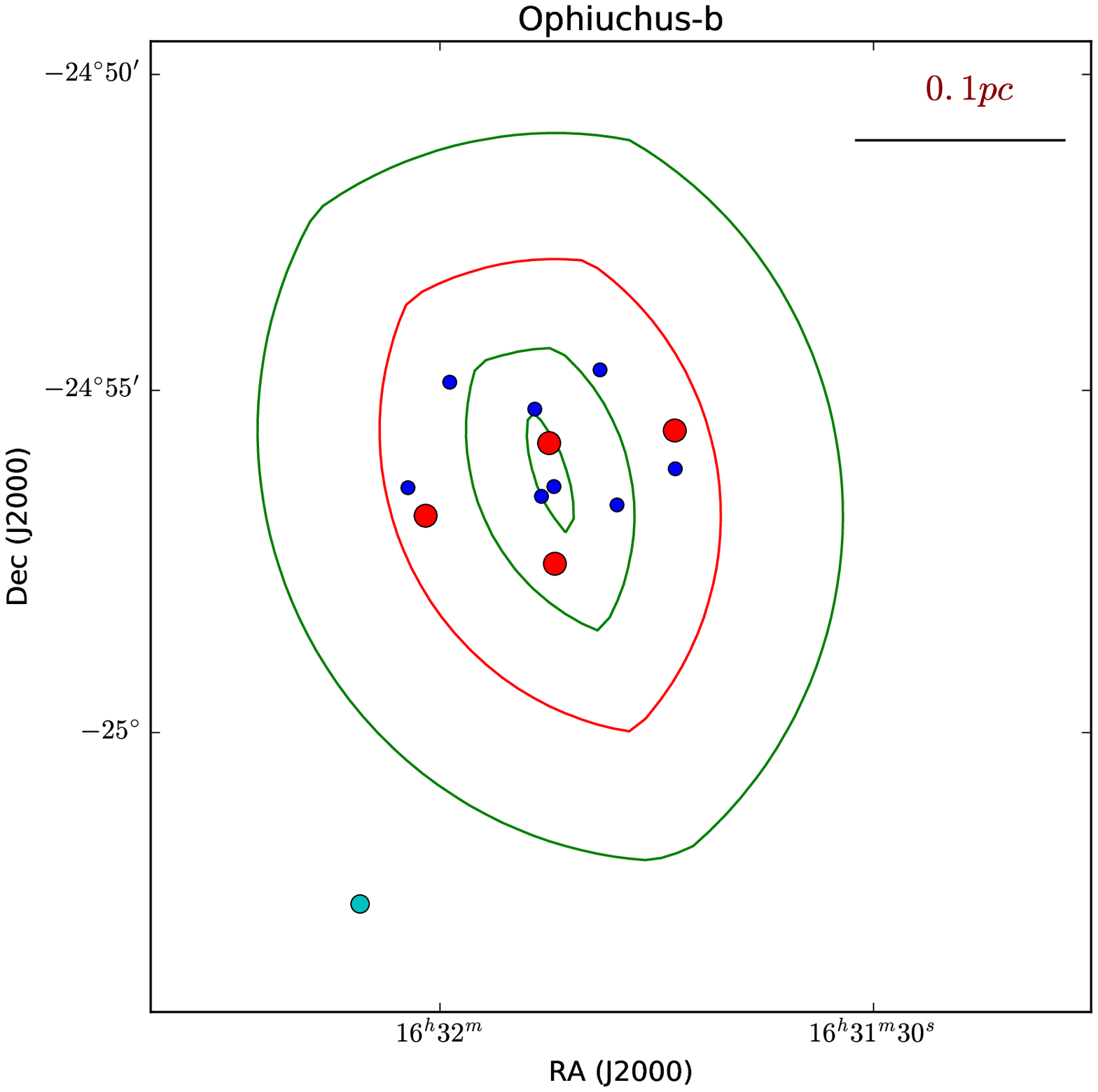}
\includegraphics[scale=0.4]{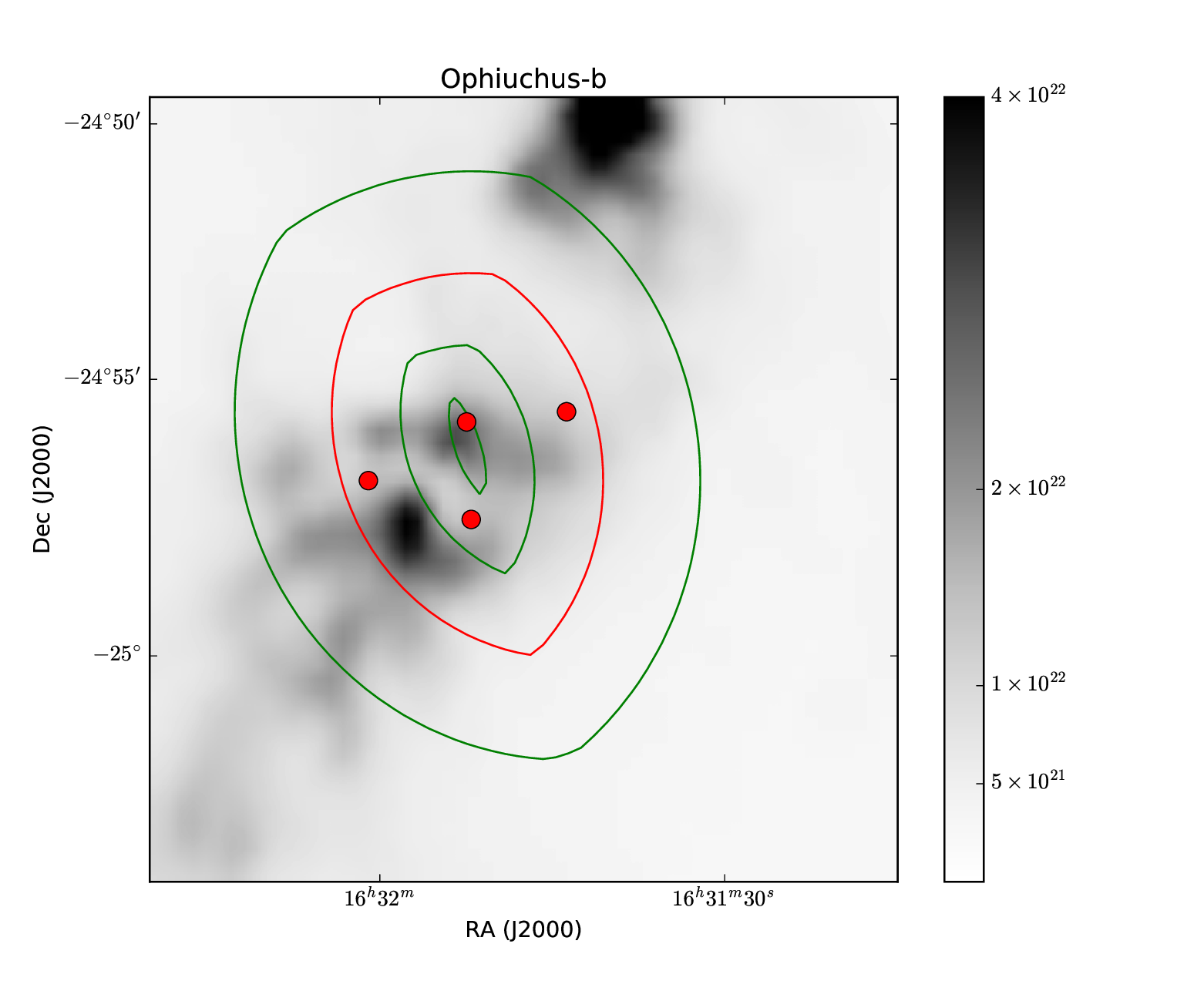}
\caption{Maps of the Ophiuchus-b region. Contours represent the PS surface density, shown at 10\%, 20\%, 40\%, and 80\% of peak value. The right panel shows the the {\it Herschel} column density map of the region with the $N=4$ surface density of PS and positions of PS objects (red dots) overlaid.}
\end{figure}

\begin{figure}[htbp]
   \centering
\includegraphics[scale=0.4]{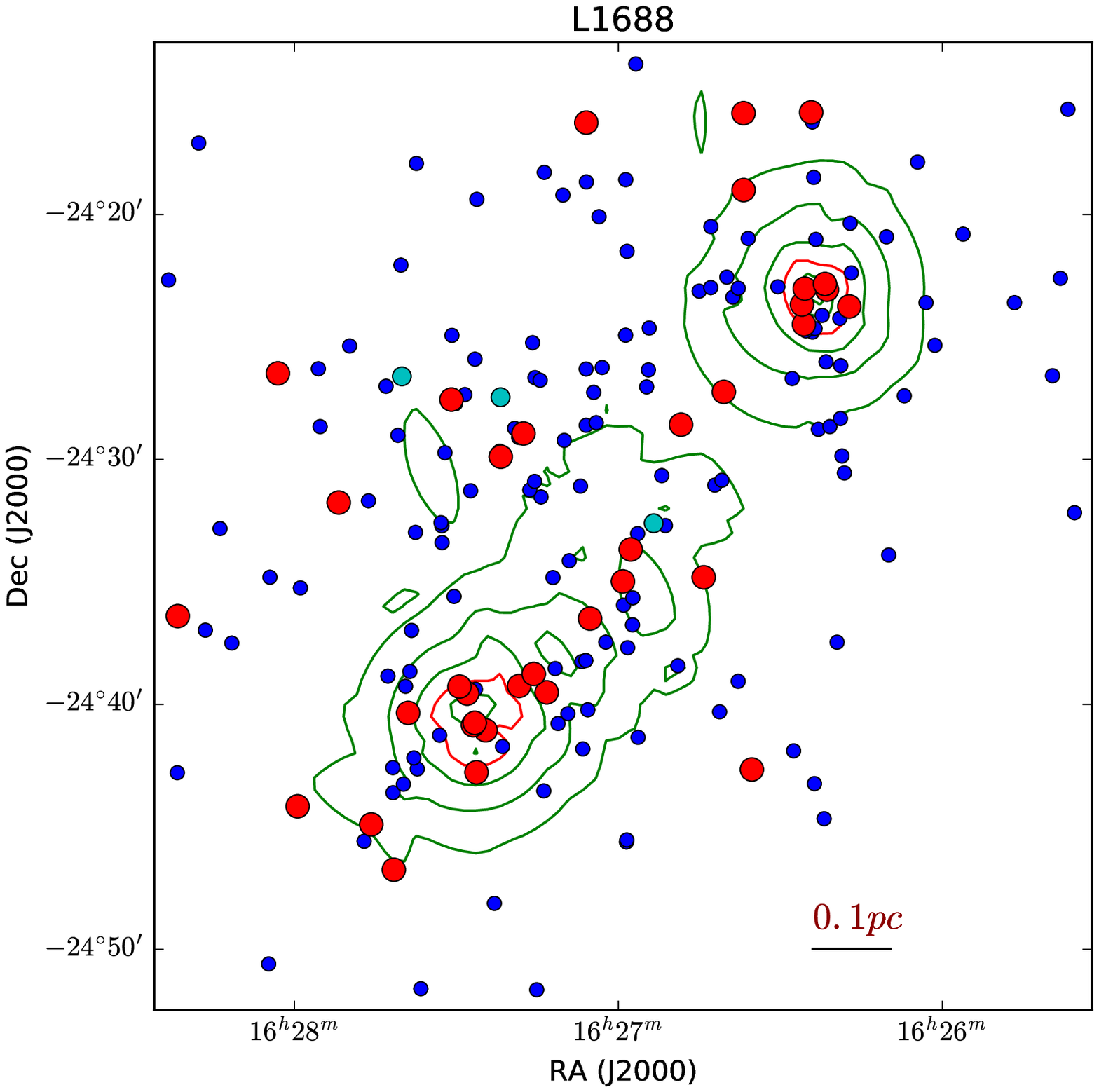}
\includegraphics[scale=0.4]{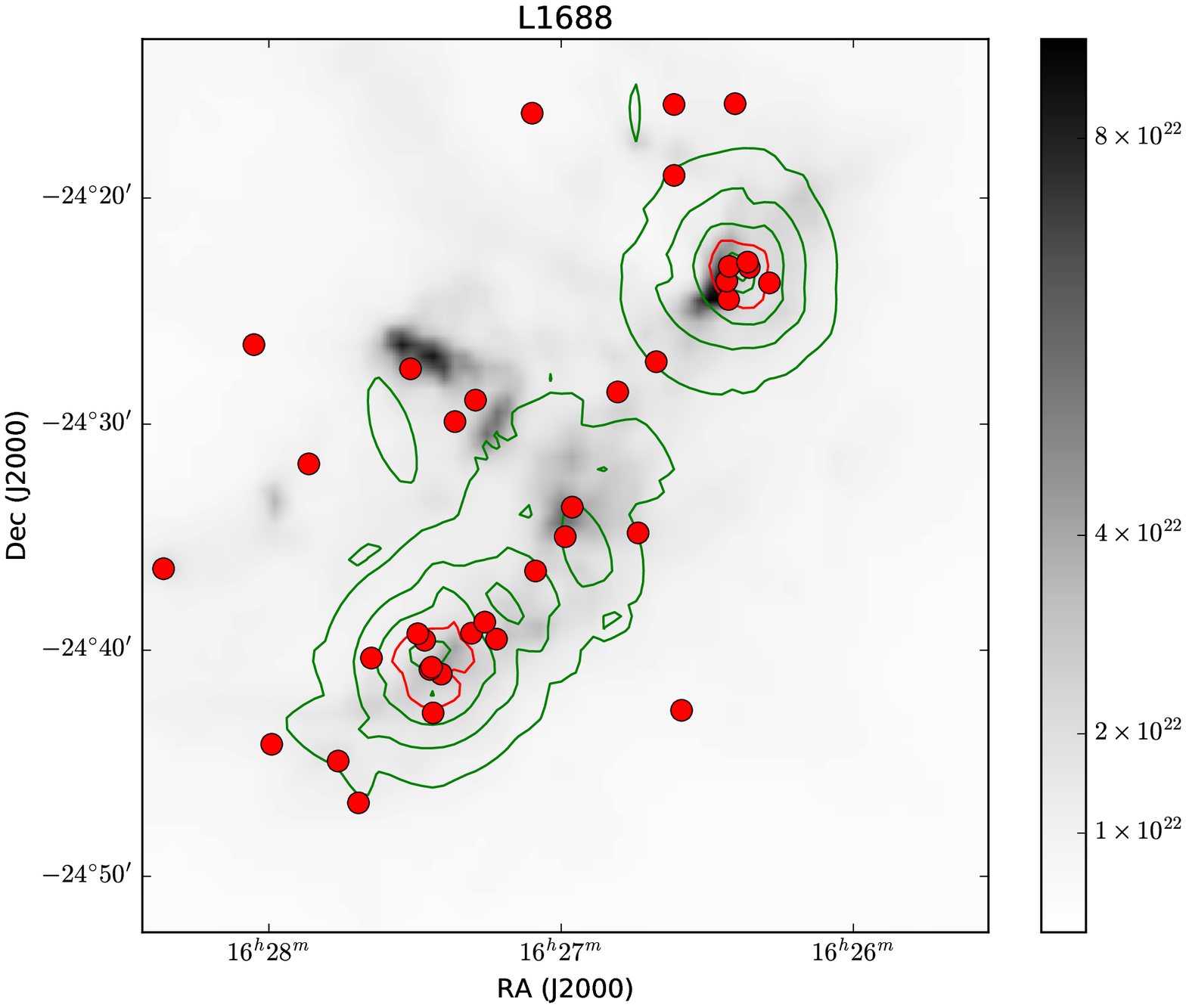}
\caption{Maps of the L1688 region. Contours represent the PS surface density, shown at 2.5\%, 5\%, 10\%, 20\%, 40\%, and 80\% of peak value. The right panel shows the the {\it Herschel} column density map of the region with the $N=4$ surface density of PS and positions of PS objects (red dots) overlaid.}
\end{figure}

\begin{figure}[htbp]
   \centering
\includegraphics[scale=0.4]{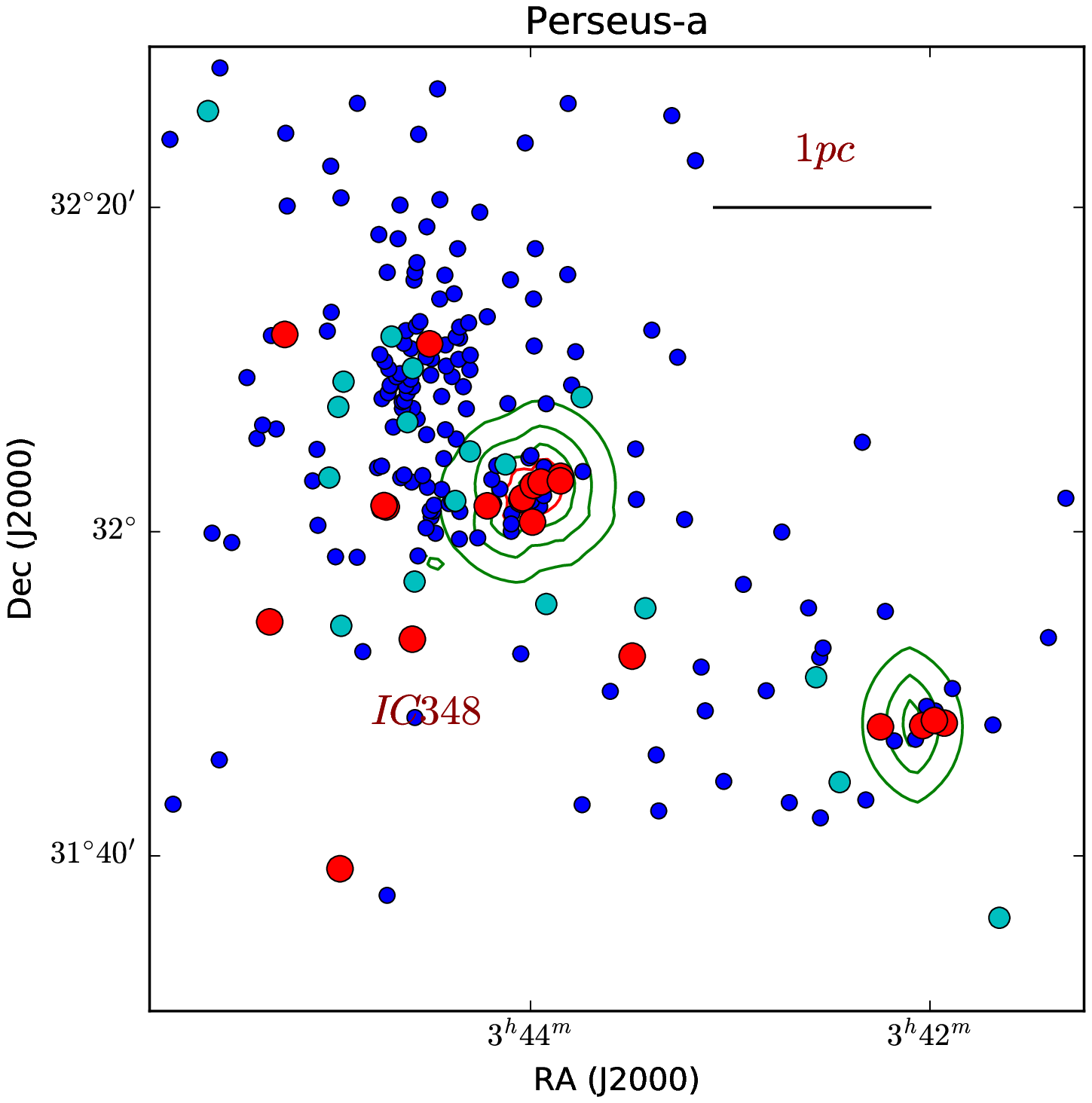}
\includegraphics[scale=0.4]{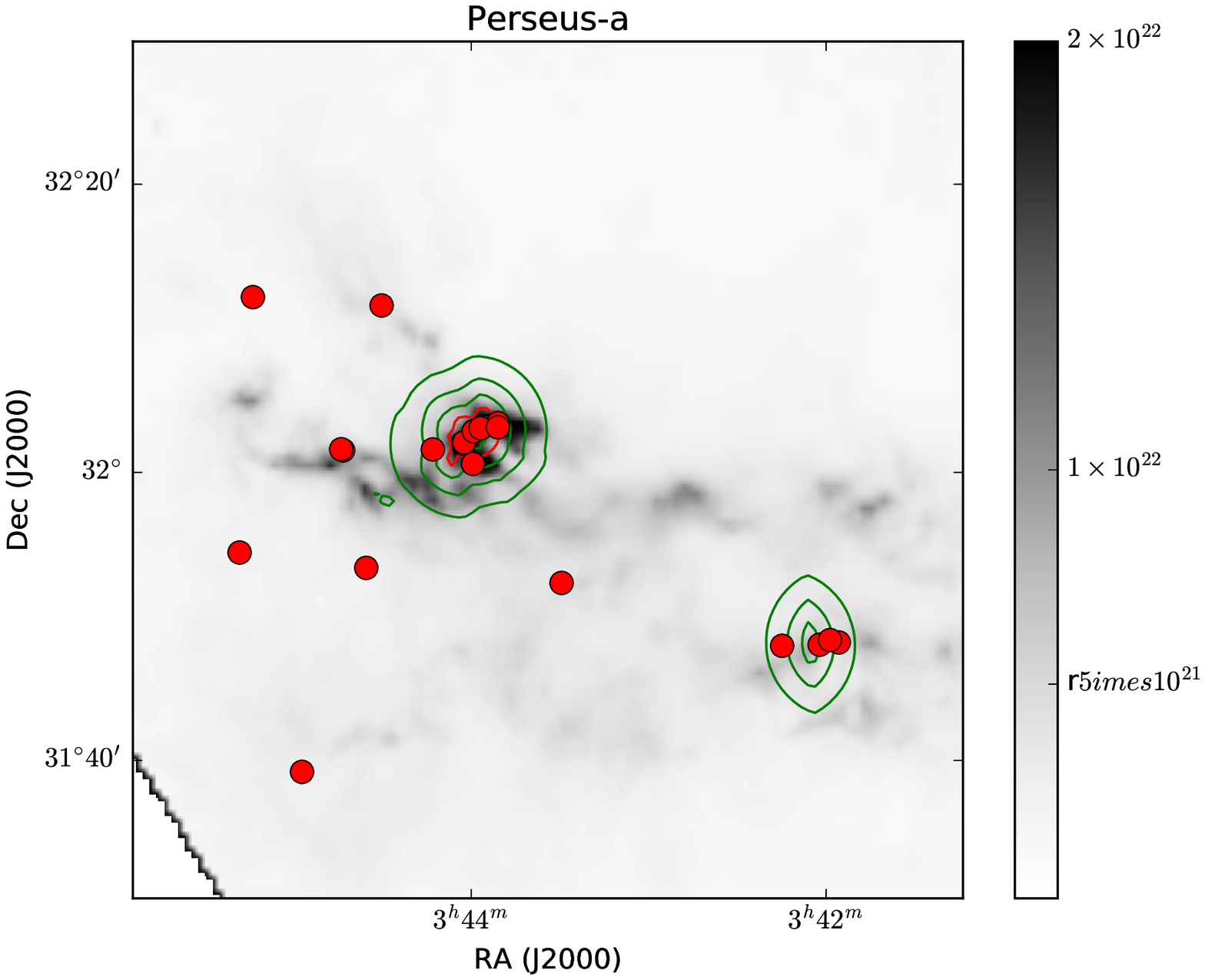}
\caption{Maps of the Perseus-a region. Contours represent the PS surface density, shown at 2.5\%, 5\%, 10\%, 20\%, 40\%, and 80\% of peak value. The right panel shows the the {\it Herschel} column density map of the region with the $N=4$ surface density of PS and positions of PS objects (red dots) overlaid.}
\end{figure}

\begin{figure}[htbp]
   \centering
\includegraphics[scale=0.4]{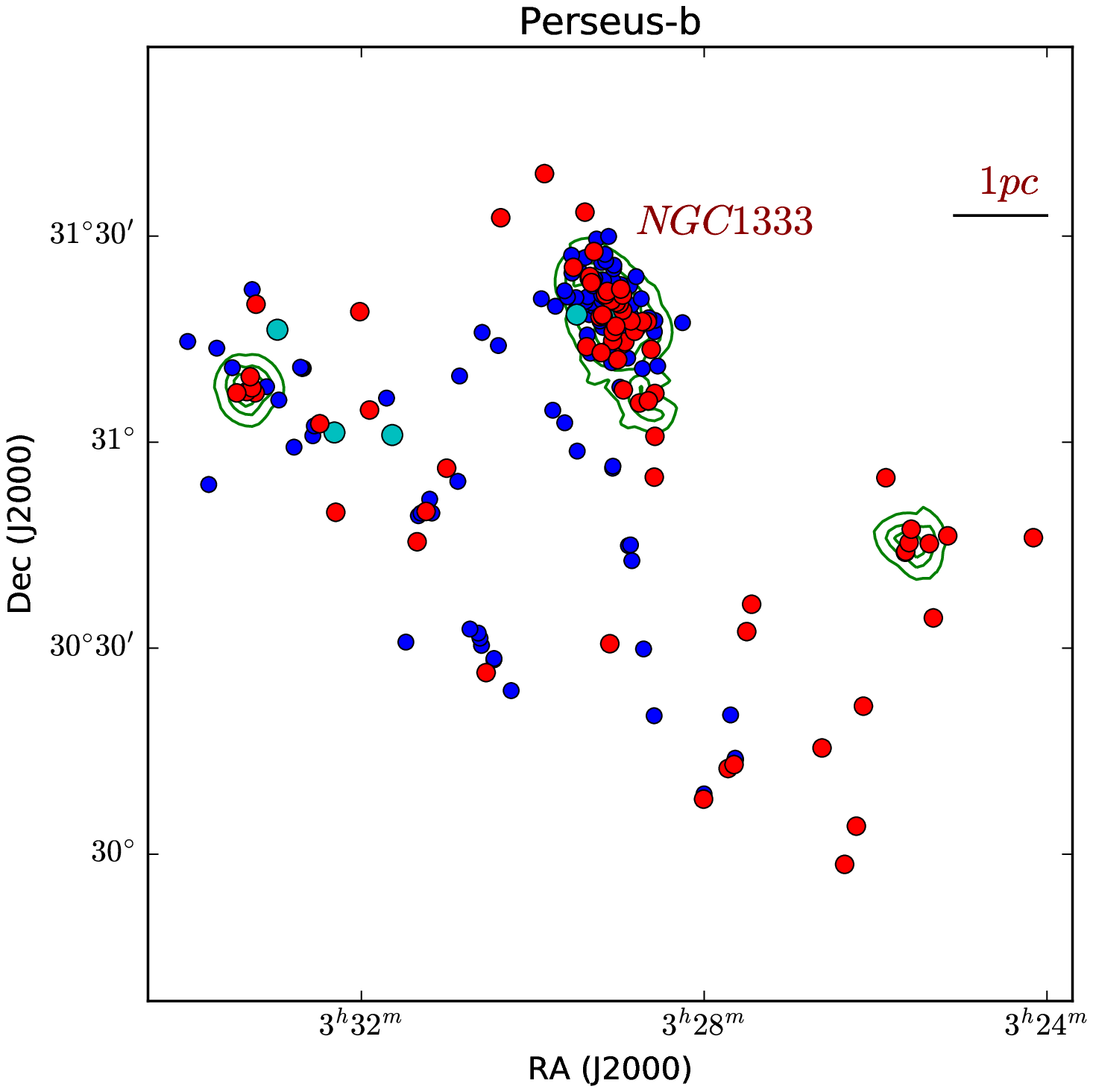}
\includegraphics[scale=0.4]{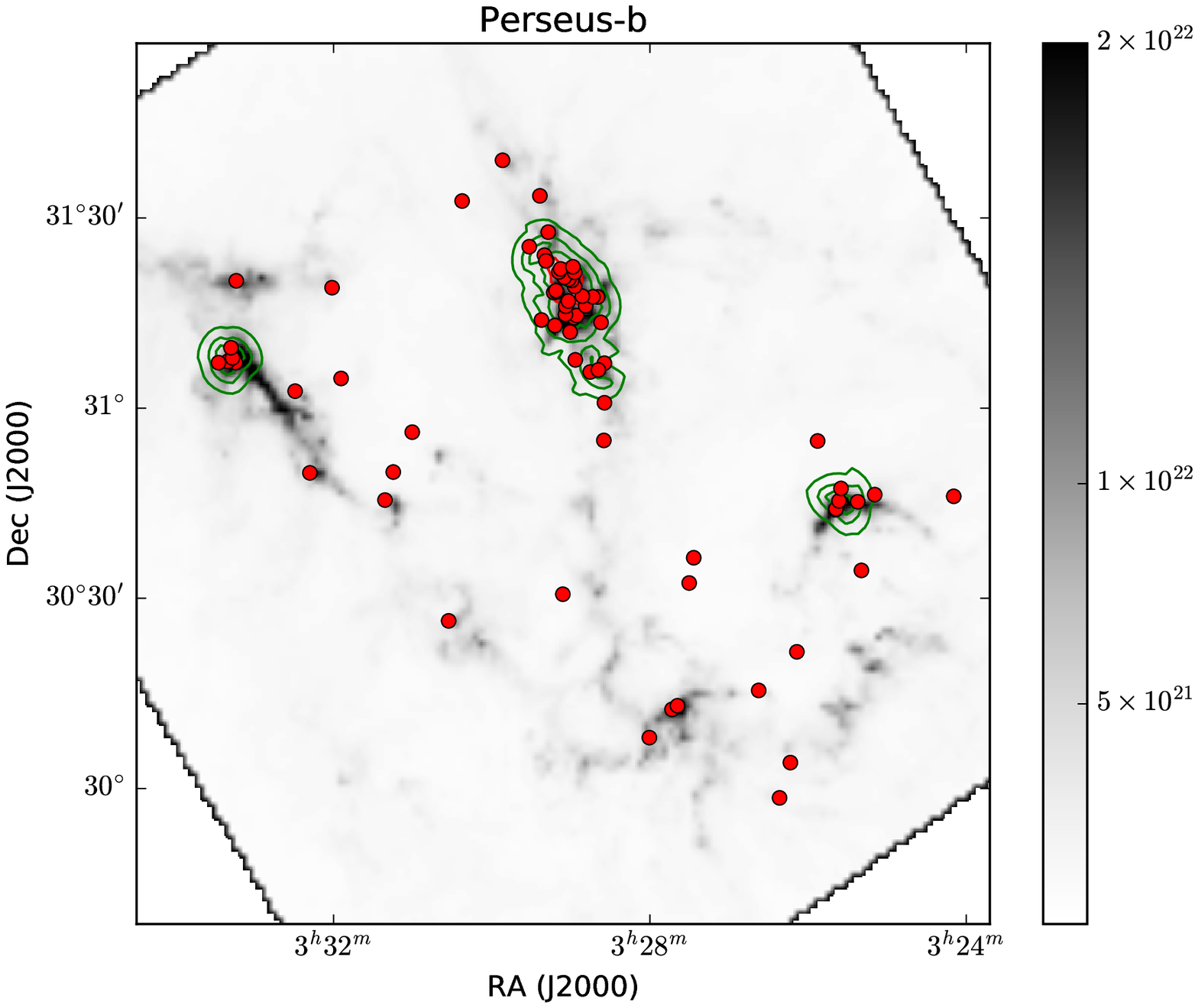}
\includegraphics[scale=0.4]{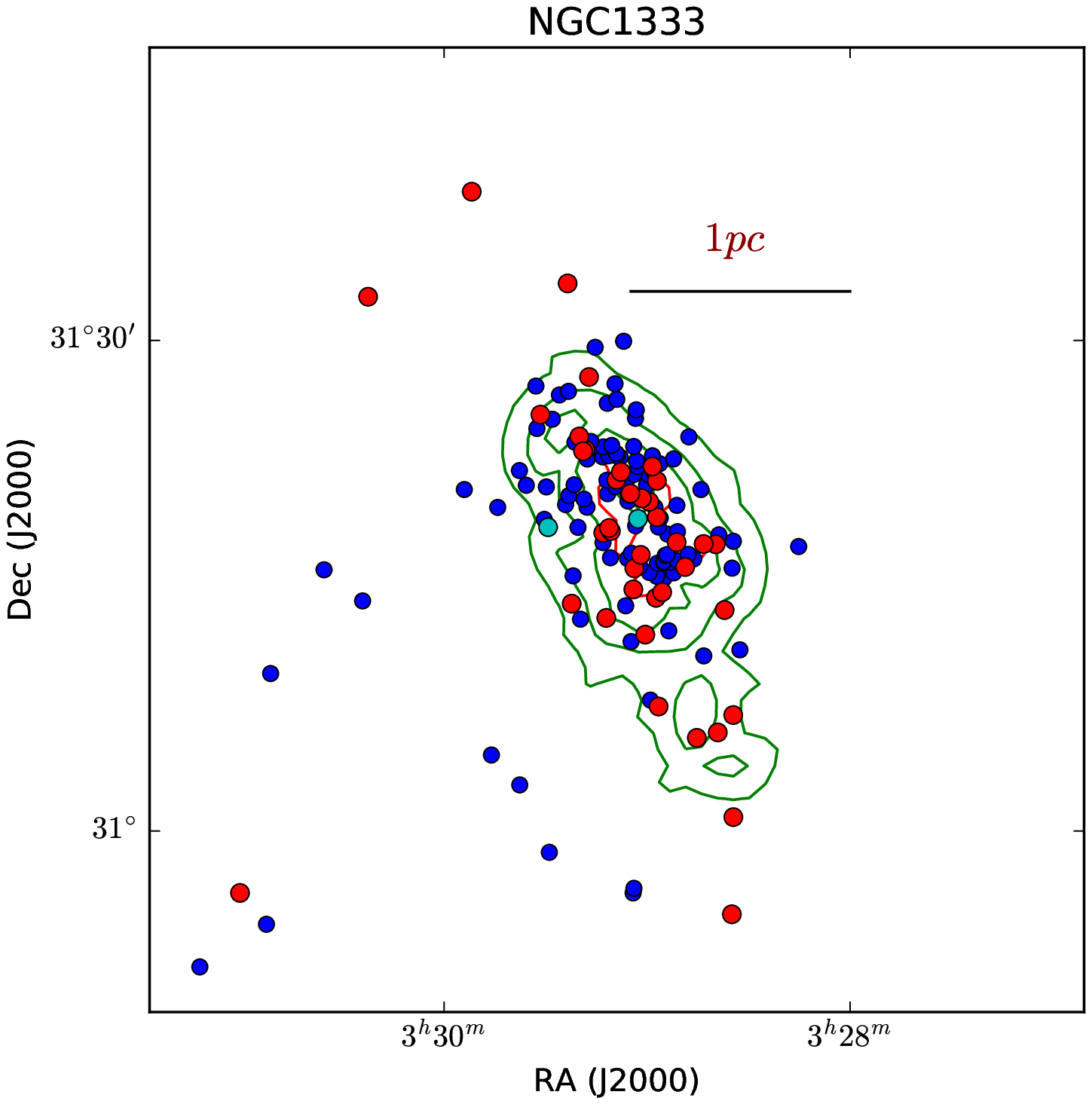}
\includegraphics[scale=0.4]{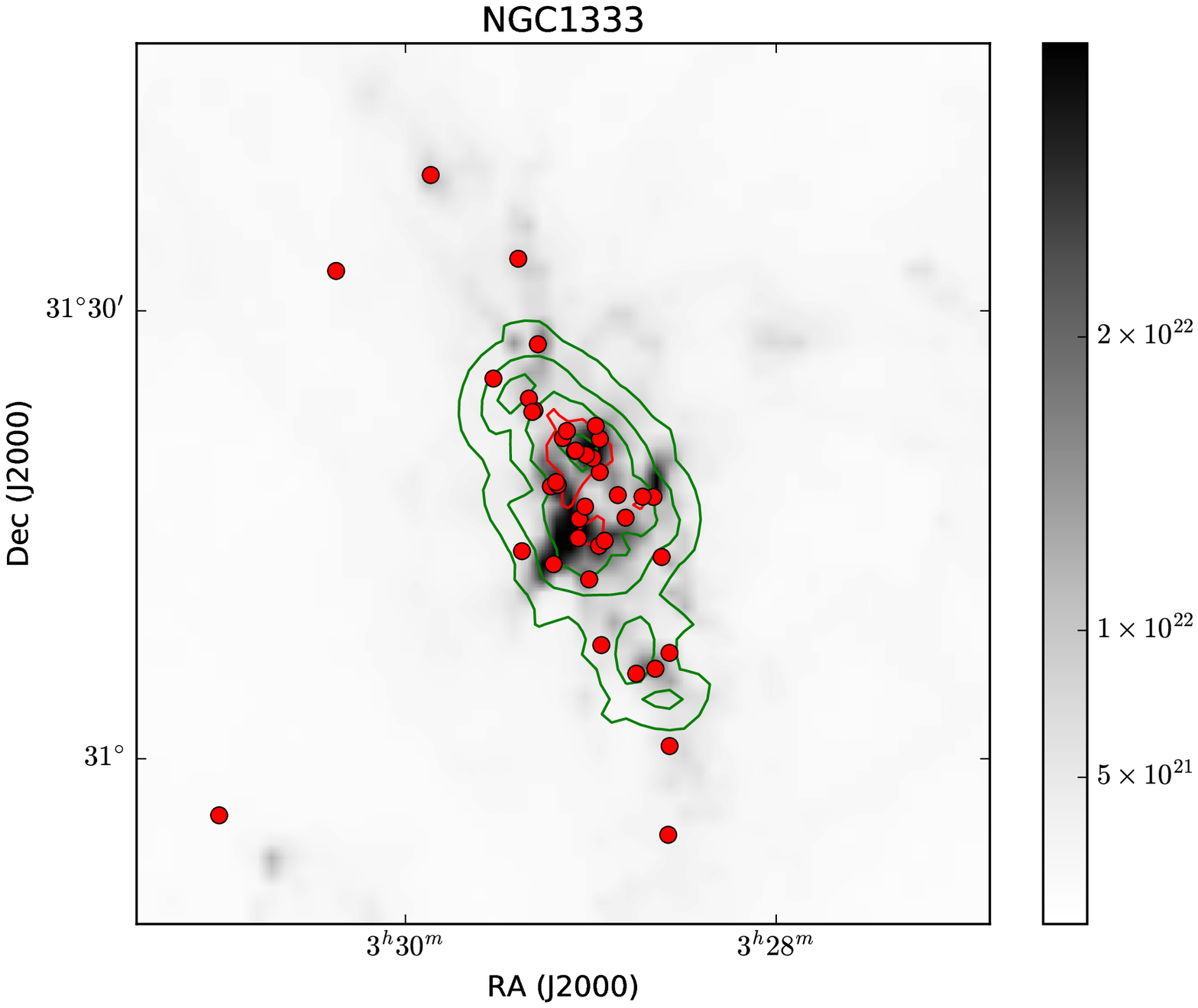}
\caption{Maps of the Perseus-b region. Contours represent the PS surface density, shown at 2.5\%, 5\%, 10\%, 20\%, 40\%, and 80\% of peak value. The right panel shows the the {\it Herschel} column density map of the region with the $N=4$ surface density of PS and positions of PS objects (red dots) overlaid.}
\end{figure}

\begin{figure}[htbp]
   \centering
\includegraphics[scale=0.4]{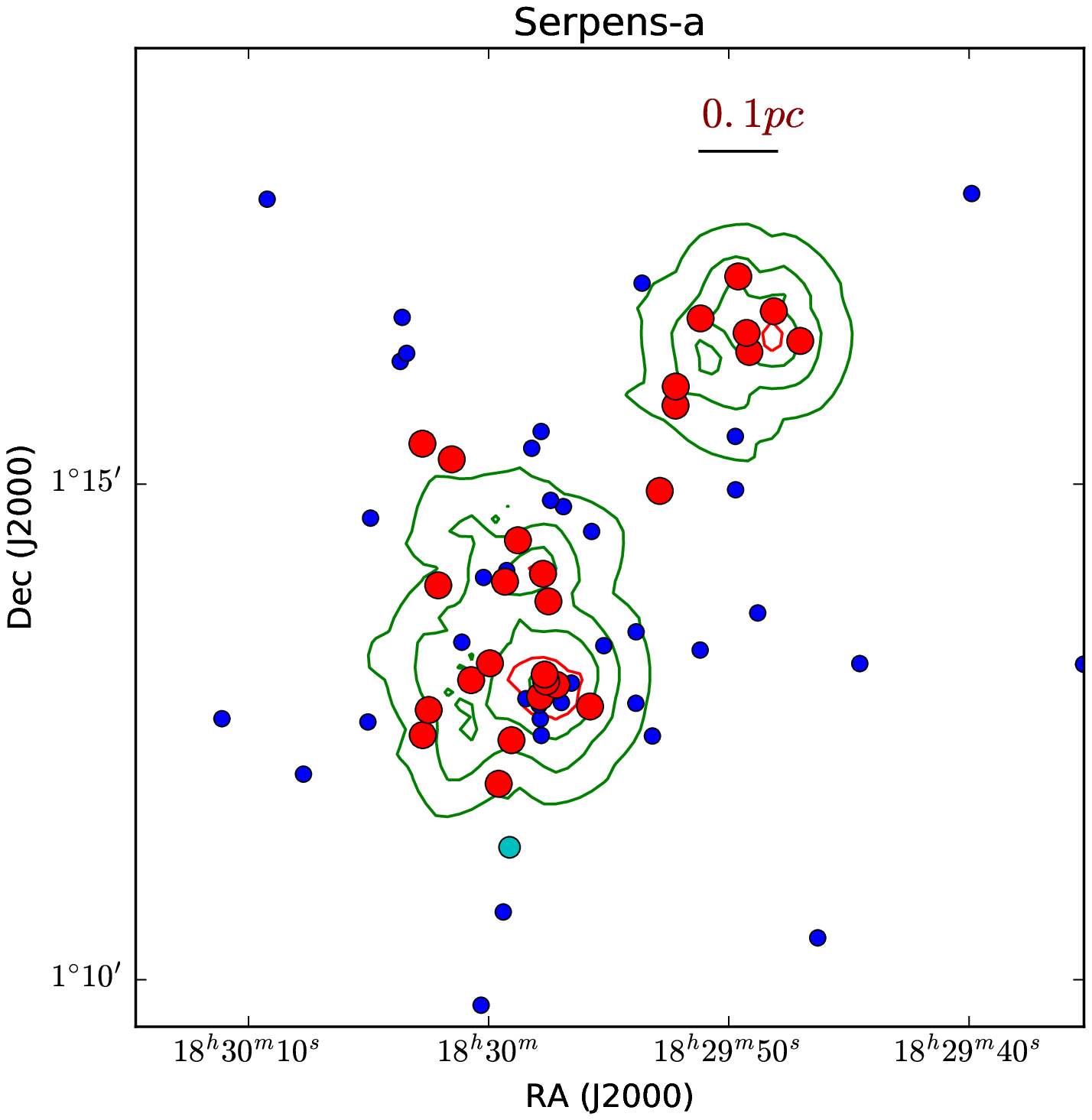}
\includegraphics[scale=0.4]{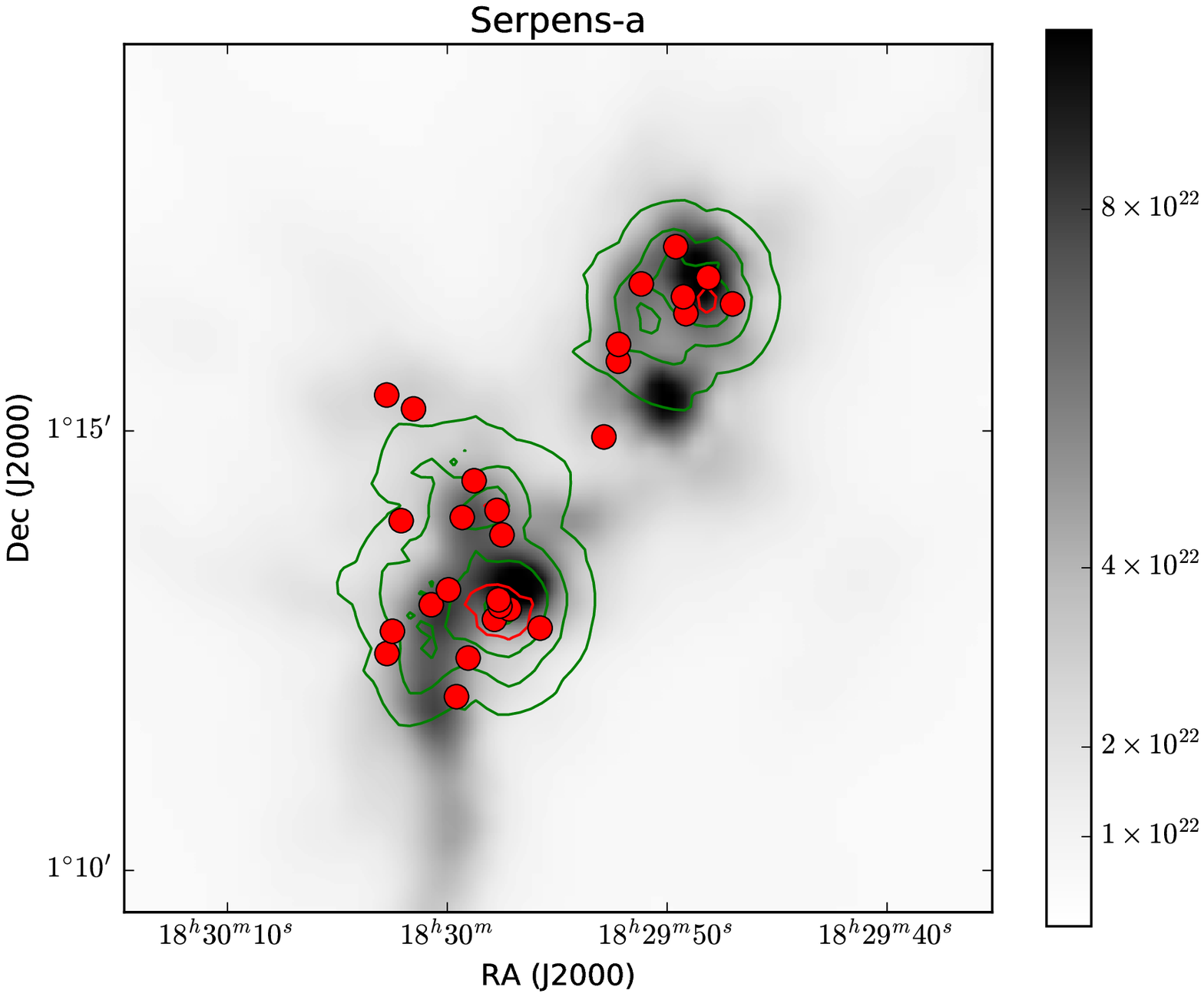}
\caption{Maps of the Serpens-a region. Contours represent the PS surface density, shown at 10\%, 20\%, 40\%, and 80\% of peak value. The right panel shows the the {\it Herschel} column density map of the region with the $N=4$ surface density of PS and positions of PS objects (red dots) overlaid.}
\end{figure}

\begin{figure}[htbp]
   \centering
\includegraphics[scale=0.4]{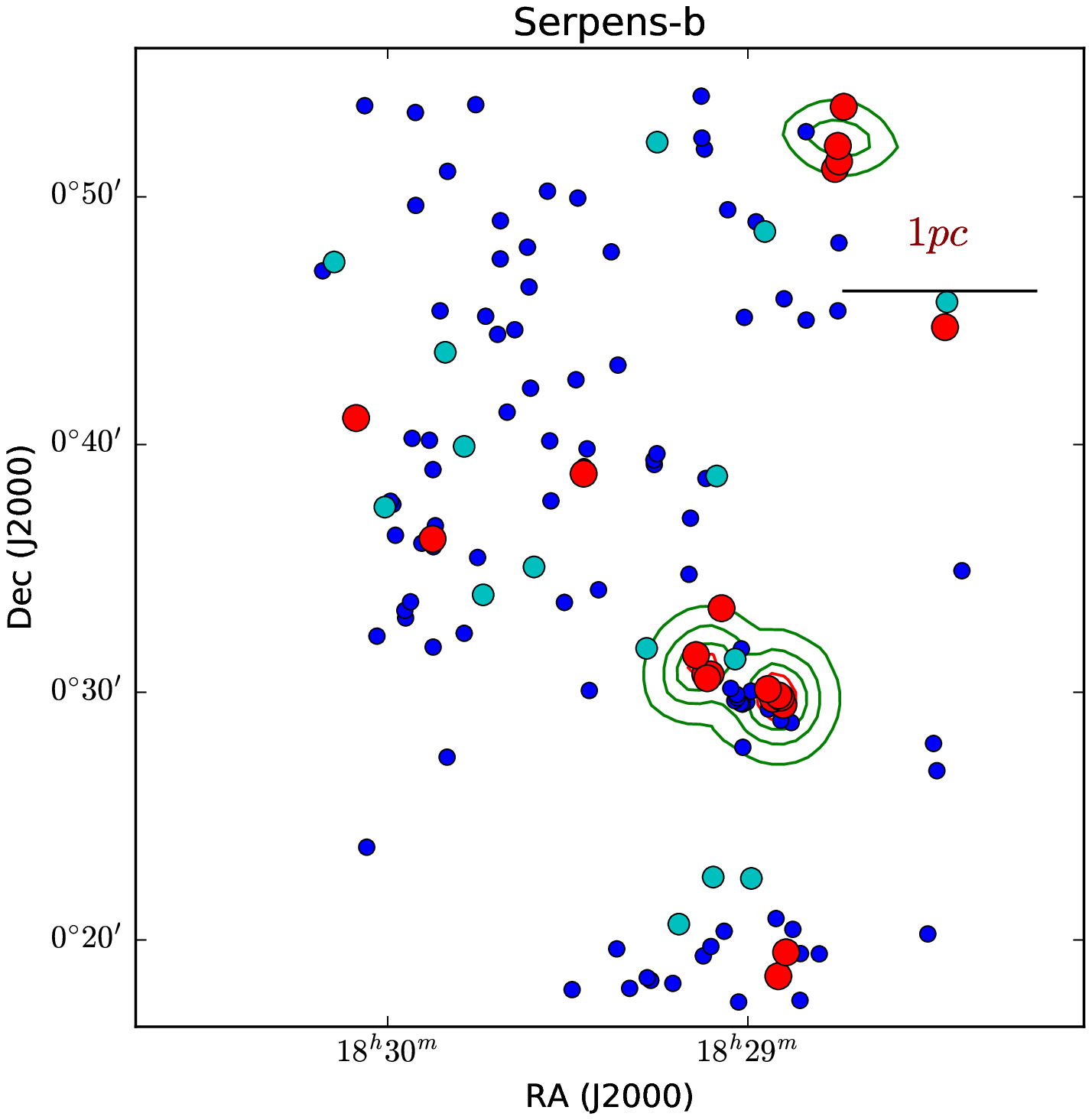}
\includegraphics[scale=0.4]{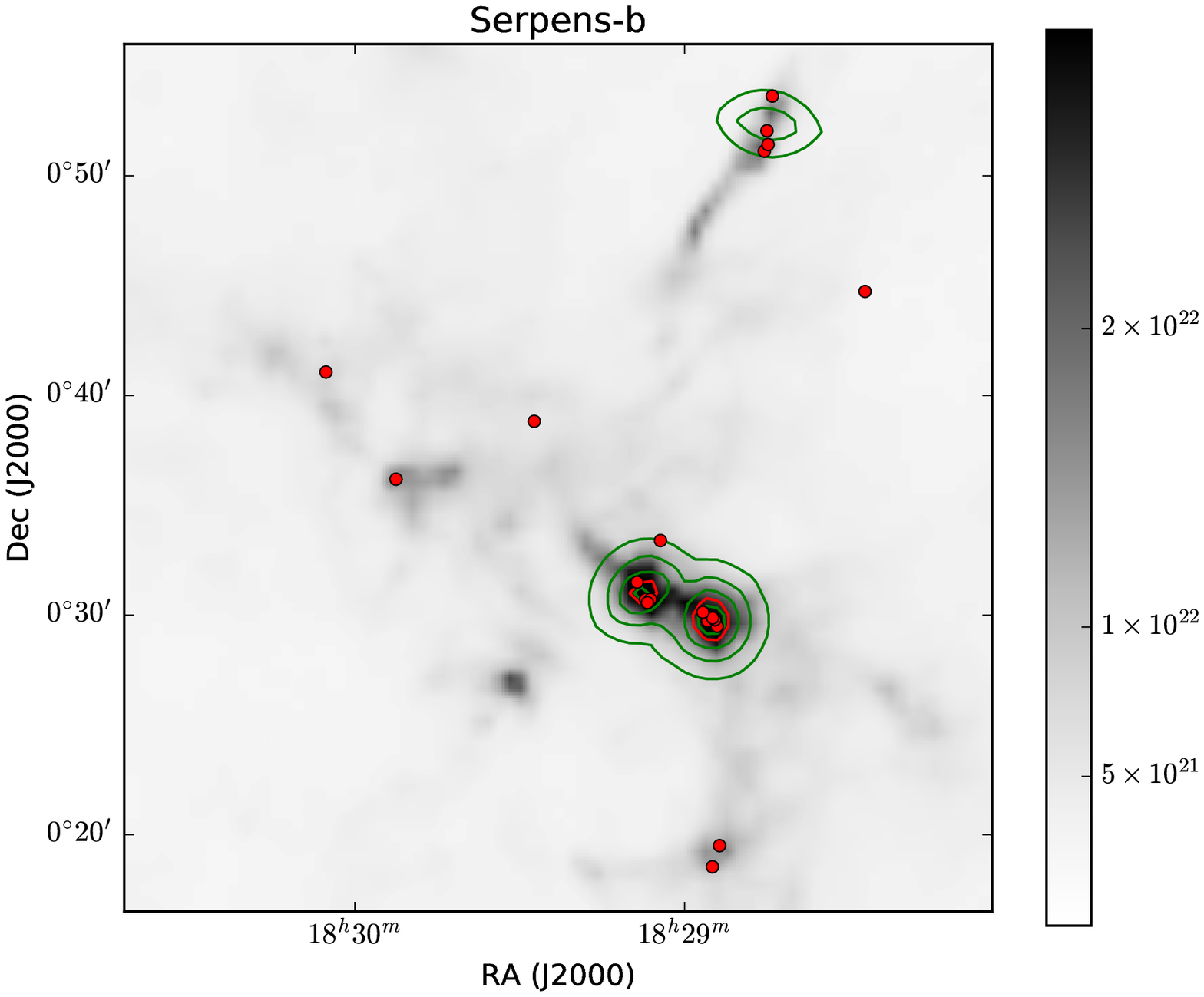}
\caption{Maps of the Serpens-b region. Contours represent the PS surface density, shown at 10\%, 20\%, 40\%, and 80\% of peak value. The right panel shows the the {\it Herschel} column density map of the region with the $N=4$ surface density of PS and positions of PS objects (red dots) overlaid.}
\end{figure}

\begin{figure}[]
\begin{center}
\includegraphics[width=4.5in]{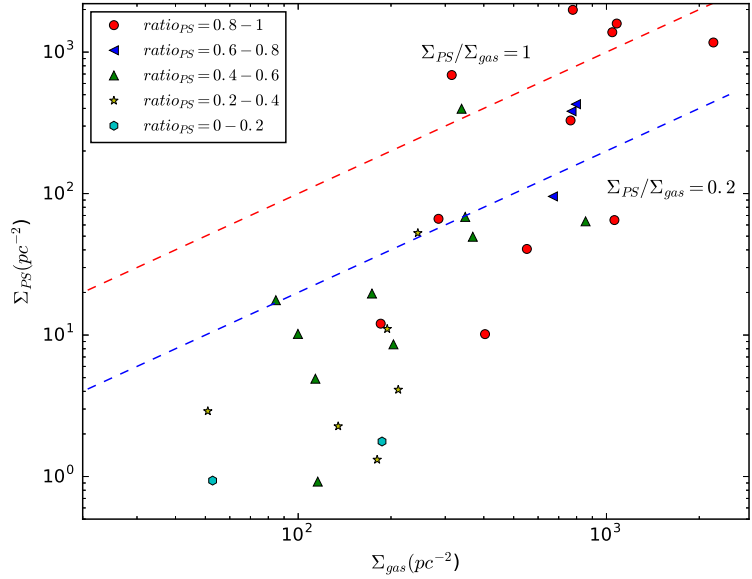}
\vspace*{0.5 cm} \caption{\label{psgas} Surface densities of PS objects versus the gas column densities. The red circles indicate clusters with PS ratios ($ratio_{PS}$) higher than 0.8. The blue triangles indicate clusters with $ratio_{PS}$ ranging from 0.6 to 0.8. The green triangles indicate clusters with $ratio_{PS}$ ranging from 0.4 to 0.6. The yellow stars indicate clusters with $ratio_{PS}$ ranging from 0.2 to 0.4. The diamonds indicate clusters with $ratio_{PS}$ below 0.2. }
\end{center}
\end{figure}

\begin{figure}[htbp]
\centering
\includegraphics[scale=0.4]{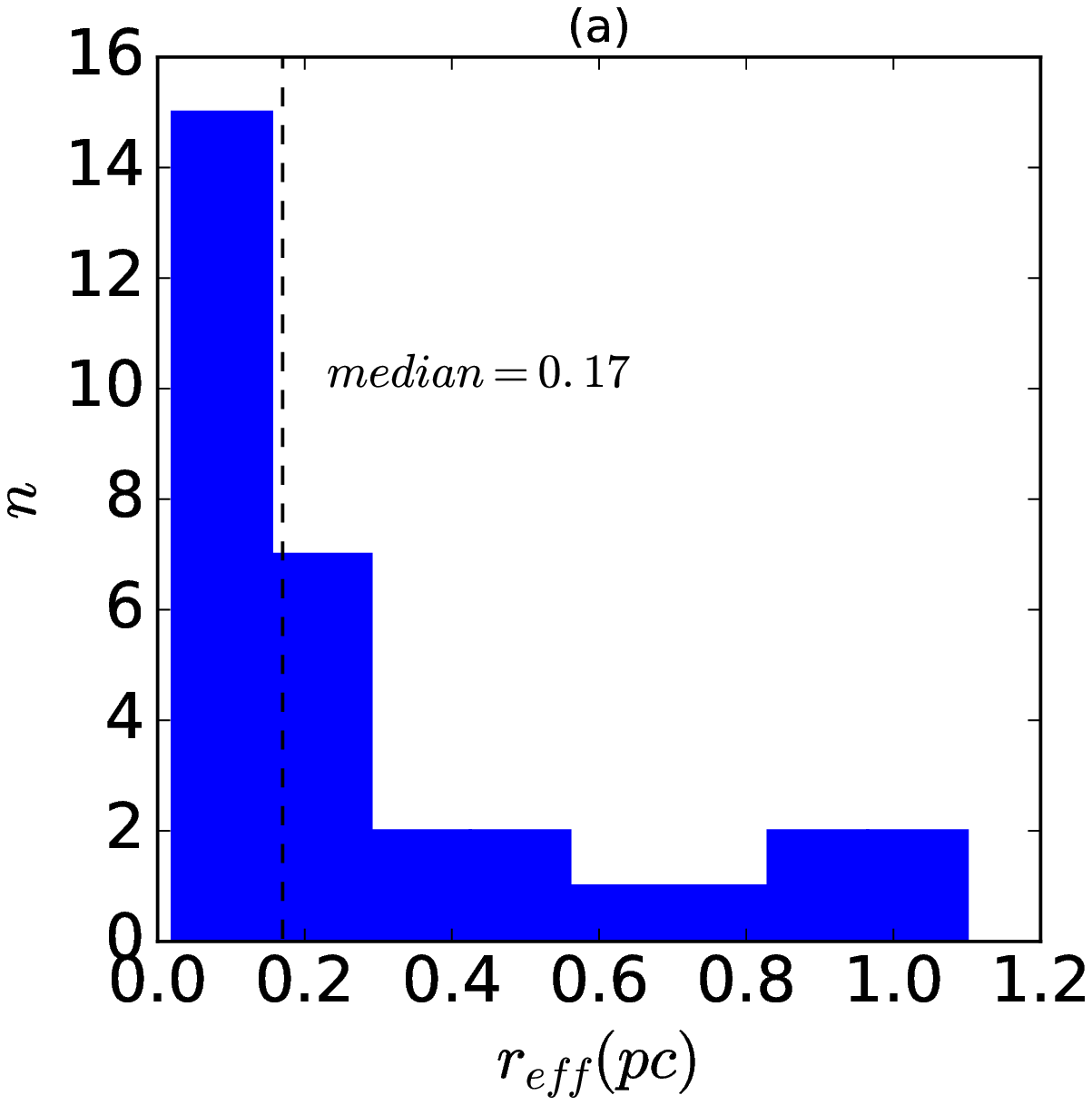}
\includegraphics[scale=0.4]{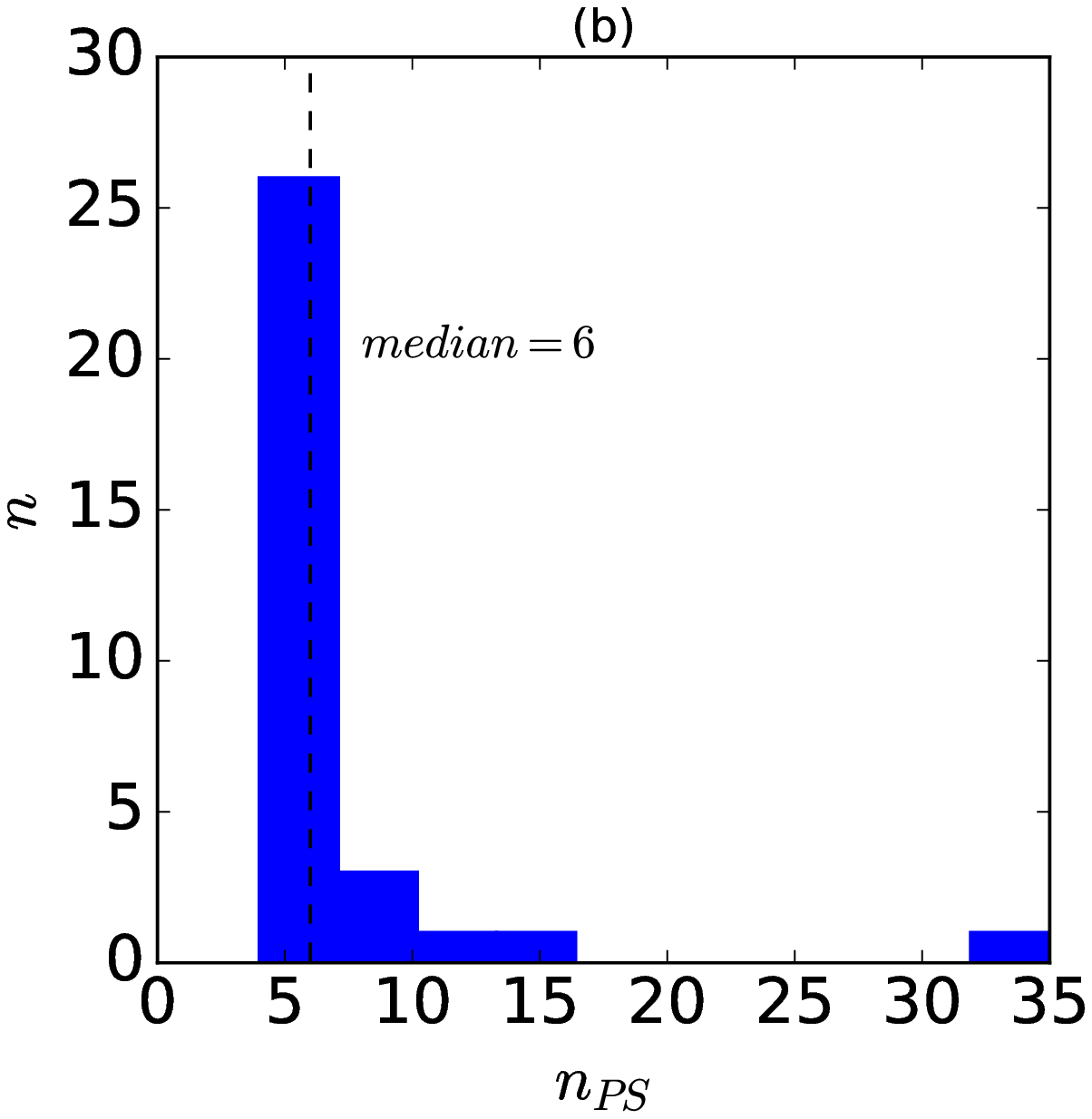}
\includegraphics[scale=0.4]{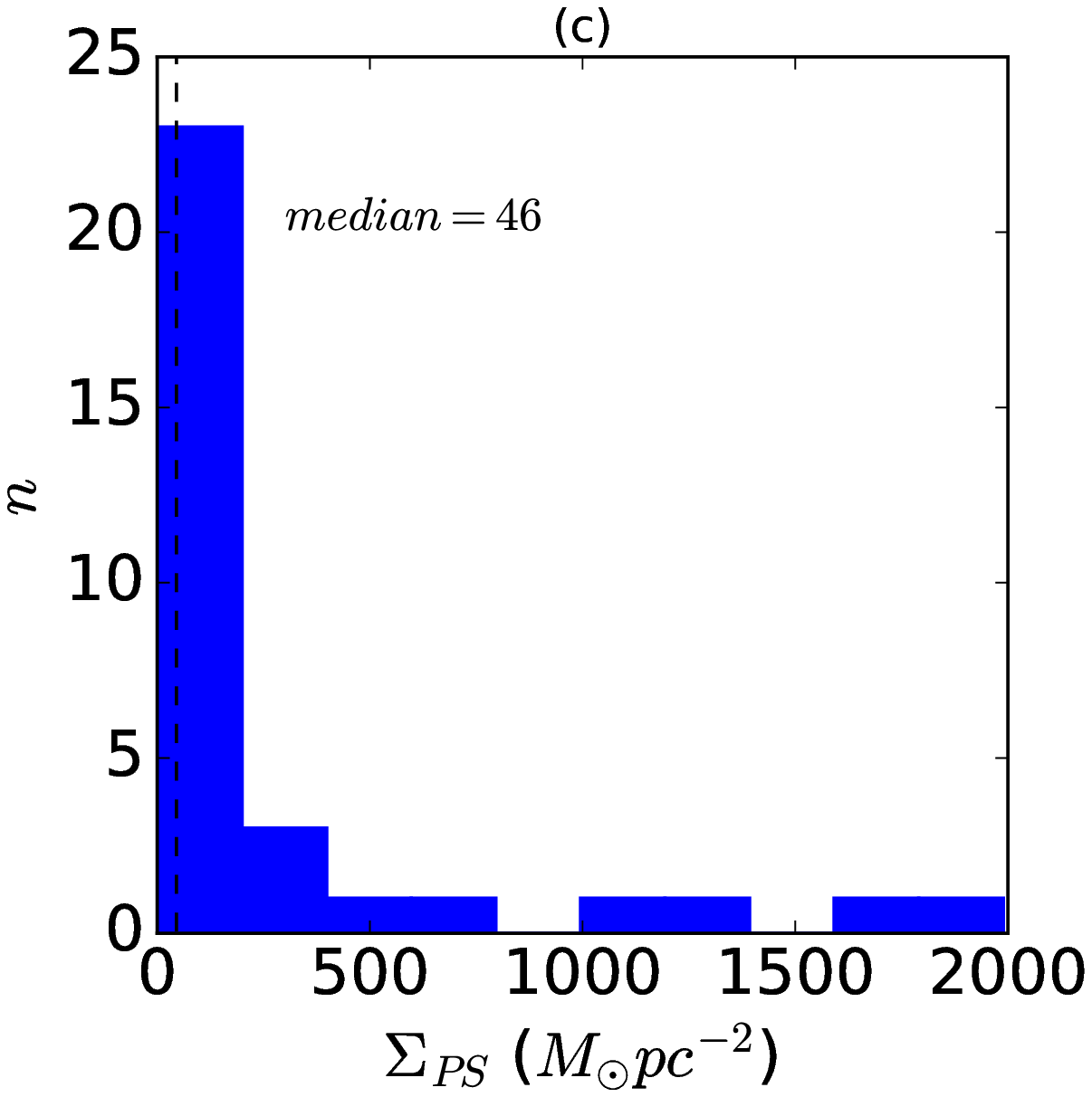}
\includegraphics[scale=0.4]{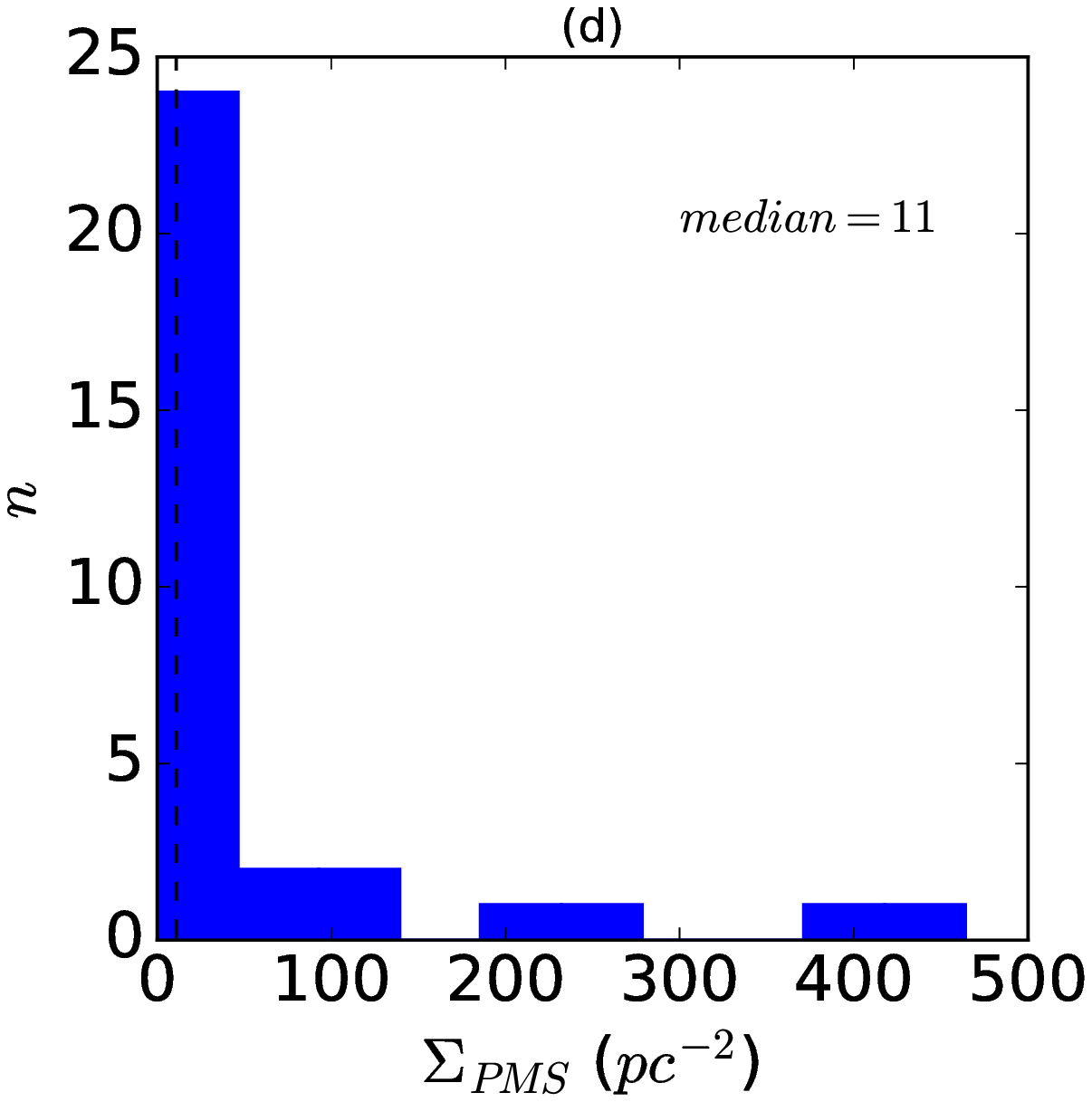}
\includegraphics[scale=0.4]{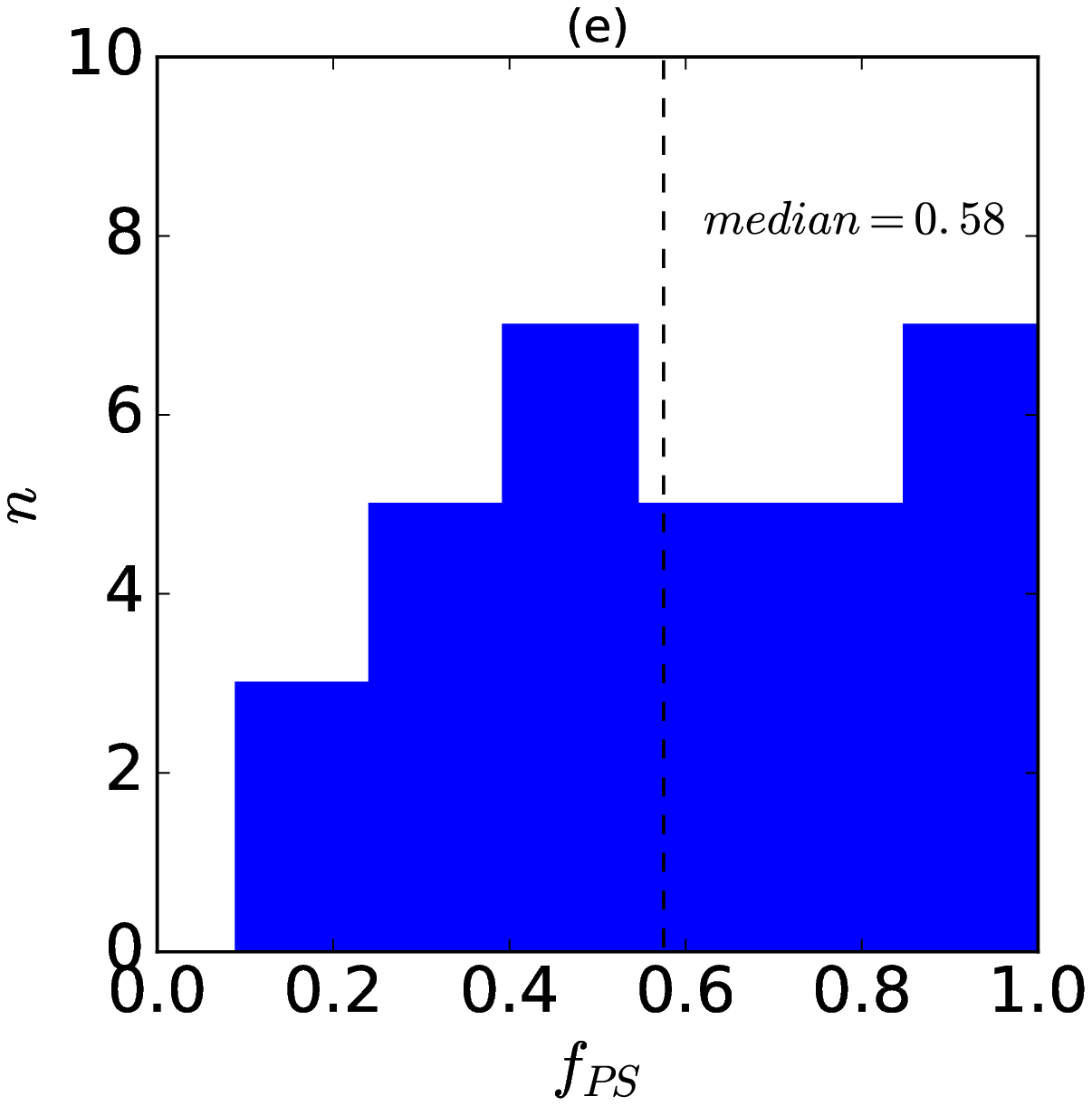}
\includegraphics[scale=0.4]{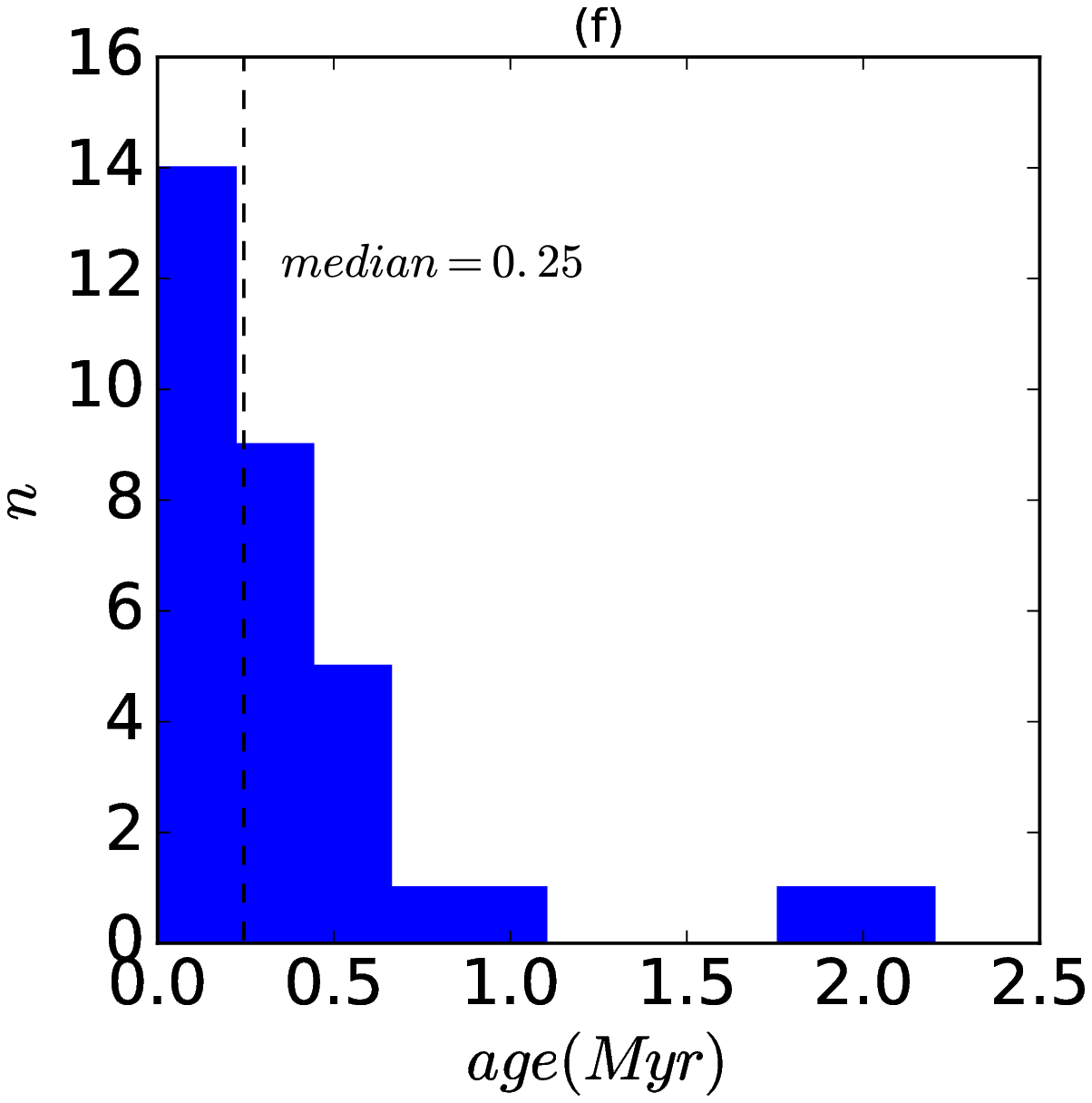}
\includegraphics[scale=0.4]{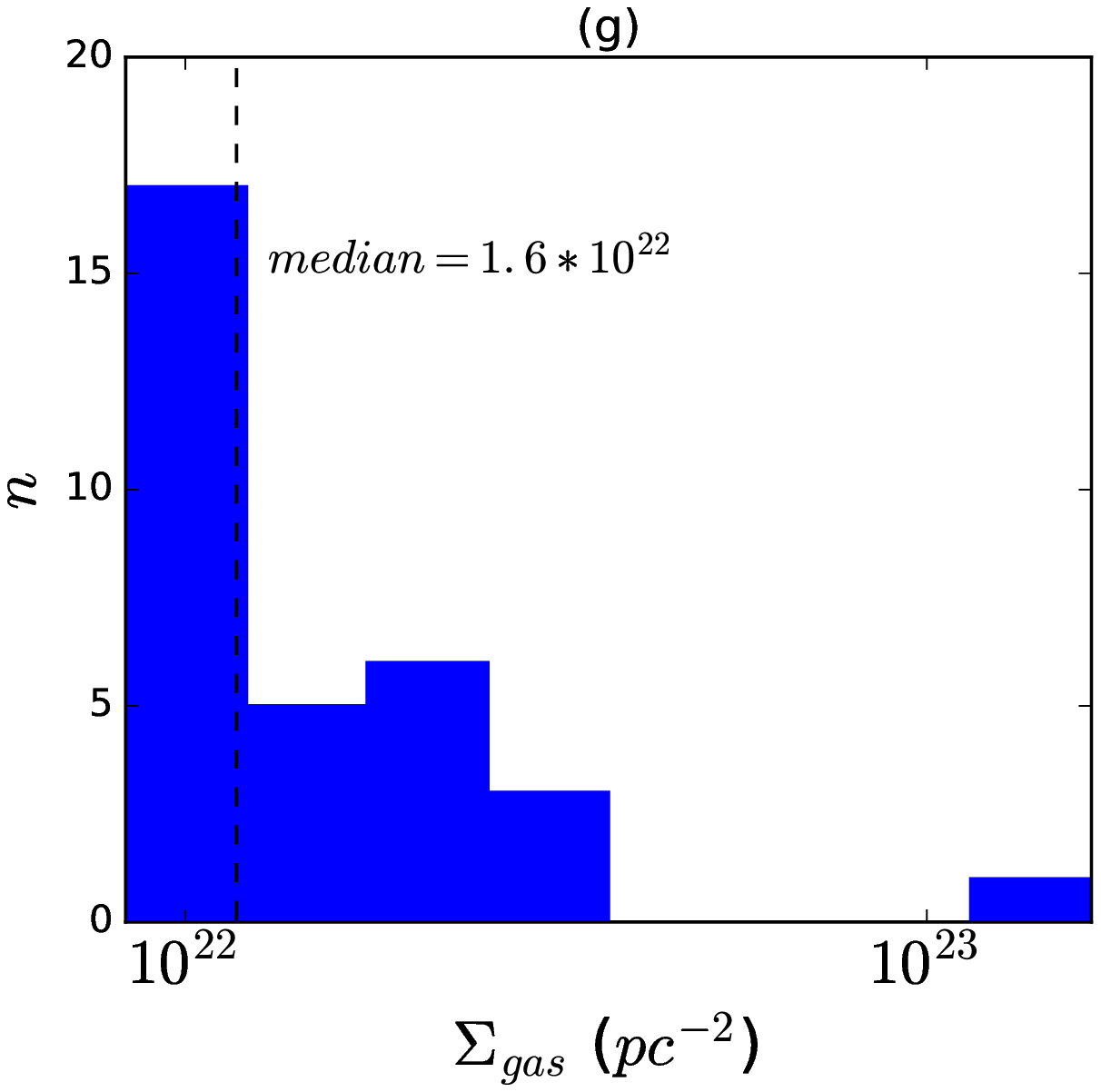}
\includegraphics[scale=0.4]{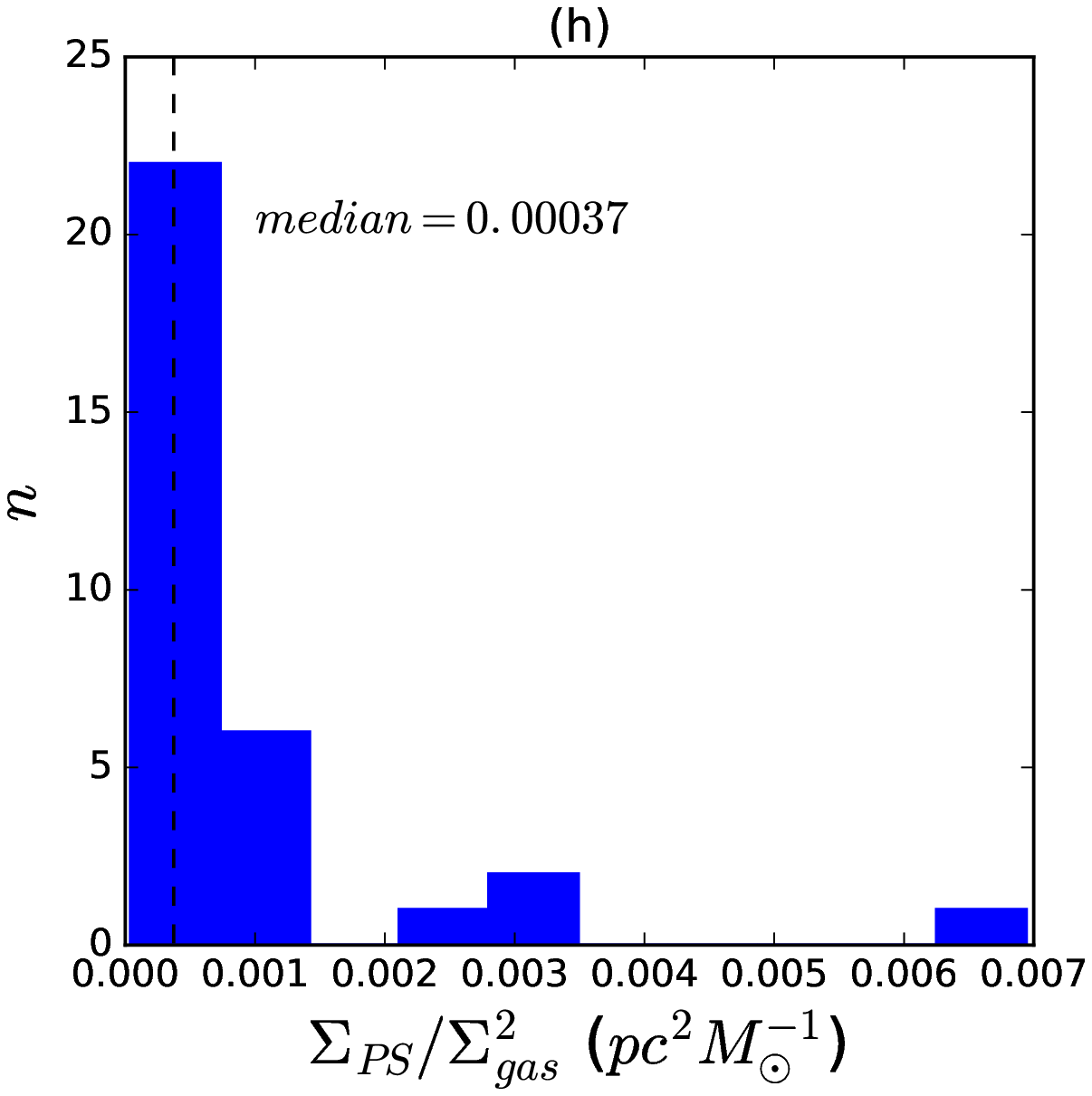}
\includegraphics[scale=0.4]{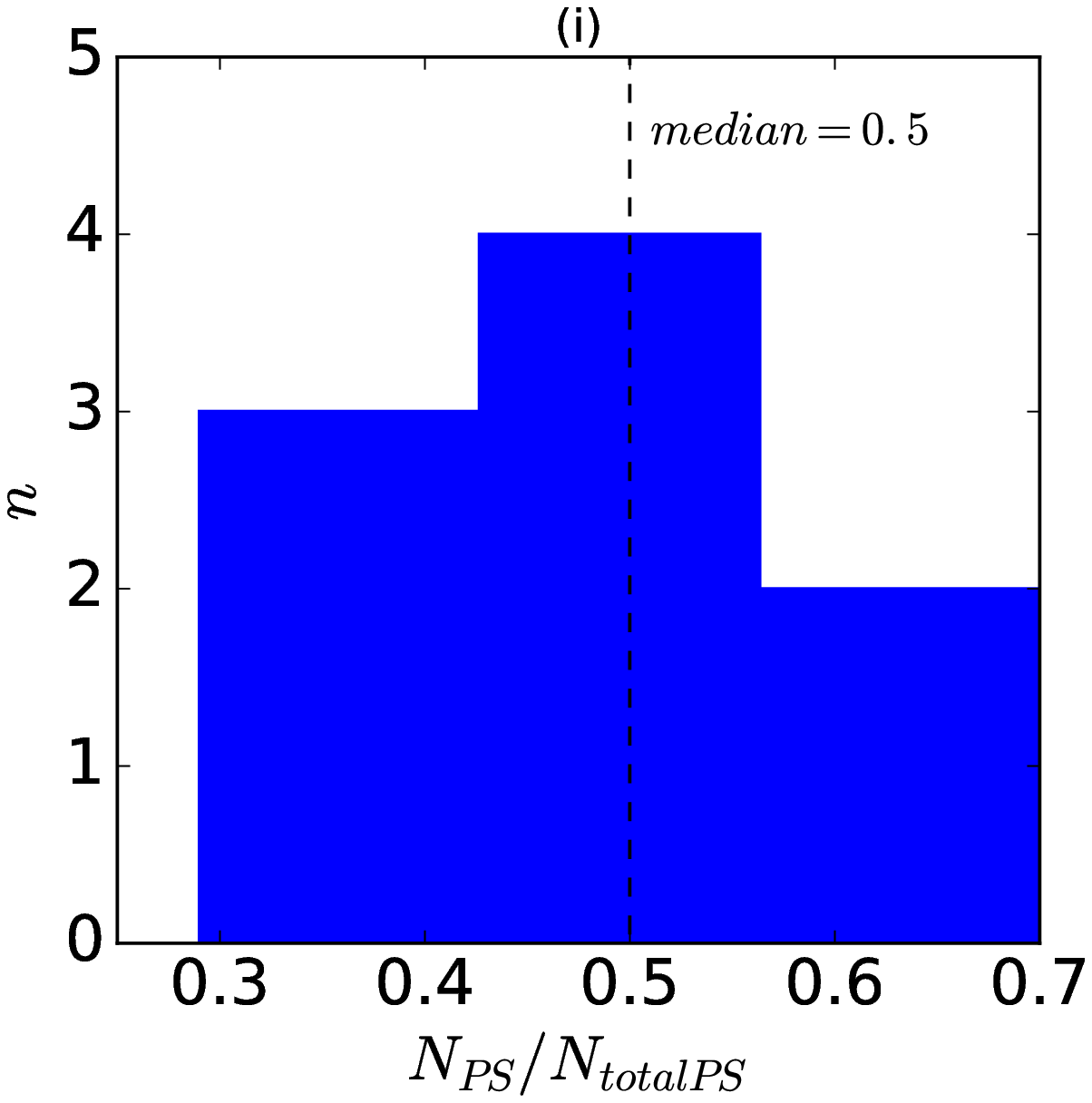}
\caption{Histogram of measured physical properties for the entire sample of 32 groups with high PS fraction. (a) Effective radius, (b) PS counts, (c) PS surface density, (d) PMS surface density, (e) ratio of PS stars ($f_{PS}$), (f) age, (g) average column density of gas, (h) $\frac{\Sigma_{PS}}{\Sigma_{gas}^2}$, (i) $N_{PS}/N_{totalPS}$.The median values are indicated by the vertical dashed line in each panel. The major outliers on the far right end of (b), (c), (d), (f), (g) and (h) are NGC 1333, Serpens-b3, Aquila-b5, Auriga/CMC-3, Aquila-b3, and Aquila-b6, respectively. } 
\end{figure}

\begin{figure}[htbp]
   \centering
\includegraphics[scale=0.45]{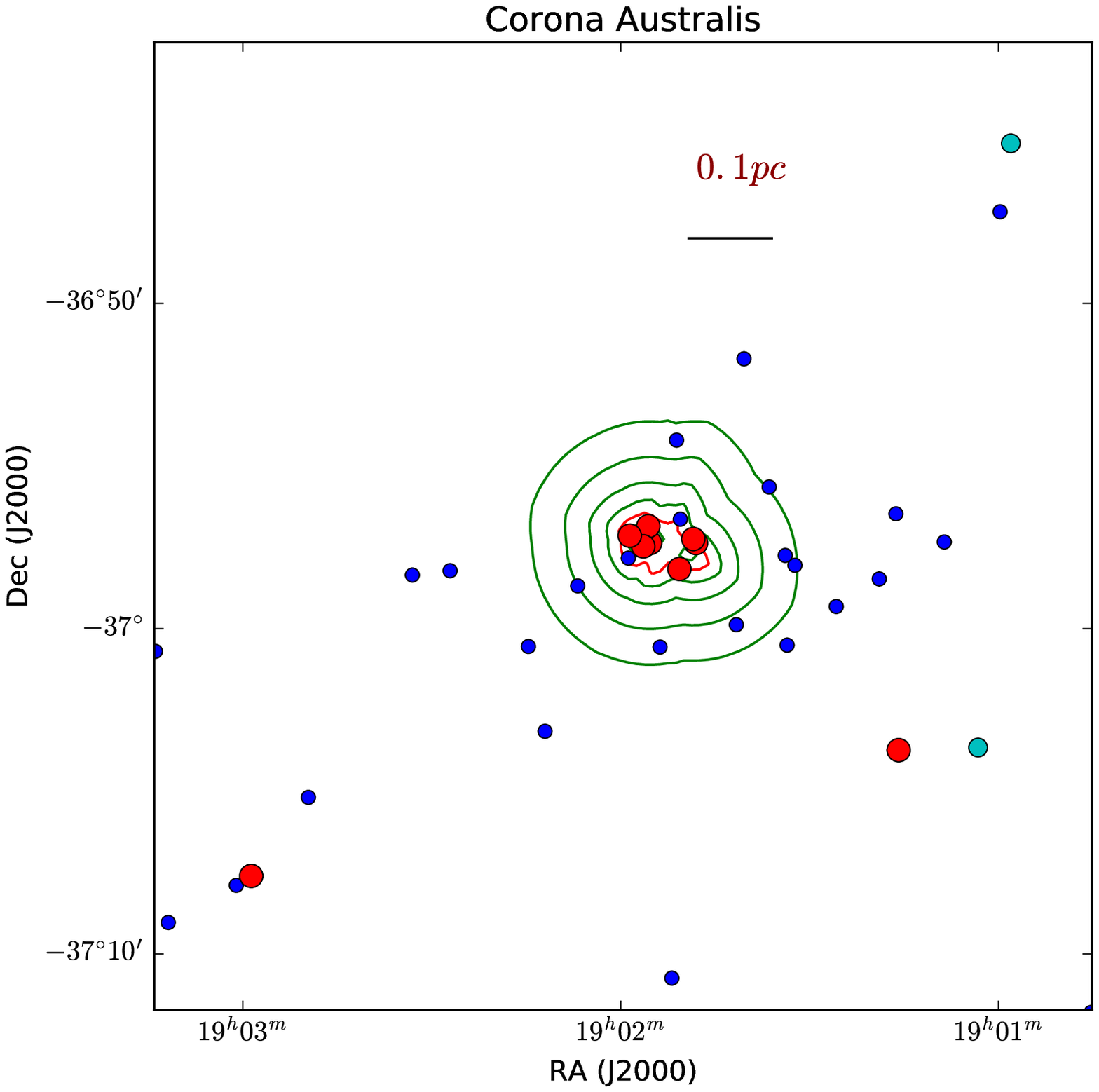}
\includegraphics[scale=0.45]{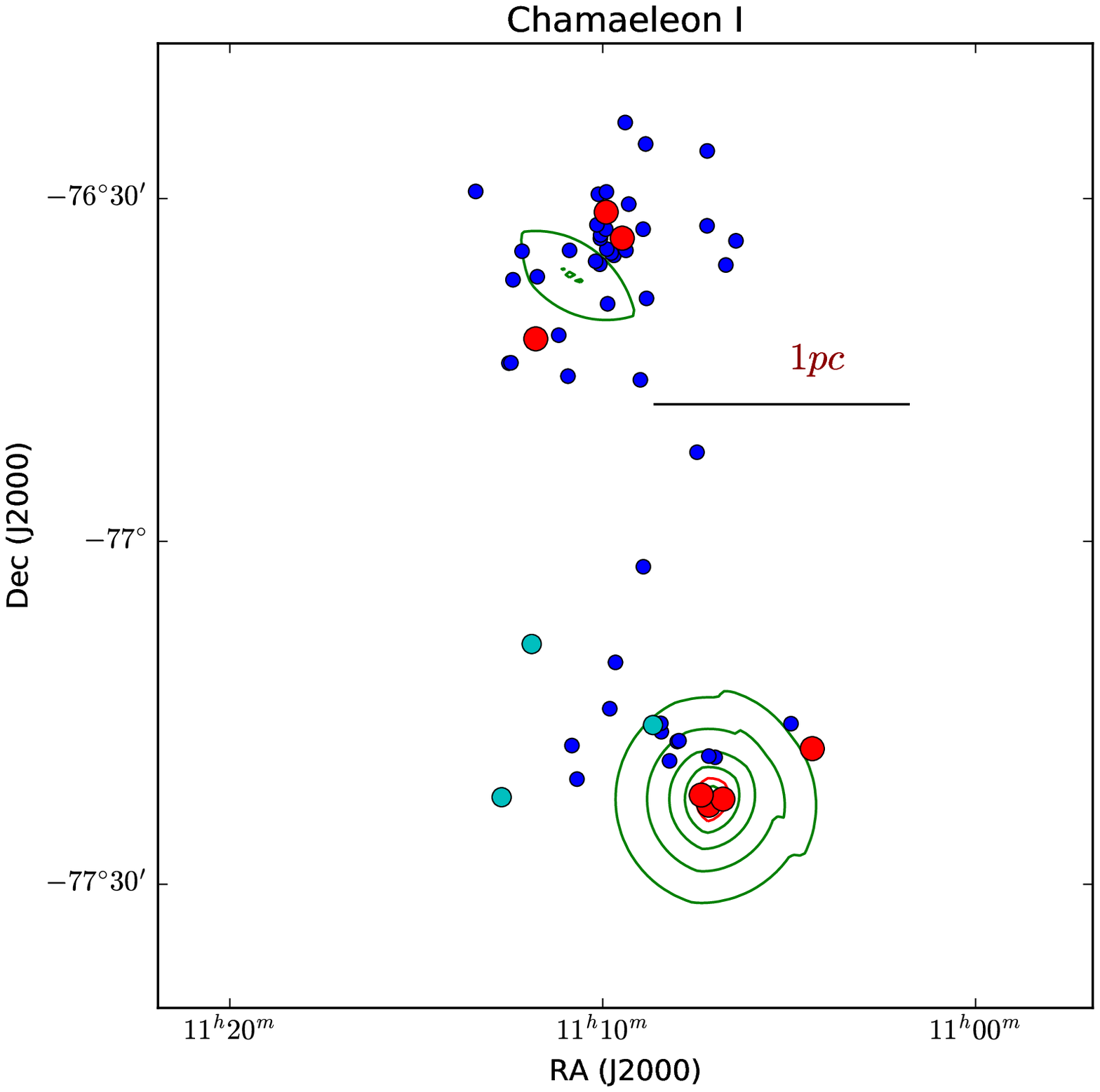}
\caption{Maps of the Corona Australis (left) and Chameleon I (right) for $N=3$ (right). Contours represent the PS surface density, shown at  1.25\%, 2.5\%, 5\%, 10\%, 20\%, 40\%, and 80\% of peak value.} 
\end{figure}

\begin{figure}[]
\begin{center}
\includegraphics[width=4.0in]{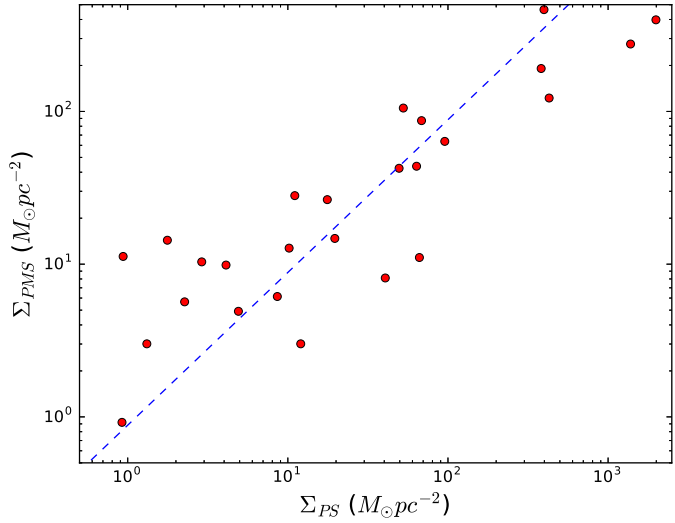}
\vspace*{0.5 cm} \caption{\label{pspms} Surface densities of PS objects versus PMS stars. The blue dashed line represents the best fit to the data: $\Sigma_{PMS}= (0.236\pm0.001) \Sigma_{PS}$ ($r_{corr}=0.78$). }
\end{center}
\end{figure}

\begin{figure}[]
\begin{center}
\includegraphics[width=4.0in]{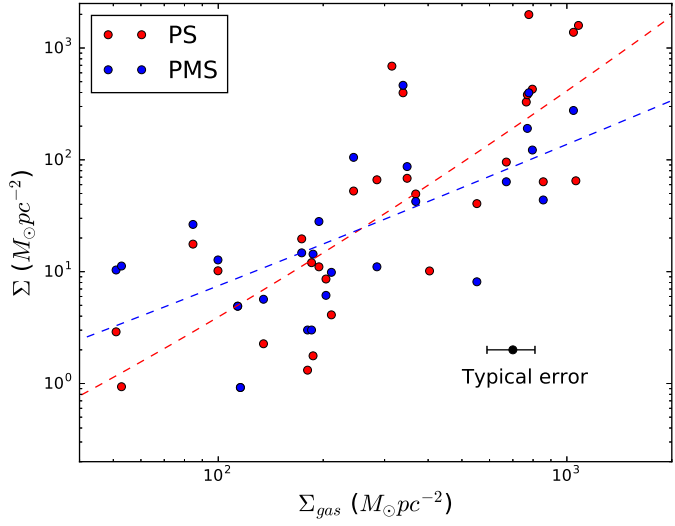}
\vspace*{0.5 cm} \caption{\label{ysogas} Surface densities of PS and PMS versus gas column densities. The red circles indicate PS objects, while the red dashed line indicates the best fit to the PS data: $log \Sigma_{PS}= log \Sigma_{gas}^ {1.40\pm0.01}- (2.05\pm0.12)$ ($r_{corr}=0.82$). The blue circles indicate PMS, while the blue dashed line indicates the best fit to the PMS data: $log \Sigma_{PMS}= log \Sigma_{gas}^ {1.13\pm0.02}- (1.31\pm0.12)$ ($r_{corr}=0.66$). }
\end{center}
\end{figure}

\begin{figure}[]
\begin{center}
\includegraphics[width=4.0in]{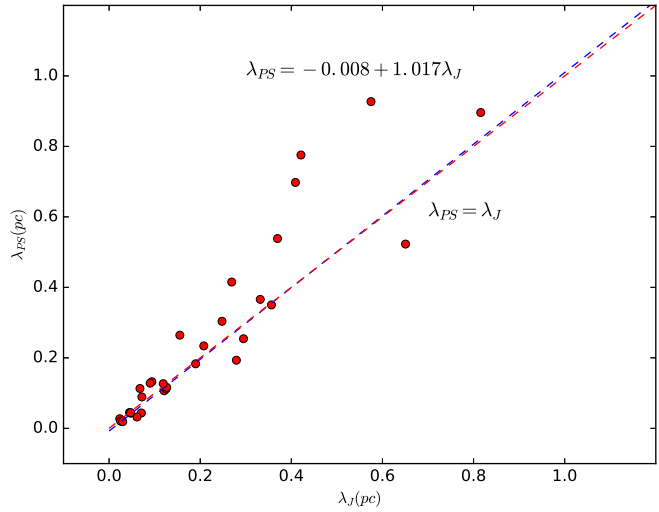}
\vspace*{0.5 cm} \caption{\label{psgas} Mean spacing of PS objects versus Jeans length. The blue dashed line represents the best fit to the data: $\lambda_{PS}=(-0.008\pm0.001)+(1.017\pm0.007)\lambda$ ($r_{corr}=0.92$). The red solid line represents $\lambda_{PS}=\lambda_J$. }
\end{center}
\end{figure}

\begin{figure}[]
\begin{center}
\includegraphics[width=4.0in]{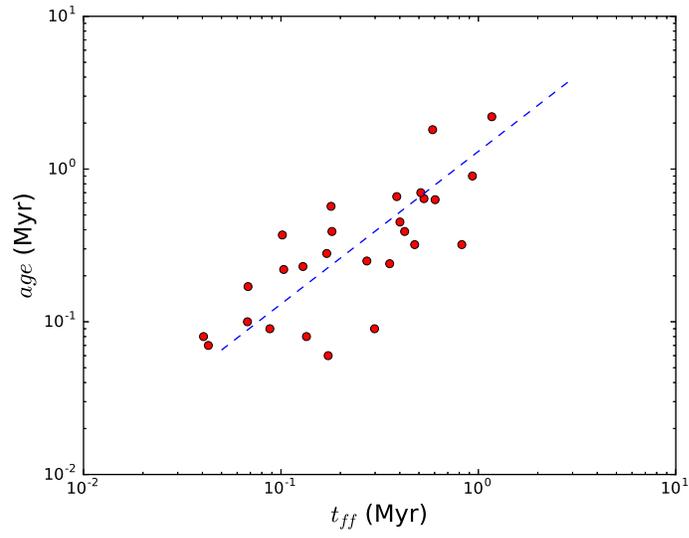}
\vspace*{0.5 cm} \caption{\label{psgas} Age of PS groups versus the gas free fall time. The blue dashed line represents the best fit to the data: $age= (1.31\pm0.02) \tau_{ff}$ ($r_{corr}=0.78$). }
\end{center}
\end{figure}

%%%%%%%%%%%%%%%%%%%%%%%%%%%%%%%%%%%%%%%%%%%%%%%%

\end{document}